\documentclass[10pt,aps,prc,floatfix,twocolumn,nofootinbib, superscriptaddress,preprintnumbers]{revtex4-1}

\usepackage{amsfonts,amsmath,amssymb,bm, xspace}
\usepackage{color,graphicx}
\usepackage{bbm}
\usepackage{dcolumn}                 
\graphicspath{{./figures/}}
\usepackage{graphicx}
\usepackage{float}
\usepackage[dvipsnames]{xcolor}
\usepackage{bbold}
\usepackage{enumitem}
\usepackage{subcaption} 
\usepackage{caption} 
\usepackage[colorlinks=true,
linkcolor=blue,anchorcolor=blue,citecolor=blue,
hypertexnames=false]{hyperref}                

\usepackage{array,physics}
\usepackage[normalem]{ulem}
\usepackage{tikz}
\usepackage{booktabs}
\usepackage[abs]{overpic} 
\newcolumntype{d}[1]{D{.}{.}{#1}}

\usepackage{xcolor}

\newcommand{\Msol}{\,\text{M}_{\odot}\xspace}

\newcommand{\MeV}{\, \text{MeV}}

\newcommand{\nsat}{\,n_{\rm sat}}
\newcommand{\gap}{\Delta_{\rm CFL}}

\renewcommand{\max}{\textrm{max}}

\newcommand{\datumi}{\mathcal{D}_i}
\newcommand{\prob}{\mathcal{P}}

\hypersetup{%
    pdfsubject=Paper,
    pdfkeywords={nuclear physics} 
    unicode=true,
    breaklinks=true,
     colorlinks   = true,
     linkcolor = blue,
     citecolor = blue,
     menucolor = blue,
     urlcolor = blue
}

\begin{document}

\title{Is the coexistence of strange quark stars and hadronic stars\\ favored by astrophysical data? A Bayesian analysis}

\author{Luca Passarella}
\email{luca\_passarella@polito.it}
\affiliation{Department of Applied Science and Technology, Politecnico di Torino, I-10129, Torino, Italy}
\affiliation{INFN, Sezione di Torino, I-10126, Torino, Italy}
\author{Mirco Guerrini}
\author{Giuseppe Pagliara}
\affiliation{Department of Physics and Earth Science, University of Ferrara, via Saragat 1, I-44122 Ferrara, Italy}
\affiliation{INFN, Sezione di Ferrara, via Saragat 1, I-44122 Ferrara, Italy}

\author{Andrea Lavagno}
\affiliation{Department of Applied Science and Technology, Politecnico di Torino, I-10129, Torino, Italy}
\affiliation{INFN, Sezione di Torino, I-10126, Torino, Italy}
\author{Alessandro Drago}


\affiliation{Department of Physics and Earth Science, University of Ferrara, via Saragat 1, I-44122 Ferrara, Italy}
\affiliation{INFN, Sezione di Ferrara, via Saragat 1, I-44122 Ferrara, Italy}
\date{\today}

\begin{abstract}
Hadronic stars and strange quark stars could coexist within the so-called two-families scenario. In this respect, hadronic matter and strange quark matter correspond to two distinct equilibrium phases described by two different equations of state. We perform here the first detailed Bayesian analysis that makes use of astrophysical and laboratory data in order to constrain the equations of state adopted within the two-families scenario for hadronic and strange quark matter. In particular, in hadronic matter we consider the possible formation of hyperons and delta resonances (beside nucleons) within a class of non linear relativistic mean field models and in quark matter we consider the possible formation of a color-superconducting phase within a bag-like model. Results of the analysis indicate that while at the moment both the one-family and the two-families scenarios are compatible with the data, by comparing the Bayes factors of both models, the two-families scenario is favored with respect to the one-family scenario. Specifically, the two-families framework naturally relieves the tension between the intermediate-density softness of the equation of state required by small-radius objects and the high-density stiffness needed to support massive pulsars. Ultimately, future detections of even more massive compact objects, very compact ordinary-mass objects, or precise measurements of two distinct masses with the same radius, will provide strong indications in favor of the two-families scenario.

\end{abstract}

\maketitle 
\section{Introduction}
Determining the equation of state (EOS) of strongly interacting dense matter is one of the central open problems in nuclear physics and astrophysics~\cite{Burgio:2021vgk,MUSES:2023hyz}. 
Different regions of the quantum chromodynamics (QCD) phase diagram are accessible to complementary theoretical and experimental probes: lattice QCD is predictive at high temperature and small baryon chemical potential~\cite{Borsanyi:2020fev,Borsanyi:2021sxv}, perturbative QCD applies at asymptotically large baryon chemical potential or temperature~\cite{Kurkela:2014vha,Komoltsev:2021jzg}, chiral effective field theory ($\chi$EFT) constrains cold matter up to around nuclear saturation density $\nsat$~\cite{Drischler:2017wtt,Drischler:2021kxf}, low-energy nuclear experiments probe matter in the vicinity of $\nsat$~\cite{Lattimer:2012xj,Lattimer:2023rpe}, and heavy-ion collisions explore matter at intermediate baryon chemical potential and finite temperature~\cite{Danielewicz:2002pu,Sorensen:2023zkk}. None of these tools, however, directly probe the cold regime at a few times $\nsat$ that governs the bulk of compact stars.

Compact stars therefore provide the only laboratory in which this intermediate-density window can be explored~\cite{Lattimer:2012nd,Oertel:2016bki}. The advent of multimessenger astronomy has turned them into quantitative probes of dense matter: the tidal-deformability constraints extracted from the binary neutron-star merger gravitational-wave signal GW170817~\cite{LIGOScientific:2018cki}, the mass--radius inferences for the pulsars PSR~J0030$+$0451 and PSR~J0740$+$6620 obtained by NICER~\cite{Miller:2019cac,Riley:2021pdl,Miller:2021qha,Choudhury:2024xbk,Vinciguerra:2023qxq}, and precise radio-timing measurements that have established the existence of compact stars with masses above $2 \Msol$~\cite{Antoniadis:2013pzd,Fonseca:2021wxt,Romani:2022jhd} now place complementary constraints on the pressure--density relation of compact-star matter. At the same time, unusually compact low-mass objects such as the central compact object in HESS~J1731--347 may probe the opposite end of the mass distribution, although their interpretation remains less settled~\cite{Doroshenko:2022nwp,DiClemente:2022wqp,Sagun:2023rzp}.

The microscopic composition of compact stars, however, remains debated. At supranuclear densities, exotic hadronic degrees of freedom are expected to become energetically favored: hyperons can soften the EOS and reduce the maximum mass, leading to the hyperon puzzle~\cite{Chatterjee:2015pua,Vidana:2018bdi,Tolos:2020aln, Lonardoni:2014bwa,Logoteta:2019utx}; $\Delta$ resonances can appear already at a few times $\nsat$ and produce a similar softening~\cite{Drago:2014oja,Li:2018qaw,Ribes:2019kno}. Whether, and at which density, quark matter is deconfined inside compact stars is likewise an open question~\cite{Annala:2019puf,Alford:2013aca,Weissenborn:2011qu,Baym:2017whm}.

A key but often implicit assumption shaping this discussion concerns the true ground state of strongly interacting matter. 
If ordinary nuclear matter is absolutely stable, deconfinement in compact stars leads to the formation of hybrid stars, characterized by a deconfined quark core and a hadronic outer part ~\cite{Han:2019bub,Annala:2019puf,Weissenborn:2011qu,Alford:2013aca,Constantinou:2023ged,Constantinou:2025wxj,Baym:2017whm}. 
The alternative, first formulated by Bodmer and Witten~\cite{Bodmer:1971we,Witten:1984rs}, posits that three-flavour strange quark matter (SQM) is absolutely stable with respect to nuclear matter, so that self-bound strange quark stars (QSs) can exist~\cite{Alcock:1986hz,Madsen:1998uh,Weber:2004kj,Haensel:1986qb}.

This hypothesis underlies the \emph{two-families scenario}~\cite{Drago:2013fsa,Drago:2015cea,Drago:2015dea,Burgio:2018yix}, in which hadronic stars (HSs) and quark stars (QSs) coexist and populate two disconnected branches of compact stars. Since hadronic matter is not the absolute ground state, the hadronic branch is only \emph{metastable} with respect to conversion into QSs. In this picture, hyperons and $\Delta$ resonances are allowed to soften the hadronic EOS, producing compact hadronic stars with relatively small radii and limiting the maximum mass of the metastable HS branch, while a sufficiently stiff SQM EOS supports the heaviest observed compact stars on an independent stable QS branch. The scenario therefore offers a natural way to reconcile the possible coexistence of very compact stars and very massive compact objects, while effectively removing the hyperon puzzle. It has also been invoked to interpret candidates at both extremes of the mass distribution, such as the $\sim(2.5$--$2.67)\,\Msol$ secondary of GW190814 as a very massive QS rather than a black hole~\cite{Bombaci:2020vgw}, and the unusually light and compact object in HESS~J1731--347 as a low-mass QS~\cite{DiClemente:2022wqp}.

The two-families scenario also leads to a characteristic dynamical phenomenology. Quark deconfinement, which triggers the conversion of a HS into a QS~\cite{Drago:2015dea,Drago:2015fpa,Drago:2015cea}, may in principle operate through several evolutionary channels, including the proto-neutron-star phase~\cite{Guerrini:2026_inpreparation}, fallback accretion following core collapse, compact-binary mergers~\cite{DePietri:2019khb}, and later spin-down or mass accretion in binaries~\cite{Becerra:2025ryy}, and it has been connected to gamma-ray bursts and other energetic transients~\cite{Berezhiani:2002ks,Drago:2015fpa,Pagliara:2013tza,Drago:2018jds}. 
In this context, the central issue is not merely whether SQM is energetically favored, but whether the metastable hadronic phase can decay on the relevant evolutionary timescale. The conversion can start only if a critical SQM droplet is nucleated, thereby triggering the subsequent macroscopic transformation of the star~\cite{Guerrini:2025mxx,Drago:2015cea,Bombaci:2016xuj}. In particular, it has long been suggested that a sufficient amount of strangeness must be produced locally in the metastable hadronic phase in order to favor the formation of the first SQM seed and make the conversion possible~\cite{Bombaci:2016xuj,Drago:2015cea,DePietri:2019khb,Bombaci:2020vgw}. Recent developments on metastability and nucleation in the two-families scenario are discussed in~\cite{Guerrini:2026_inpreparation}.

In parallel with these theoretical issues, Bayesian inference has become a standard framework for constraining the compact-star EOS from multimessenger observations~\cite{Cartaxo:2025jpi,Dietrich:2020efo,Huth:2021bsp,Pang:2022rzc,Gorda:2022jvk}. Agnostic and phenomenological analyses have used piecewise-polytropic representations~\cite{Raaijmakers:2019qny,Raaijmakers:2021uju}, speed-of-sound parametrizations~\cite{Raaijmakers:2021uju,Brandes:2023hma}, and non-parametric reconstructions~\cite{Landry:2018prl,Legred:2021hdx}. More physics-informed analyses have instead used relativistic mean-field models~\cite{Traversi:2020aaa,Passarella:2025zqb,Char:2020utj,Malik:2022zol,Huang:2023grj}, non-relativistic Skyrme-type effective interactions~\cite{Beznogov:2024vcv}, and nuclear metamodels based on empirical parameters around saturation~\cite{Margueron:2017eqc,Montefusco:2026jlq}. Within this broader effort, targeted Bayesian studies have examined the role of non-nucleonic degrees of freedom, including hyperons~\cite{Malik:2022jqc,Char:2025nli,Huang:2024rvj}, $\Delta$ resonances~\cite{Parmar:2025csx}, and deconfined quark matter in hybrid stars~\cite{Albino:2024ymc,Albino:2025puc,Ayriyan:2025rub,Shahrbaf:2021cjz,Takatsy:2023xzf,Kopp:2025aez,Pang:2023dqj,Annala:2023cwx,Grundler:2025mcz}.

These analyses provide important constraints on dense matter, but they all assume that ordinary hadronic matter is absolutely stable. Thus, when quark matter is included, it appears as a deconfined core inside hybrid stars, which may form either a single connected sequence or, for sufficiently strong first-order transitions, disconnected twin-star branches (in which, however, all branches still arise from the same equilibrium EOS).
Bayesian analyses of self-bound QSs, and therefore of absolutely stable SQM, have also been performed~\cite{Li:2020wbw,Miao:2021nuq,Traversi:2021fad}. These works, however, constrain the QS EOS under the assumption that the relevant observed sources are QSs.
On the other hand, the two-families scenario has not yet been explored within a Bayesian framework, and this is the goal of the present work. We aim to constrain simultaneously the parameters of the metastable hadronic EOS and of the absolutely stable SQM EOS, while testing whether present data are better described by a single hadronic population or by a mixed population of HSs and QSs.

We contrast two physically distinct scenarios on the same footing.
In the \emph{one-family} (1F) scenario, all observed compact stars are described as hadronic stars belonging to a single relativistic mean-field (RMF) EOS including nucleons, hyperons, and $\Delta$ resonances.
In the \emph{two-families} (2F) scenario, the same metastable hadronic sector coexists with a self-bound SQM branch, described by a color-flavor-locked (CFL) bag-model EOS satisfying the absolute-stability condition for strange quark matter.
The two hypotheses are explored within the same Bayesian framework, using up to 14 EOS parameters: 11 parameters for the hadronic sector and, in the 2F case, three additional parameters for the SQM sector. We combine astrophysical constraints from electromagnetic and gravitational-wave observations, laboratory information from heavy-ion collision experiments, and low-density theoretical constraints from ab-initio calculations.
We emphasize that hybrid stars are not included as an additional 1F hypothesis in the present analysis and are left to future work.

The paper is organized as follows. Section~\ref{sec:EOS} introduces the hadronic RMF model with hyperons and $\Delta$ resonances and the CFL bag-model description of SQM. Section~\ref{sec:method} presents the Bayesian framework, including the parameter space and priors, the likelihoods built from electromagnetic, radio, gravitational-wave, $\chi$EFT, and heavy-ion constraints, the marginalization over family labels in the 2F scenario, and the numerical setup. Section~\ref{sec:results} discusses the posterior constraints on the EOS parameters, the stellar properties, and the Bayes factor comparing the 1F and 2F hypotheses. Section~\ref{sec:conclusions} summarizes our main findings and their implications. Technical details on the EOS and likelihood construction are collected in the Appendices.

We will use natural units $\hbar=c=1$, $G=1$ and $k_B=1$.

\section{Equations of state}
\label{sec:EOS}
In this work we compare two physical scenarios for compact stars: the one-family (1F) scenario, in which all observed compact stars are hadronic HSs, and the two-families (2F) scenario, in which hadronic HSs and QSs coexist as two disconnected branches. 
The two scenarios require, respectively, a single hadronic EOS and a pair of independent EOSs, one describing the metastable hadronic matter and one for absolutely stable SQM. 
This section introduces the two models adopted for computing both EOSs. The explicit field equations, the equations for the thermodynamical variables and the mapping between the couplings in the Lagrangian and the nuclear empirical parameters (NEP) are collected in
Appendix~\ref{app:EOS_details}.

\subsection{Hadronic sector}
\label{sec:EOS_had}

For the hadronic phase we adopt a non-linear relativistic mean-field (RMF) model~\cite{Glendenning:1991es}, in which baryons interact through the exchange of the $\sigma$, $\omega$, $\rho$, and $\phi$ mesons.
The Lagrangian density reads:
\begin{widetext}
\begin{eqnarray}
\mathcal{L}_H &=& \sum_{b} \bar{\psi}_{b}
\Big[
i\gamma_\mu\partial^\mu
- m_b
+ g_{\sigma b}\sigma
- g_{\omega b}\gamma_\mu\omega^\mu
- g_{\phi b}\gamma_\mu\phi^\mu
- g_{\rho b}(n_B)\,\gamma_\mu \boldsymbol{I}_b\!\cdot\!\boldsymbol{\rho}^{\,\mu}
\Big]\psi_{b}
\nonumber\\
&&+
\sum_{d}
\bar{\psi}_{d\nu}
\Big[
i\gamma_\mu\partial^\mu
- m_d
+ g_{\sigma d}\sigma
- g_{\omega d}\gamma_\mu\omega^\mu
- g_{\rho d}(n_B)\,\gamma_\mu \boldsymbol{I}_d\!\cdot\!\boldsymbol{\rho}^{\,\mu}
\Big]\psi_d^\nu
\nonumber\\
&&+
\frac{1}{2}
\left(
\partial_\mu\sigma\,\partial^\mu\sigma
- m_\sigma^2\sigma^2
\right)
-\frac{1}{3}\,b\,m_N(g_{\sigma N}\sigma)^3
-\frac{1}{4}\,c\,(g_{\sigma N}\sigma)^4
\nonumber\\
&&-
\frac{1}{4}\,\omega_{\mu\nu}\omega^{\mu\nu}
+\frac{1}{2}\,m_\omega^2\,\omega_\mu\omega^\mu
-\frac{1}{4}\,\boldsymbol{\rho}_{\mu\nu}\!\cdot\!\boldsymbol{\rho}^{\mu\nu}
+\frac{1}{2}\,m_\rho^2\,\boldsymbol{\rho}_\mu\!\cdot\!\boldsymbol{\rho}^\mu
\nonumber\\
&&-
\frac{1}{4}\,\phi_{\mu\nu}\phi^{\mu\nu}
+\frac{1}{2}\,m_\phi^2\,\phi_\mu\phi^\mu \,,
\label{eq:Lagrangian}
\end{eqnarray}
\end{widetext}
where $b\in\{n,p,\Lambda,\Sigma^{+},\Sigma^{0},\Sigma^{-},\Xi^{0},\Xi^{-}\}$ runs over the baryon octet and $d\in\{\Delta^{++},\Delta^{+},\Delta^{0},\Delta^{-}\}$ over the $\Delta$ quartet. The spin-$1/2$ baryons are described by Dirac spinors $\psi_b$, while the spin-$3/2$ $\Delta$ resonances are described by Rarita--Schwinger spinors $\psi_d^\nu$~\cite{Drago:2013fsa,Drago:2014oja}.
The operators $\boldsymbol{I}_b$ and $\boldsymbol{I}_d$ denote the isospin generators in the
appropriate representation for each baryon species. In homogeneous matter and within the mean-field approximation only the time-like third component of the $\rho$ field survives, so that the interaction reduces to the standard form which is proportional to $I_{3i}\rho_3^0$. The $\phi$ meson, being an $\bar{s}s$ state, couples only to strange baryons, therefore $g_{\phi b}=0$ for nucleons and $\Delta$ resonances.

The mesonic sector contains the kinetic and mass terms of the $\sigma$, $\omega$, $\rho$, and $\phi$ fields, together with the non-linear self-interactions of the $\sigma$ meson. The latter are controlled by the parameters $b$ and $c$ and are introduced to reproduce the empirical incompressibility of symmetric nuclear matter. Note that Eq.~\eqref{eq:Lagrangian} builds on the nucleonic RMF framework adopted in~\cite{Passarella:2025zqb} and extends it to include hyperons and $\Delta$ resonances, following~\cite{Drago:2013fsa,Drago:2014oja}.

A key feature of the parametrization is the density dependence of the hadrons-$\rho$ coupling, which is introduced to control the slope of the symmetry energy without affecting the isoscalar sector at saturation~\cite{Typel:1999yq,Typel:2005ba,Typel:2009sy,Drago:2014oja,Drago:2015cea}:
\begin{equation}
g_{\rho i}(n_B)=x_{\rho i}\,g_{\rho N}(\nsat)\,
\exp\!\left[-a_\rho\!\left(\frac{n_B}{\nsat}-1\right)\right],
\label{eq:grho}
\end{equation}
where $i$ runs over the baryon octet and delta resonances, $x_{\rho i}\equiv g_{\rho i}/g_{\rho N}$, $n_B$ is the baryon number density and $\nsat$ is the nuclear saturation density.
In the absence of this density dependence, the model predicts values of the symmetry-energy slope parameter $L_\mathrm{sym}$ of order $90$ MeV~\cite{Drago:2014oja}, significantly larger than the empirically favored range $40\lesssim L_\mathrm{sym}\lesssim 62$ MeV~\cite{Lattimer:2012xj,Lattimer:2023rpe}. The additional parameter $a_\rho$ therefore provides a minimal and efficient way to tune $L_\mathrm{sym}$ while preserving the remaining isoscalar saturation properties. 

The choice of this hadronic framework is motivated by two considerations. 
First, its parameterization can be expressed in terms of standard nuclear empirical parameters at saturation, making it convenient for Bayesian inference, since physically motivated priors can be assigned directly to quantities with a clear interpretation in nuclear physics.
Second, the model is flexible enough to span the range of stiffness allowed by present nuclear and astrophysical constraints, while remaining simple enough to be used in repeated EOS evaluations as required in Bayesian analyses.

The inclusion of the $\Delta$ quartet is especially relevant in this context. Once $L_\mathrm{sym}$ is restricted to the empirically favored range, the threshold density of the $\Delta^-$ decreases to $\sim (2$--$3)\nsat$, namely the same density range in which hyperons are expected to appear, and the $\Delta^-$ may even occur before the $\Lambda$~\cite{Drago:2014oja}.
The early onset of $\Delta$ resonances softens the hadronic EOS at intermediate densities and leads to small stellar radii, a feature that is particularly important for the hadronic branch in the two-families scenario~\cite{Drago:2013fsa,Drago:2015cea}.

The hadronic EOS is fully specified by six nuclear empirical parameters defined at saturation in symmetric nuclear matter: the saturation density $\nsat$, the binding energy per baryon $E_\mathrm{sat}$, the incompressibility $K_\mathrm{sat}$, the symmetry energy $E_\mathrm{sym}$, its slope $L_\mathrm{sym}$, and the Dirac effective mass ratio $m^*/m$. In the nucleonic sector these quantities are in a one-to-one correspondence with the six independent couplings $(g_{\sigma N},g_{\omega N},g_{\rho N}(\nsat),b,c,a_\rho)$~\cite{Passarella:2025zqb}.

The hyperon couplings to the vector mesons are fixed according to SU(6) symmetry relations (see e.g.~\cite{Fortin:2017dsj}), while the scalar hyperon couplings are determined from the phenomenological hyperon optical potentials in symmetric nuclear matter at saturation, $U_{\Lambda}$, $U_\Sigma$, and $U_\Xi$. For the $\Delta$ sector, the scalar and vector coupling ratios
\begin{equation}
x_{\sigma\Delta}\equiv \frac{g_{\sigma\Delta}}{g_{\sigma N}},
\qquad
x_{\omega\Delta}\equiv \frac{g_{\omega\Delta}}{g_{\omega N}},
\end{equation}
are treated as free parameters in the inference, while
\begin{equation}
x_{\rho\Delta}\equiv \frac{g_{\rho\Delta}}{g_{\rho N}}=1
\end{equation}
is kept fixed throughout.

For each parameter set we compute the hadronic EOS under the relevant physical conditions: for stellar-structure calculations, we use cold charge-neutral matter in weak equilibrium, including electrons and muons as free relativistic Fermi gases. The uniform-matter core EOS is matched at $n_B\simeq 0.08~\mathrm{fm}^{-3}$ to the Baym--Pethick--Sutherland crust~\cite{Baym:1971pw}. For the comparison with low-density nuclear theory and heavy-ion data, we instead compute pure neutron matter and symmetric nuclear matter with zero strangeness, no leptons, and fixed electric charge fraction $Y_C\equiv n_C/n_B=0$ and $Y_C=0.5$, respectively. Additional technical details are reported in Appendix~\ref{app:EOS_details}.


\subsection{Quark sector}
\label{sec:EOS_quark}

In the two-families scenario, the second branch of compact stars is identified with self-bound QSs composed of absolutely stable SQM. At asymptotically large baryon density and low temperature, QCD predicts color superconducting color-flavor-locked (CFL) phase \cite{Alford:2007xm}. At the densities relevant for compact-star interiors, however, the structure of the phase diagram is unknown. Nevertheless, since pairing lowers the free energy of three-flavor quark matter, a color-superconducting SQM phase is a natural candidate for the absolutely stable state required by the Bodmer--Witten hypothesis.

In this work we describe the deconfined phase by means of the phenomenological CFL bag-model EOS presented in \cite{Alford:2007xm,Alford:2001zr,Weissenborn:2011qu} and used, e.g. in~\cite{Bombaci:2020vgw,Becerra:2025ryy}.
Using the average quark chemical potential $\mu_q\equiv\mu_B/3$, the grand thermodynamic potential reads
\begin{equation}
\Omega_Q(\mu_q)=
-\frac{3a_4}{4\pi^2}\mu_q^4
+\frac{3}{4\pi^2}\left(m_s^2-4\gap^2\right)\mu_q^2
+B.
\label{eq:OmegaCFL_expanded}
\end{equation}
Here $m_s$ is the strange-quark mass, $\gap$ is the CFL pairing gap, $B$ is an effective bag constant accounting for the energy density difference between the non–perturbative-QCD vacuum and the perturbative one, and $a_4$ parametrizes the leading perturbative-QCD correction to the free fermions term. 

The thermodynamic quantities follow from Eq.~\eqref{eq:OmegaCFL_expanded} by standard thermodynamic identities and are reported in Appendix~\ref{app:EOS_details}.

To be compatible with the Bodmer--Witten hypothesis, the quark EOS must satisfy the standard stability window at zero pressure. On the one hand, bulk three-flavor quark matter must be absolutely stable, so that SQM is energetically favored with respect to ordinary nuclei. On the other hand, bulk two-flavor quark matter must not be absolutely stable, which ensures that ordinary nuclei do not dissolve into deconfined up-down quark matter. We stress that this second condition refers to non-strange two-flavor quark matter, which cannot be in the CFL phase: CFL pairing requires the simultaneous participation of $u$, $d$, and $s$ quarks and, in bulk, implies equal quark number densities and no electrons.

The pairing gap lowers the grand potential and therefore makes SQM more stable. This is an important feature in the present context. The parameters $B$ and $\gap$ play different and complementary roles in the stellar sequence: $B$ mainly sets the zero-pressure point and the characteristic self-bound energy density, thus affecting the overall compactness of the star, while $\gap$ increases the pressure through the pairing contribution and is especially effective in raising the maximum mass. By varying $B$ and $\gap$ simultaneously, one can tune both the mass and radius of the QS sequence. In the standard unpaired bag model this additional freedom is absent. 
The parameter $a_4$ provides a further correction to the stiffness of the EOS in the $p-\mu_B$ plane, but values that are too small may lead to rehadronization, namely to a regime in which hadronic matter becomes energetically favored over SQM at high density. 

In the present work we fix $m_s=100~\mathrm{MeV}$, while $a_4$, $B$, and $\gap$ are treated as free parameters in the Bayesian analysis.

For each parameter set, we compute stellar equilibrium sequences by solving the TOV equations for a cold, $\beta$-equilibrated bare QS, namely a self-bound configuration composed entirely of homogeneous quark matter at $T=0$. In principle, QSs may possess a crust, such as an outer crystalline color-superconducting layer~\cite{Alford:2007xm,Anglani:2013gfu}. Since our goal here is to isolate the impact of the bulk quark EOS and to keep the model minimal, we neglect crustal structures in the baseline stellar calculations.
Furthermore, when evaluating the tidal deformability, we incorporate the required boundary correction at the star's surface to properly account for the sharp energy density discontinuity characteristic of bare configurations, following the prescription described in ~\cite{Han:2018mtj,Takatsy:2020bnx}.

\section{Method}
\label{sec:method}

We aim to infer the posterior distribution of the EOS parameters within a Bayesian framework, considering two alternative hypotheses for the compact-star population: the one-family (1F) and two-families (2F) scenarios. According to Bayes' theorem, the posterior probability distribution of the parameter set $\theta$, given the data set $\{\datumi\}$ and conditioned on the hypothesis $\mathcal{H}\in\{\mathrm{1F},\mathrm{2F}\}$, is
\begin{equation}
\prob\!\left(\theta\mid \{\datumi\},\mathcal{H}\right)=\frac{\mathcal{L}^{(\mathcal{H})}\!\left(\theta\right)\,\pi^{(\mathcal{H})}(\theta)}{\mathcal{Z}^{(\mathcal{H})}\!\left(\{\datumi\}\right)} ,
\label{eq:bayes_theorem}
\end{equation}
where
\begin{equation}
\mathcal{L}^{(\mathcal{H})}\!\left(\theta\right)\equiv \prob\!\left(\{\datumi\}\mid \theta,\mathcal{H}\right)
\end{equation}
is the likelihood, $\pi^{(\mathcal{H})}(\theta)\equiv \prob (\theta\mid \mathcal{H})$ is the prior on the EOS parameters, and
\begin{equation}
\mathcal{Z}^{(\mathcal{H})}\!\left(\{\datumi\}\right)\equiv \int d\theta\;\mathcal{L}^{(\mathcal{H})}\!\left(\theta\right)\,\pi^{(\mathcal{H})}(\theta)
\label{eq:Z}
\end{equation}
is the Bayesian evidence. The evidence plays no role in parameter estimation at given $\mathcal{H}$, but it is a key quantity for model comparison between the 1F and 2F hypotheses.

Assuming that the observational data $\datumi$ are statistically independent, the total likelihood factorizes as:
\begin{equation}
\mathcal{L}^{(\mathcal{H})}\!\left(\theta\right)=\prod_i \mathcal{L}^{(\mathcal{H})}_i(\theta),
\label{eq:likelihood_factorization}
\end{equation}
where
\begin{equation}
\mathcal{L}_i^{(\mathcal{H})}(\theta)\equiv \prob\!\left(\datumi\mid \theta,\mathcal{H}\right)
\end{equation}
is the contribution of the datum $\datumi$. In practice we evaluate the log-likelihood,
\begin{equation}
\ln \mathcal{L}^{(\mathcal{H})}\!\left(\theta\right)=\sum_i \ln \mathcal{L}_i^{(\mathcal{H})}(\theta),
\end{equation}
which is numerically more stable than the direct product of many small likelihood factors (see Appendix~\ref{app:numerics}).

For each parameter set $\theta$\footnote{Throughout the paper, the same symbol $\theta$ denotes the full parameter vector of the hypothesis under consideration: $\theta=\theta_H$ in the 1F case and $\theta=(\theta_H,\theta_Q)$ in the 2F case.}, solving the TOV equations yields the theoretical mass--radius and mass--tidal-deformability relations for all stable stellar branches
\begin{equation}
R_f^{\rm TOV}(M;\theta), \qquad \Lambda_f^{\rm TOV}(M;\theta),
\end{equation}
where $f=\mathrm{HS},\mathrm{QS}$ labels the stellar family. In the 1F scenario only the hadronic branch is present, whereas in the 2F scenario both HS and QS branches may contribute. 

The purpose of the analysis is therefore twofold: first, to infer the EOS parameters favored by the data within each hypothesis; second, to determine whether the data prefer the 1F or 2F scenario through the corresponding Bayesian evidences.

\subsection{Parameter space and priors}
\label{sec:priors_theta}

It is convenient to separate the hadronic and the quark parameter spaces as follows:
\begin{align}
\theta_H \equiv \{&\nsat, E_{\rm sat}, K_{\rm sat}, E_{\rm sym}, L_{\rm sym}, m^*/m, \\&U_{\Lambda}, U_{\Sigma}, U_{\Xi}, x_{\sigma\Delta}, x_{\omega\Delta}\}\notag,
\end{align}
and
\begin{equation}
\theta_Q \equiv \{a_4,B,\gap\}.
\end{equation}
The full parameters vector depends on the hypothesis under consideration.
In the 1F scenario it reduces to the hadronic sector alone, $\theta=\theta_H$.
In the 2F scenario it includes both the hadronic and quark parameters, $\theta=(\theta_H,\theta_Q)$.

For our analysis we adopt broad uniform priors on the hadronic parameters.
The prior ranges are chosen to span the whole regions allowed by present nuclear and hypernuclear data.

The priors on $\nsat$, $E_{\rm sat}$, $K_{\rm sat}$, $E_{\rm sym}$, $L_{\rm sym}$, and $m^*/m$ are guided by recent compilations of nuclear empirical parameters and symmetry-energy constraints~\cite{Lattimer:2023rpe,MUSES:2023hyz,Drischler:2024ebw}.
In particular, we adopt broad ranges for $L_{\rm sym}$ and $m^*/m$. 
For the former indeed, different experiments suggest significantly different values (see for example parity-violating electron-scattering measurements of the neutron skin in $^{208}{\rm Pb}$ by PREX-II and in $^{48}{\rm Ca}$ by CREX~\cite{PREX:2021umo,CREX:2022kgg,Reed:2021nqk, Reinhard:2021utv, Zhang:2022bni,Zhao:2024gjz}).
Similarly, the broad range for $m^*/m$ reflects the fact that our RMF parameter is the Dirac effective mass, while phenomenological constraints usually refer to nonrelativistic or Landau effective masses associated with the single-particle spectrum or the density of states. The relation between these effective masses is model dependent, so we keep this prior deliberately broad.

For the strange-baryon sector, we use data on hypernuclei to define broad flat priors on the single-particle potentials at saturation~\cite{Gal:2016boi,Tolos:2020aln}.
The $\Lambda$ potential is relatively well established and attractive, the $\Xi$ potential is weakly attractive, whereas the $\Sigma$ sector is the most uncertain and is often inferred to be repulsive.
The adopted windows of variations, reported in Table~\ref{tab:priors_vs_posteriors_wide}, are chosen in order to take into account the large uncertainties that still affect those potentials.

For the $\Delta$ sector we fix $x_{\rho\Delta}=1$ and sample $x_{\sigma\Delta}$ and $x_{\omega\Delta}$ from broad uniform priors.
Since terrestrial data constrain mainly the $\Delta$ potential in nuclear matter and specific combinations of the couplings, rather than the individual couplings separately, we further restrict the prior support by discarding parameter sets that violate the phenomenological bounds discussed in ~\cite{Drago:2014oja,Ribes:2019kno,Li:2018qaw}:
\begin{align}
    0\lesssim x_{\sigma\Delta}-x_{\omega\Delta}&\lesssim 0.2\label{eq:deltaprior1}\\
        V_N(\nsat)-30\MeV \lesssim V_{\Delta}(\nsat) &\lesssim V_{N}(\nsat),\label{eq:deltaprior2}
\end{align}
Here $V_N(\nsat)$ and $V_{\Delta}(\nsat)$ denote, respectively, the nucleon and $\Delta$ isoscalar single-particle potentials in cold symmetric nuclear matter at saturation density $\nsat$.

For the quark-sector parameters $(a_4,B,\gap)$, we assign broad, weakly informative flat priors spanning the range commonly explored in CFL bag-model studies of QSs~\cite{Alford:2007xm,Bombaci:2020vgw,Weissenborn:2011qu}.
Because these parameters are not directly constrained by terrestrial experiments at densities relevant for compact stars, they are restricted only by the astrophysical likelihoods and by the imposed strange-matter stability conditions.
In the 2F scenario, we require the Bodmer--Witten hypothesis to hold, namely:
\begin{equation}
\mu_Q(P=0)=\left.\frac{\varepsilon_Q}{n_{B}}\right|_{P=0}<930~{\rm MeV},
\label{eq:BW_3f}
\end{equation}
and, to exclude rehadronization after deconfinement, we impose over the pressure range relevant for the stellar configurations considered
\begin{equation}
\mu_Q(P)<\mu_H(P)\label{eq:no_rehadronization},
\end{equation} 
where $\mu_Q$ and $\mu_H$ are the Gibbs energy per baryon of the SQM and hadronic phases, and correspond to the baryon chemical potentials in weak equilibrium, see Appendix \ref{app:EOS_details}.

EOS parameter sets that violate either of these conditions are excluded from the prior support.

Finally, we impose the basic physical requirements of mechanical stability and causality, namely
\begin{equation}
0<c_s^2\leq 1,
\end{equation}
and discard all parameter sets that violate them.

\subsection{Data and likelihoods}
\label{sec:data_likelihoods}

The likelihood construction depends on the type of observational constraint. In this work we consider four classes of data. A list of the used data is reported in Table~\ref{tab:summary_data}

\begin{table*}[t]
\centering
\begin{tabular}{l|c|l|l}
\hline
\hline
Category & Label & Info & Likelihood representation \\
\hline
$M$ & PSR~J0952--0607 &
$M = 2.24 \pm 0.17\,\Msol$ \cite{Romani:2022jhd}\footnote{Since this pulsar rotates rapidly, the measured mass $M=(2.35\pm0.17)\,\Msol$ may exceed the mass of the corresponding non-rotating configuration. We therefore use the correction proposed in Ref.~\cite{Ayriyan:2025rub}, obtaining $M=(2.24\pm0.17)\,\Msol$.}  &
Gaussian in $M$ 
\\
$M$ & GW~190814 &
$M = 2.59 \pm 0.05\,\Msol$ \cite{LIGOScientific:2020zkf} &
Gaussian in $M$ 
\\
\hline
$M$-$R$ & PSR~J0030+0451 &
$M = 1.44^{+0.15}_{-0.14}\,\Msol,\;
R=13.02^{+1.24}_{-1.06}\,\mathrm{km}$ \cite{Miller:2019cac}&
$q(M,R)$ KDE from posterior samples \cite{miller_2019_3473466}
\\
$M$-$R$ & PSR~J0740+6620 &
$M = 2.08 \pm 0.07\,\Msol,\;
R=13.7^{+2.6}_{-1.5}\,\mathrm{km}$ \cite{Miller:2021qha}&
$q(M,R)$ KDE from posterior samples \cite{miller_2021_4670689}
\\
$M$-$R$ & PSR~J0614--3329 &
$M = 1.44^{+0.06}_{-0.07}\,\Msol,\;
R=10.29^{+1.01}_{-0.86}\,\mathrm{km}$ \cite{Mauviard:2025dmd} &
$q(M,R)$ KDE from posterior samples \cite{mauviard_2025_17380576} 
\\
$M$-$R$ & HESS~J1731--347 &
$M = 0.77^{+0.20}_{-0.17}\,\Msol,\;
R = 10.4^{+0.86}_{-0.78}\,\mathrm{km}$ \cite{Doroshenko:2022nwp}&
$q(M,R)$ KDE from posterior samples \cite{doroshenko_2022_8232233}
\\
\hline
$M$-$\Lambda$ & GW~170817 &
$\mathcal{M}=1.186\pm0.001$ \; $\tilde{\Lambda}=300^{+420}_{-230}$ \cite{LIGOScientific:2018hze}&
$q(M_1,M_2,\Lambda_1,\Lambda_2)$ KDE from posterior samples \cite{gw170817data,LIGOScientific:2018cki} 
\\
\hline
$n_B$-$E_H^{\rm PNM}$ & $\chi$EFT &
PNM band for $E_H^{\rm PNM}(n_B)$, $n_B \le \nsat$ &
soft band likelihood from \cite{Passarella:2025zqb,Tews:2015ufa,Drischler:2017wtt,Drischler:2020hwi} 
\\
$n_B$-$P_H^{\rm SNM}$ & HIC &
SNM band for $P_H^{\rm SNM}(n_B)$, $n_B \le 3\nsat$ &
soft band likelihood from \cite{Danielewicz:2002pu} 
\\
\hline
\hline
\end{tabular}
\caption{Summary of the observational and laboratory inputs used in the analysis. Mass-only constraints are implemented as Gaussian likelihoods in the gravitational mass. Mass--radius constraints are reconstructed by kernel density estimation (KDE) from the released posterior samples. For GW170817 we use the full joint posterior in $(M_1,M_2,\Lambda_1,\Lambda_2)$, reconstructed through a four-dimensional KDE. Here $\mathcal{M} \equiv (M_1 M_2)^{3/5}(M_1+M_2)^{-1/5}$ denotes the chirp mass, and $\tilde{\Lambda} \equiv \frac{16}{13}\frac{(M_1+12M_2)M_1^4\Lambda_1+(M_2+12M_1)M_2^4\Lambda_2}{(M_1+M_2)^5}$ the effective tidal deformability.  The $\chi$EFT and HIC inputs are implemented as soft likelihood factors based on the adopted low-density pure-neutron-matter and symmetric-matter bands, respectively. For PSR~J0952--0607 we use the rotation-corrected non-rotating equivalent mass proposed in Ref.~\cite{Ayriyan:2025rub}.}
\label{tab:summary_data}
\end{table*}

\subsubsection{Mass--radius joint observations}
\label{sec:mr_obs}

Recent NICER analyses provide joint constraints on the masses and radii of several pulsars. In this work we use the posteriors released for PSR~J0740+6620 \cite{Miller:2021qha}, PSR~J0030+0451 \cite{Miller:2019cac}, and PSR~J0614--3329 \cite{Mauviard:2025dmd}. In addition, we include the mass--radius constraint inferred from X-ray observations of the central compact object in HESS~J1731--347 \cite{Doroshenko:2022nwp}. We choose to disregard sources for which currently available analyses yield mutually inconsistent constraints for the same object, such as PSR~J1231--1411 \cite{Salmi:2024bss,Qi:2025mpn} and PSR~J0437--4715 \cite{Miller:2025qfq,Choudhury:2024xbk}.

For each source $i$, we reconstruct from the released samples a smooth estimate of the posterior density
\begin{equation}
q_i(M,R)\equiv \prob(M,R\mid \datumi)
\end{equation}
using kernel density estimation (KDE).

In the 1F scenario, the observed object is assumed to belong to the single hadronic family. The corresponding likelihood associated with the observation $\datumi$ is obtained by integrating the reconstructed posterior density along the theoretical mass--radius curve determined by the TOV solution, while marginalizing over the unknown stellar mass:
\begin{equation}
\mathcal{L}_i^{(\mathrm{1F})}(\theta)\propto
\int_{M_{\rm HS,min}^{\rm (1F)}(\theta)}^{M_{\rm HS,max}^{\rm (1F)}(\theta)} dM\,
q_i\!\left[M,R_{\rm HS}^{\rm TOV}(M;\theta)\right].
\label{eq:Li_1F_MR_final}
\end{equation}

Here we assume a constant mass prior between the limiting masses $M_{\rm HS,min}^{\rm (1F)}(\theta)$ and $M_{\rm HS,max}^{\rm (1F)}(\theta)$, and a vanishing prior outside this interval. A derivation of Eq.~\eqref{eq:Li_1F_MR_final}, together with the underlying assumptions, is provided in Appendix~\ref{app:likelihood}. The choices for the integration limits $M_{\rm HS,min}^{\rm (1F)}(\theta)$ and $M_{\rm HS,max}^{\rm (1F)}(\theta)$ are discussed in Sec.~\ref{sec:priors_mass_distributions}.

As discussed in Sec.~\ref{sec:bayes_factor} and Appendix~\ref{app:likelihood}, the normalization of the mass prior must be included explicitly in order to compute the Bayes factor correctly.
Equation~\eqref{eq:Li_1F_MR_final} should therefore be multiplied by
\begin{equation}
\left[M_{\rm HS,max}^{\rm (1F)}(\theta)-M_{\rm HS,min}^{\rm (1F)}(\theta)\right]^{-1}.\label{eq:Mnorm_1F}
\end{equation}

In the 2F scenario, the family label is not known a priori and must be marginalized over together with the mass. The likelihood associated with the observation $\datumi$ is therefore:
\begin{widetext}
\begin{equation}
\mathcal{L}_i^{(\mathrm{2F})}(\theta)\propto
\sum_{f\in\{\mathrm{HS},\mathrm{QS}\}}
\int_{M_{f,\rm min}^{\rm (2F)}(\theta)}^{M_{f,\rm max}^{\rm (2F)}(\theta)} dM\,
q_i\!\left[M,R_f^{\rm TOV}(M;\theta)\right]\,
\eta_f^{(\mathrm{2F})}(M;\theta)
\equiv
\mathcal{L}_{{\rm HS},i}^{(\mathrm{2F})}(\theta)+\mathcal{L}_{{\rm QS},i}^{(\mathrm{2F})}(\theta),
\label{eq:Li_2F_MR_final}
\end{equation}
\end{widetext}
where
\begin{equation}
\eta_f^{(\mathrm{2F})}(M;\theta)\equiv \prob(f\mid M,\theta,\mathrm{2F}),
\qquad f\in\{\mathrm{HS},\mathrm{QS}\},
\end{equation}
encodes the relative population weight of each family at fixed mass. 

As in the 1F case, we assume a constant mass prior between the limiting masses $M_{f,\rm min}^{\rm (2F)}(\theta)$ and $M_{f,\rm max}^{\rm (2F)}(\theta)$, and a vanishing prior outside this interval. Again, the proper normalization of the mass prior must be included explicitly, see Sec.~\ref{sec:priors_mass_distributions}. Therefore, both contributions $f=\mathrm{HS},\mathrm{QS}$ must be multiplied by
\begin{equation}
\left[M_{f,\rm max}^{\rm (2F)}(\theta)-M_{f,\rm min}^{\rm (2F)}(\theta)\right]^{-1}.\label{eq:Mnorm_2F}
\end{equation}

A derivation of Eq.~\eqref{eq:Li_2F_MR_final}, together with the underlying assumptions, is provided in Appendix~\ref{app:likelihood}. The choices for the integration limits and for $\eta_f^{(\mathrm{2F})}(M;\theta)$ are discussed in Sec.~\ref{sec:priors_mass_distributions}.

Note that the total likelihood in the 2F scenario is a \emph{sum} of the likelihood contributions from the HS and QS branches. This does \emph{not} mean that a given observed star is simultaneously a HS and a QS. Rather, for each source there is a single true family assignment, but that assignment is not known from the observation alone. The statistically correct procedure is therefore to treat the family label as a latent discrete variable and marginalize over it. As a result, if the data are compatible with either branch, both possibilities contribute to the total likelihood, weighted by the corresponding population priors. In this respect, the sum reflects our uncertainty about the family assignment of the source, not the physical coexistence of the two interpretations.

\subsubsection{Mass-only observations}
\label{sec:max_mass}

We now consider observations that constrain only the gravitational mass of a compact object. In this work, we use this category of observations primarily to impose a lower bound on the maximum mass supported by the EOS. In particular, we include the black widow pulsar PSR~J0952--0607 \cite{Romani:2022jhd}, and we also examine, for comparison, the possibility that the secondary component in GW190814 \cite{Bombaci:2020vgw,LIGOScientific:2020zkf} was a compact star rather than a black hole.

We do not include upper bounds on $M_{\rm TOV}^{\rm max}$ inferred from multimessenger interpretations of GW170817~\cite{Rezzolla:2017aly,Shibata:2019ctb} as likelihood constraints, because their applicability to the 2F scenario is not straightforward\footnote{These bounds are derived under a standard 1F interpretation of GW170817. In the 2F scenario, the remnant could be a QS, having in general different rotational increase of the maximum mass, collapse threshold, angular-momentum and energy losses during the differentially rotating phase, and ejecta properties. A dedicated multimessenger interpretation of GW170817 within the 2F scenario is left to future work.}.
We will nevertheless discuss the compatibility of our posterior predictions with these constraints in the results.

For a mass-only datum $\datumi\equiv M_i\pm \sigma_i$, we reconstruct a one-dimensional probability density
\begin{equation}
q_i(M)=\frac{1}{\sqrt{2\pi\sigma_i^2}}
\exp\!\left[-\frac{(M-M_i)^2}{2\sigma_i^2}\right].
\label{eq:q_mass_gaussian}
\end{equation}

Assuming a constant prior on the object mass within an interval of each branch and a vanishing prior outside it, the likelihood contribution associated with the observation $\datumi$ in the 1F scenario is

\begin{equation}
\mathcal{L}_i^{(\mathrm{1F})}(\theta)
\propto
\int_{M_{\rm HS,min}^{\rm (1F)}(\theta)}^{M_{\rm HS,max}^{\rm (1F)}(\theta)} dM\,
q_i(M).
\label{eq:L1F_mass_final}
\end{equation}

In the 2F scenario, the family of the observed object is not known a priori and must be marginalized over. Assuming that the mass-only observation does not further distinguish between the HS and QS interpretations at fixed $M$, the likelihood becomes
\begin{equation}
\mathcal{L}_i^{(\mathrm{2F})}(\theta)
\propto
\sum_{f\in\{{\rm HS},{\rm QS}\}}
\int_{M_{f,\rm min}^{\rm (2F)}(\theta)}^{M_{f,\rm max}^{\rm (2F)}(\theta)} dM\,
q_i(M)\,
\eta_f^{(\mathrm{2F})}(M;\theta),
\label{eq:L2F_mass_final}
\end{equation}

As in the mass--radius case, the total likelihood in the 2F scenario is therefore a sum of the HS and QS contributions, reflecting the fact that the family assignment of the source is not known a priori.

The formal derivation of Eqs.~\eqref{eq:L1F_mass_final} and \eqref{eq:L2F_mass_final} follows the same argument as for the mass--radius likelihood discussed in Appendix~\ref{app:likelihood}.
Moreover, for the computation of the Bayes factor, the proper normalization of the mass prior must be included explicitly, namely Eq.~\eqref{eq:L1F_mass_final} and each term of Eq.~\eqref{eq:L2F_mass_final} should be multiplied by Eq.~\eqref{eq:Mnorm_1F} and Eq.~\eqref{eq:Mnorm_2F} respectively.

\subsubsection{GW170817 mass--tidal joint observations}
\label{sec:gw170817}

Gravitational-wave observations of binary compact-star mergers constrain the component masses and tidal deformabilities, $(M_1,\Lambda_1)$ and $(M_2,\Lambda_2)$, where conventionally $M_1\geq M_2$. The signal does not determine the properties of the two stars independently; rather, it constrains combinations of binary parameters, namely the chirp mass $\mathcal{M}$ and the effective tidal deformability $\tilde{\Lambda}$. As a consequence, the inferred values of $(M_1,M_2,\Lambda_1,\Lambda_2)$ are correlated and must be described through a single joint probability density rather than through two independent single-star constraints.

In this work we use the posterior samples released for GW170817 \cite{LIGOScientific:2018cki}. From these samples we reconstruct, by means of KDE, a smooth four-dimensional density that we denote by
\begin{equation}
q_{\rm GW170817}(M_1,M_2,\Lambda_1,\Lambda_2).
\end{equation}

In the 1F scenario, both components are assumed to belong to the single hadronic family. Assuming a constant prior on the component masses within a chosen interval and a vanishing prior outside it, the likelihood contribution associated with GW170817 reads:
\begin{widetext}
\begin{equation}
\mathcal{L}_{\rm GW170817}^{(\mathrm{1F})}(\theta)
\propto
\int_{M_{\rm HS,min}^{\rm (1F)}(\theta)}^{M_{\rm HS,max}^{\rm (1F)}(\theta)} dM_1
\int_{M_{\rm HS,min}^{\rm (1F)}(\theta)}^{M_{\rm HS,max}^{\rm (1F)}(\theta)} dM_2\,
q_{\rm GW170817}\!\left[
M_1,M_2,
\Lambda_{\rm HS}^{\rm TOV}(M_1;\theta),
\Lambda_{\rm HS}^{\rm TOV}(M_2;\theta)
\right].
\label{eq:L_GW170817_1F}
\end{equation}
\end{widetext}
The formal derivation follows the same logic discussed in Appendix~\ref{app:likelihood}.

In the 2F scenario, the family labels of the two components are latent variables.
The likelihood is therefore obtained by summing over the four possible binary-family assignments,
\begin{equation}
\mathcal{C}=\{\mathrm{HS\text{-}HS},\mathrm{HS\text{-}QS},
\mathrm{QS\text{-}HS},\mathrm{QS\text{-}QS}\}.
\end{equation}
\begin{widetext}
\begin{equation}
\mathcal{L}_{\rm GW170817}^{(\mathrm{2F})}(\theta)
\propto
\sum_{(f_1,f_2)\in\mathcal{C}}
\int_{M_{f_1,\min}^{(2F)}}^{M_{f_1,\max}^{(2F)}} dM_1
\int_{M_{f_2,\min}^{(2F)}}^{M_{f_2,\max}^{(2F)}} dM_2\,
q_{\rm GW170817}
\!\left[
M_1,M_2,
\Lambda_{f_1}^{\rm TOV}(M_1;\theta),
\Lambda_{f_2}^{\rm TOV}(M_2;\theta)
\right]\,
\eta_{f_1f_2}^{(2F)}(M_1,M_2;\theta).
\label{eq:L_GW170817_2F_compact}
\end{equation}
\end{widetext}

where $\eta_{i\text{-}j}^{\rm (2F)}$ is the prior probability that the heavier star belongs to family $i$ and the lighter one to family $j$, at fixed $(M_1,M_2)$. 

A simple factorized choice is
\begin{equation}
\eta_{i\text{-}j}^{(\mathrm{2F})}(M_1,M_2;\theta)
=
\eta_i^{(\mathrm{2F})}(M_1;\theta)\,
\eta_j^{(\mathrm{2F})}(M_2;\theta),
\label{eq:eta_norm}
\end{equation}
The expanded form of Eq.~\eqref{eq:L_GW170817_2F_compact}, together with the numerical KDE implementation, is reported in Appendix~\ref{app:numerics}.

Again, for the computation of the Bayes factor, Eq.~\eqref{eq:L_GW170817_1F} and each integral of Eq.~\eqref{eq:L_GW170817_2F_compact} should be multiplied by Eq.~\eqref{eq:Mnorm_1F} and Eq.~\eqref{eq:Mnorm_2F} respectively.

Moreover, one may wish to impose additional prior information on the allowed binary compositions.
If some configurations are disfavored or excluded on the basis of external considerations, this can be implemented as a hard prior by setting the corresponding $\eta_{i\text{-}j}^{(\mathrm{2F})}$ to zero and renormalizing the remaining weights so that they still sum to unity over the allowed set.
For instance, if QS--QS mergers are excluded as a possible source of  GW170817, one sets
\begin{equation}
\eta_{\rm QS\text{-}QS}^{(\mathrm{2F})}(M_1,M_2;\theta)=0
\end{equation}
and renormalizes the remaining $\eta_{ij}^{(\mathrm{2F})}$ over the three surviving channels.
The same procedure applies also if one excludes (for some physical reason) HS--HS mergers, or any other subset of binary configurations.

A possible motivation for such a restriction comes from the interpretation of GW170817 within the two-families scenario. In that framework, and for the soft hadronic branch, GW170817 is more naturally interpreted as a mixed HS-QS merger than as HS-HS or QS-QS. Indeed, an HS-HS binary with total mass comparable to that of GW170817 would likely undergo a prompt collapse to a black hole, making it difficult to account for the bright kilonova AT2017gfo and the sGRB ~\cite{Drago:2017bnf,DePietri:2019khb}.
On the other hand, QS-QS mergers are expected to produce a strongly suppressed kilonova signal~\cite{Miao:2024qik}.  
By contrast, a HS-QS system is more naturally compatible with the production of substantial ejecta and with the formation of a sufficiently long-lived hypermassive remnant~\cite{DePietri:2019khb,Burgio:2018yix}.
At the same time, the detailed interpretation of specific GW events in the two-families scenario, especially concerning the remnant evolution and the associated kilonova, are still not well known.
For this reason, in the present work we do not impose a unique astrophysical interpretation of GW170817, but simply test whether our qualitative results are affected by suppressing some channels, such as HS--HS and/or QS--QS, through hard priors on the corresponding $\eta_{i\text{-}j}^{(\mathrm{2F})}$.

\subsubsection{Chiral effective field theory}
\label{sec:chieft}

At low baryon density, microscopic calculations based on chiral effective field theory ($\chi$EFT) provide the natural benchmark for the hadronic EOS of cold pure neutron matter. In the present work we follow the same approach adopted in~\cite{Passarella:2025zqb}: rather than imposing $\chi$EFT as a hard prior or cut, we include it as a likelihood factor that rewards EOSs which are consistent with the low-density band and smoothly suppresses those that deviate from it. 

More explicitly, for each sampled hadronic EOS we compare the predicted energy per baryon in cold pure neutron matter with the adopted discretized $\chi$EFT band over the density interval covered by the microscopic calculations. EOSs lying inside the band receive no penalty, whereas EOSs outside it are exponentially down-weighted according to their distance from the band boundary. The explicit discretization and the analytic form of the likelihood are given in Appendix~\ref{app:chieft_hic}.

\subsubsection{Constraints from heavy-ion collisions}
\label{sec:HIC}

Complementary information on the hadronic EOS above saturation density is provided by collective-flow observables in intermediate-energy heavy-ion collisions (HIC), which constrain the pressure of symmetric nuclear matter.
In the present work we use the flow band of~\cite{Danielewicz:2002pu} as a phenomenological constraint on cold symmetric matter, but we deliberately implement it as a soft likelihood rather than as a hard cut, since the extraction of the band relies on transport-model simulations and is therefore not fully model independent~\cite{Danielewicz:2002pu}, and since the band is not provided as a posterior distribution.

In~\cite{Huth:2021bsp}, the HIC likelihood is built primarily from the more recent FOPI and ASY-EOS constraints~\cite{LeFevre:2016vpp,Russotto:2016ucm}. The flow constraint of~\cite{Danielewicz:2002pu} is used only as supplementary information for symmetric nuclear matter in the range $2$--$3\nsat$, namely in the overlapping ASY-EOS-sensitive range.
In the same spirit, in the present work we use the band of~\cite{Danielewicz:2002pu} only up to $3\nsat$. We account phenomenologically for the residual model dependence of this constraint by allowing smooth excursions outside the band, that is, by exponentially down-weighting EOS models according to their distance from the band boundary in the pressure--baryon-density plane of symmetric nuclear matter. The explicit reconstruction of the band and the corresponding likelihood function are reported in Appendix~\ref{app:chieft_hic}.

\subsection{Mass and family-assignment priors}
\label{sec:priors_mass_distributions}

We now discuss the priors associated with the stellar masses entering the likelihood functions. In the present work, we separate this prior information into two conceptually distinct ingredients. The first one is the support in mass space, specified by the integration limits
$M_{f,\rm min}^{(\mathcal{H})}(\theta)$ and $M_{f,\rm max}^{(\mathcal{H})}(\theta)$, with
$f\in\{{\rm HS},{\rm QS}\}$ and $\mathcal{H}\in\{{\rm 1F},{\rm 2F}\}$. The second one is the function $\eta_f^{(\mathcal{H})}(M;\theta)$, which encodes the relative probability that an object of gravitational mass $M$ belongs to the family $f$, given the EOS parametrization $\theta$.

The integration limits set the mass interval in which a configuration of a given family is admitted in the hypothesis under consideration. More explicitly, they define the support of the prior in mass space and can be represented by an indicator function (with the proper normalization, that corresponds to Eqs. (\ref{eq:Mnorm_1F}) and (\ref{eq:Mnorm_2F}))
\begin{equation}
\frac{\mathbf{1}\!\left[M_{f,\rm min}^{(\mathcal{H})}(\theta)\le M \le M_{f,\rm max}^{(\mathcal{H})}(\theta)\right]}{M_{f,\rm max}^{(\mathcal{H})}(\theta)-M_{f,\rm min}^{(\mathcal{H})}(\theta)},
\label{eq:proper_norm}
\end{equation}
which assign equal prior weight to all masses within the corresponding interval and zero weight outside it. This simple choice avoids introducing additional assumptions on the mass probability distribution of given family branch, which would be difficult to constrain given the limited number of available observations.

A zeroth-order choice is to identify the available mass interval of each family with the full stable branch obtained from the TOV solutions. In that case one may set
\begin{equation}
M_{f,\rm min}^{(\mathcal{H})}(\theta)\simeq 0,
\qquad
M_{f,\rm max}^{(\mathcal{H})}(\theta)=M_{f,\rm TOV}^{\rm max}(\theta),
\end{equation}
where the lower limit can in practice be replaced by the minimum mass returned by the numerical TOV solver. Since none of the observational constraints used in this work probes the very low-mass region, the precise value of this technical lower bound has negligible impact on the results. This corresponds to a flat prior support over the full stable branch of each family.

Beyond this minimal choice, the integration limits can be used to incorporate astrophysically motivated prior information. On the hadronic side, current core-collapse supernova calculations do not provide a robust formation channel for neutron stars with masses substantially below $\sim 1.17$--$1.19\Msol$ \cite{Suwa:2018uni,Muller:2024aod}. Motivated by these studies, we adopt as an astrophysically informed lower bound:
\begin{equation}
M_{\rm HS,min}^{(\mathcal{H})}(\theta)=1.17\Msol.
\end{equation}
By contrast, viable formation channels for low-mass QSs have been at least qualitatively discussed in the literature; see, e.g.~\cite{DiClemente:2022wqp}. For this reason, a natural baseline choice is
\begin{equation}
M_{\rm QS,min}^{(\mathrm{2F})}(\theta)\simeq 0.
\end{equation}
Following this reasoning, in the 2F scenario we enforce the $1.17\Msol$ lower bound for HSs, relying on the proposed QS formation mechanisms to account for exceptionally low-mass objects such as HESS J1731-347. In contrast, for the 1F scenario, we will use $\simeq0$ as default; otherwise, a purely hadronic framework would be unable to explain the HESS measurement. However, we will later perform consistency tests using different choices to test their impact on the final results.

The upper integration limits can also be refined beyond the purely structural stability condition. In the two-families scenario, HSs are metastable with respect to conversion into QSs, and the conversion probability is expected to increase with stellar mass because the central density and strangeness content increase, thereby reducing the free-energy barrier against deconfinement \cite{Drago:2015cea,Bombaci:2016xuj,Guerrini:2026_inpreparation}. This suggests a microphysics-motivated prior on the maximum mass effectively accessible to the hadronic branch in the 2F scenario. Instead of identifying the upper limit with the maximum mass of the full stable TOV sequence, one may set
\begin{equation}
M_{\rm HS,max}^{(\mathrm{2F})}(\theta)=M_{\rm conv}(\theta),
\end{equation}
where $M_{\rm conv}(\theta)$ denotes the mass at which quark-matter nucleation becomes efficient in the HS center, triggering the conversion. In this way, hadronic configurations with $M>M_{\rm conv}$ are assumed to convert into QSs and are therefore removed from the prior support of the hadronic branch.

As a simple phenomenological prescription, we adopt the criterion proposed in~\cite{DePietri:2019khb} and subsequently used in~\cite{Bombaci:2020vgw,Becerra:2025ryy}, namely that conversion is triggered when the strangeness density in the center of the star reaches the nuclear saturation density, $n_{S,c}=\nsat$.
This criterion is motivated by the idea that strange quarks in the hadronic phase must be sufficiently close to interact and seed the first drop of deconfined matter.
Accordingly, we set
\begin{equation}
M_{\rm conv}(\theta)=M\bigl(n_{S,c}=\nsat\bigr).
\end{equation}
We stress that this should be regarded only as a first phenomenological implementation of a microphysically motivated prior. A fully consistent determination of the maximum HS mass effectively reachable in nature within the two-families scenario would require following the formation and evolution of the star along all relevant astrophysical channels, including the proto-neutron-star phase and the conditions for quark-matter nucleation along the corresponding evolutionary paths (see e.g. \cite{Guerrini:2026_inpreparation}).

Within those intervals, the relative family weight is encoded in the function $\eta_f^{(\mathcal{H})}(M;\theta)$. In the 1F scenario this quantity is trivial, since only the hadronic branch is present. One may therefore either omit it altogether, or equivalently set:
\begin{equation}
\eta_{\rm HS}^{(\mathrm{1F})}(M;\theta)=1,
\qquad
\eta_{\rm QS}^{(\mathrm{1F})}(M;\theta)=0.
\end{equation}
In the 2F scenario, by contrast, the family assignment is nontrivial and we define
\begin{align}
\eta_{\rm HS}^{(\mathrm{2F})}(M;\theta)
&\equiv
\prob(f={\rm HS}\mid M,\theta,\mathrm{2F}),
\\
\eta_{\rm QS}^{(\mathrm{2F})}(M;\theta)
&\equiv
\prob(f={\rm QS}\mid M,\theta,\mathrm{2F}),
\end{align}
with the normalization condition
\begin{equation}
\eta_{\rm HS}^{(\mathrm{2F})}(M;\theta)+\eta_{\rm QS}^{(\mathrm{2F})}(M;\theta)=1.
\end{equation}

In the absence of detailed information on the relative abundance of HSs and QSs at fixed mass, we adopt as a simple baseline choice
\begin{widetext}
\begin{equation}
\eta_{\rm HS}^{(\mathrm{2F})}(M;\theta)=
\begin{cases}
1, & M \in [M_{\rm HS,min}^{(\mathrm{2F})}(\theta), M_{\rm HS,max}^{(\mathrm{2F})}(\theta)]
\ \text{and}\
M \notin [M_{\rm QS,min}^{(\mathrm{2F})}(\theta), M_{\rm QS,max}^{(\mathrm{2F})}(\theta)]
\\[0.4em]
1/2, & M\in [M_{\rm HS,min}^{(\mathrm{2F})}(\theta),M_{\rm HS,max}^{(\mathrm{2F})}(\theta)]
\ \text{and}\
M\in [M_{\rm QS,min}^{(\mathrm{2F})}(\theta), M_{\rm QS,max}^{(\mathrm{2F})}(\theta)]
\\[0.4em]
0, & M\notin [M_{\rm HS,min}^{(\mathrm{2F})}(\theta),M_{\rm HS,max}^{(\mathrm{2F})}(\theta)]
\ \text{and}\
M\in [M_{\rm QS,min}^{(\mathrm{2F})}(\theta), M_{\rm QS,max}^{(\mathrm{2F})}(\theta)]
\end{cases}
\label{eq:etaHS_baseline}
\end{equation}
\end{widetext}
The corresponding QS weight is then fixed by normalization,
\begin{equation}
\eta_{\rm QS}^{(\mathrm{2F})}(M;\theta)=1-\eta_{\rm HS}^{(\mathrm{2F})}(M;\theta).
\end{equation}
Thus, equal prior weight is assigned to the two families only in the mass range where both are allowed, while outside the overlap region the full weight is assigned to the only family that is available.

This simple prescription is sufficient for the exploratory analysis presented here. It separates the role of structural stability and astrophysical accessibility, encoded in the mass intervals, from the relative assignment of a given mass to one family or the other, encoded in $\eta_f^{(\mathrm{2F})}$. More sophisticated choices, based for instance on population-synthesis calculations or on explicit conversion probabilities, can be implemented straightforwardly within the same framework.

%
\subsection{Bayes factor: one family versus two families}
\label{sec:bayes_factor}

The global comparison between the one-family (1F) and two-families (2F) scenarios is quantified by the Bayes factor
\begin{equation}
\mathcal{B}_{\mathrm{2F},\mathrm{1F}}
\equiv
\frac{\mathcal{Z}^{(\mathrm{2F})}(\{\datumi\})}{\mathcal{Z}^{(\mathrm{1F})}(\{\datumi\})},
\label{eq:bayes_factor}
\end{equation}
where the Bayesian evidence $\mathcal{Z}^{(\mathcal{H})}(\{\datumi\})$ associated with the hypothesis $\mathcal{H}\in\{\mathrm{1F},\mathrm{2F}\}$ has been defined in Eq.~\eqref{eq:Z}. Equivalently,
\begin{equation}
\ln \mathcal{B}_{\mathrm{2F},\mathrm{1F}}
=
\ln \mathcal{Z}^{(\mathrm{2F})}(\{\datumi\})
-
\ln \mathcal{Z}^{(\mathrm{1F})}(\{\datumi\}).
\end{equation}
The Bayes factor quantifies how the data update the prior odds between the two hypotheses. In particular,
\begin{equation}
\frac{\prob(\mathrm{2F}\mid \{\datumi\})}{\prob(\mathrm{1F}\mid \{\datumi\})}
=
\mathcal{B}_{\mathrm{2F},\mathrm{1F}}
\frac{\prob(\mathrm{2F})}{\prob(\mathrm{1F})}.
\end{equation}
Therefore, $\mathcal{B}_{\mathrm{2F},\mathrm{1F}}>1$ means that the data favor the 2F scenario over the 1F scenario only in the sense of likelihood updating; if equal prior probabilities are assigned to the two hypotheses, then $\mathcal{B}_{\mathrm{2F},\mathrm{1F}}>1$ also implies posterior preference for 2F.

The parameter spaces have different dimensionalities under the two hypotheses. This is fully consistent with Bayesian model comparison: the evidence automatically accounts for the different prior volumes of the competing hypotheses, so that a more flexible model is favored only if the improvement in likelihood compensates for the larger parameter space over which the likelihood is averaged.

In the present analysis, the evidences entering Eq.~\eqref{eq:bayes_factor} are computed using the likelihood construction described in Sec.~\ref{sec:data_likelihoods}, together with the EOS parameter priors and the mass-support assumptions discussed in Sec.~\ref{sec:priors_mass_distributions}. In particular, all multiplicative factors that depend on the EOS parameters $\theta$ or differ between the two hypotheses must be retained in the numerical implementation.

The proportionality signs appearing in the likelihood constructions of Sec.~\ref{sec:data_likelihoods} are harmless only if the omitted factors are independent of $\theta$ and common to both hypotheses. In that case, they factor out of the evidence integrals and cancel exactly in the Bayes factor. 
In the present analysis, the proportionality factors left implicit in the likelihood expressions arise only from source-dependent normalizations of the reconstructed $q_i$ and are independent of both $\theta$ and $\mathcal{H}$, so that they cancel exactly in the Bayes factor.


\subsection{Numerical setup}
\label{sec:setup}
We perform the Bayesian inference with the \texttt{dynesty} package~\cite{Speagle:2019ivv,sergey_koposov_2025_17268284}, using dynamic nested sampling.
This choice is well suited to the present problem because it provides posterior samples and Bayesian evidences within the same run. For each hypothesis, the sampler uses a prior transform from the unit hypercube to the physical parameter vector and a log-likelihood function returning $\ln\mathcal{L}^{(\mathcal{H})}(\theta)$.
We use multi-ellipsoidal bounding and slice sampling, and continue the baseline exploration until the estimated remaining contribution to the evidence satisfies $\Delta\ln\mathcal{Z}<0.01$.

\section{Results}
\label{sec:results}
In this section we discuss how the combined astrophysical and laboratory inputs constrain the EOS parameters and how these constraints map into macroscopic stellar properties. 

\subsection{Parameter posteriors}
\label{sec:results_posteriors}

\begin{figure*}[t]
    \centering
    \includegraphics[width=1\linewidth]{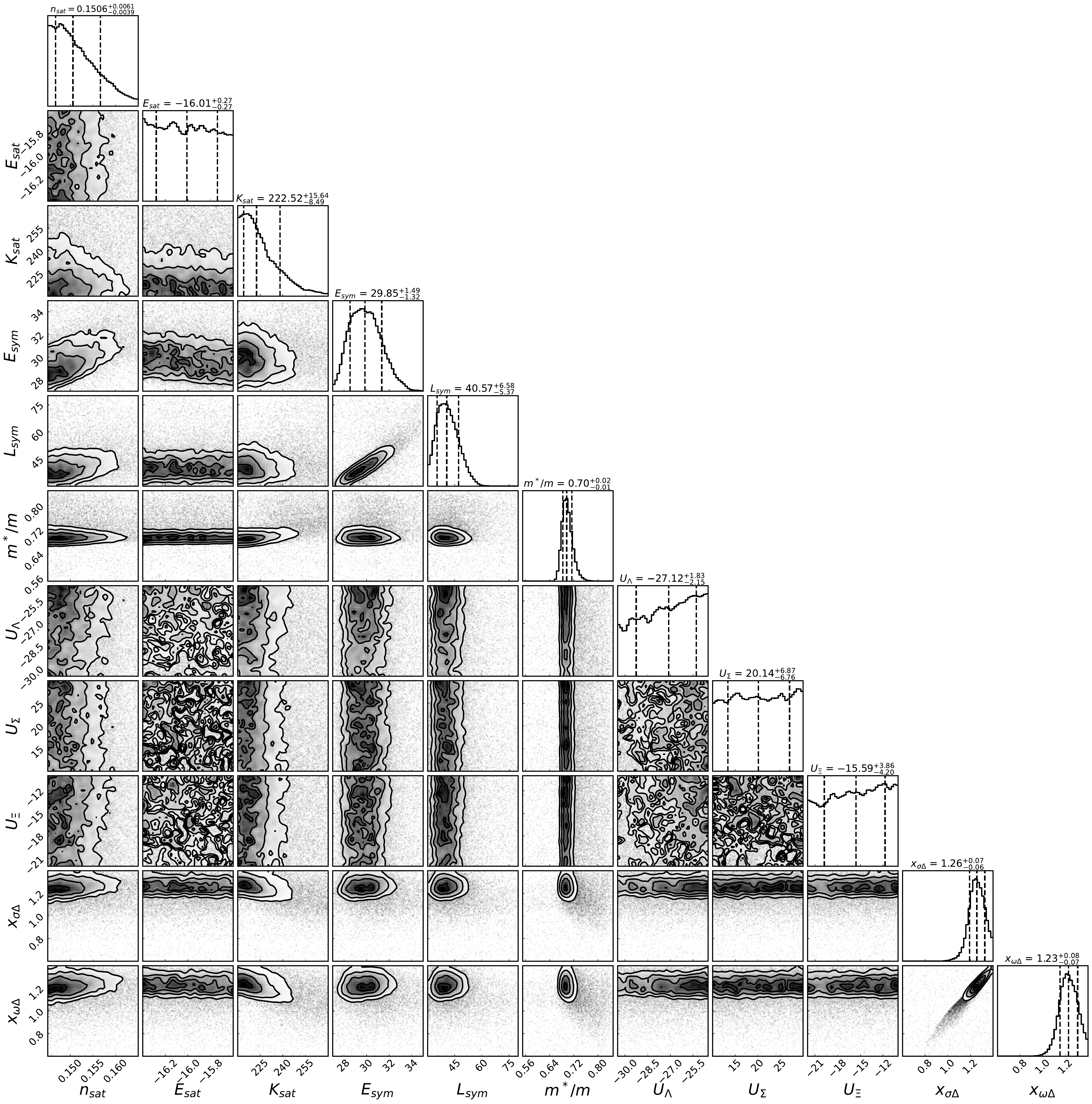}
    \caption{Corner plot showing the posterior probability distributions and correlations for the one-family scenario. The inference is restricted solely to the hadronic sector parameters, evaluating the compatibility of a single hadronic EOS with the combined astrophysical and experimental dataset. The hadronic sector includes the nuclear empirical parameters ($\nsat,\ E_{sat},\ K_{sat},\ E_{sym},\ L_{sym},\ m^*/m$), hyperon potentials ($U_\Lambda,\ U_\Sigma,\ U_\Xi$), and $\Delta$-resonance coupling ratios ($x_{\sigma\Delta},\ x_{\omega\Delta}$). Diagonal panels display the 1D marginal posteriors with dashed lines indicating the median and 1$\sigma$ credible intervals, while off-diagonal panels illustrate the 2D joint posterior distributions. 
    }
    \label{fig:corner1F}
\end{figure*}

\begin{figure*}[t]
    \centering
    \includegraphics[width=1\linewidth]{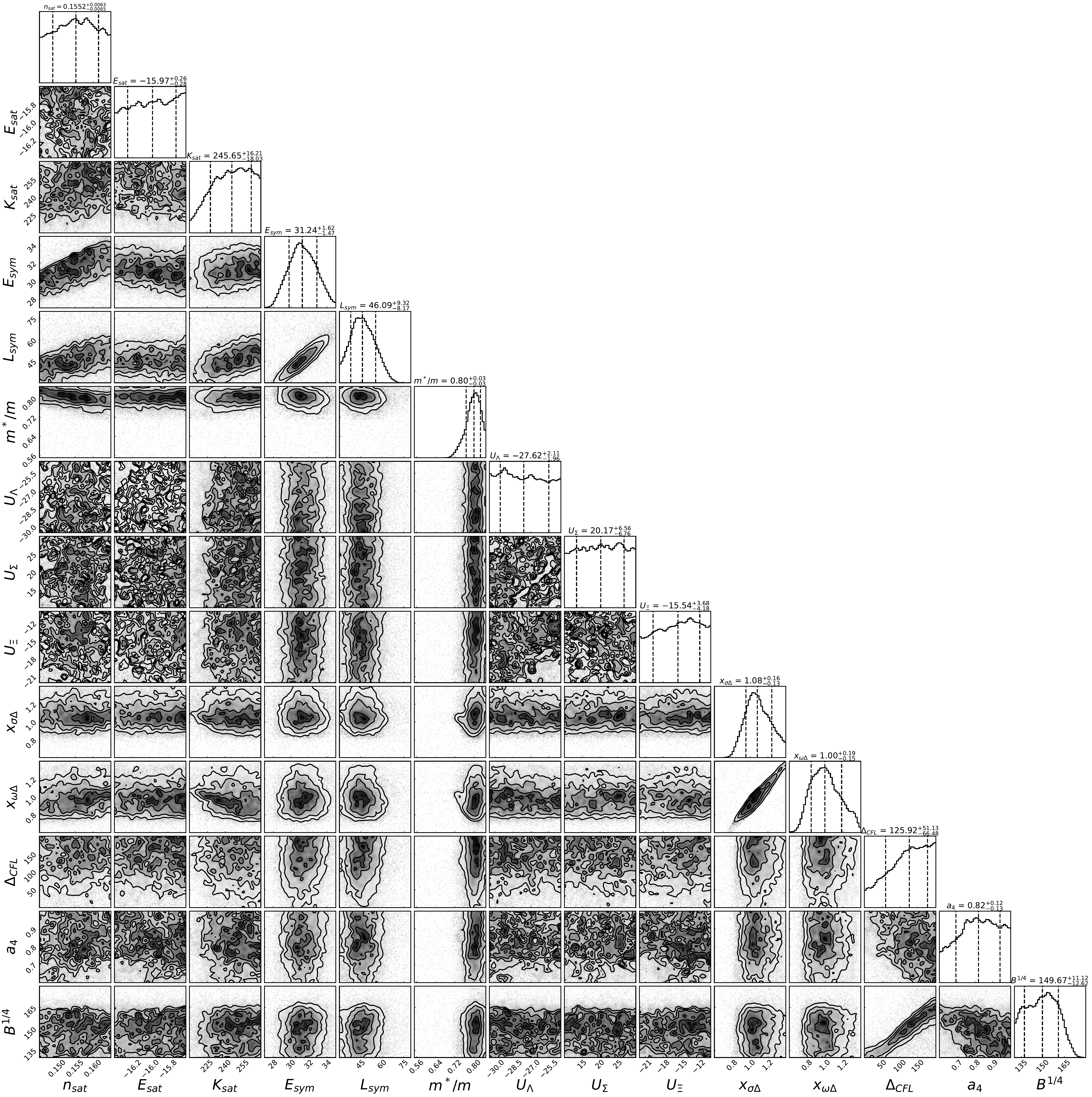}
    \caption{Same as Fig.~\ref{fig:corner1F} but for the two-families scenario: corner plot with posterior probability distributions and correlations. }
    \label{fig:corner2F}
\end{figure*}

The posterior distributions of the EOS parameters for the 1F and 2F scenarios are shown in Figs.~\ref{fig:corner1F} and~\ref{fig:corner2F}, respectively, while the corresponding marginal credible intervals for the parameters and the parameter set associated with the maximum total likelihood (which, given the flat priors adopted in this work, also coincides with the maximum-a-posteriori point) are summarized in Table~\ref{tab:priors_vs_posteriors_wide}.

\begin{table*}[t]
\centering
\small
\begin{tabular*}{\textwidth}{@{\extracolsep{\fill}}lccccc}
\hline
\hline
Parameter & Prior  & Marg. Posterior 1F  & Max. Likelihood 1F & Marg. Posterior 2F  & Max. Likelihood  2F \\
 & [min, max] &  (Median $\pm 1\sigma$) &  & (Median $\pm 1\sigma$) & \\
\hline
$\nsat\,(\mathrm{fm^{-3}})$          & $[0.145, 0.165]$ & $0.1506_{-0.0039}^{+0.0061}$ & $0.1499$ & $0.1552_{-0.0065}^{+0.0063}$ & $0.1575$ \\
$E_{\mathrm{sat}}\,(\mathrm{MeV})$   & $[-16.4, -15.6]$ & $-16.01_{-0.27}^{+0.27}$     & $-16.24$ & $-15.97_{-0.28}^{+0.26}$     & $-15.71$ \\
$K_{\mathrm{sat}}\,(\mathrm{MeV})$   & $[210.0, 270.0]$ & $222.52_{-8.49}^{+15.64}$    & $214.22$ & $245.65_{-18.03}^{+16.22}$   & $256.11$ \\
$E_{\mathrm{sym}}\,(\mathrm{MeV})$   & $[27.0, 35.0]$   & $29.85_{-1.32}^{+1.49}$      & $28.89$  & $31.24_{-1.47}^{+1.62}$      & $32.29$ \\
$L_{\mathrm{sym}}\,(\mathrm{MeV})$   & $[30.0, 80.0]$   & $40.57_{-5.37}^{+6.58}$      & $35.69$  & $46.09_{-8.17}^{+9.32}$      & $53.46$ \\
$m^*/m$                              & $[0.55, 0.85]$   & $0.70_{-0.01}^{+0.02}$       & $0.68$   & $0.80_{-0.03}^{+0.03}$       & $0.79$ \\
$U_{\Lambda}\,(\mathrm{MeV})$        & $[-30.5, -24.5]$ & $-27.12_{-2.15}^{+1.83}$     & $-26.18$ & $-27.62_{-1.96}^{+2.11}$     & $-25.21$ \\
$U_{\Sigma}\,(\mathrm{MeV})$         & $[10.0, 30.0]$   & $20.15_{-6.76}^{+6.87}$      & $13.89$  & $20.17_{-6.77}^{+6.56}$      & $28.44$ \\
$U_{\Xi}\,(\mathrm{MeV})$            & $[-22.0, -10.0]$ & $-15.59_{-4.20}^{+3.86}$     & $-11.00$ & $-15.54_{-4.18}^{+3.68}$     & $-20.67$ \\
$x_{\sigma\Delta}$                   & $[0.6, 1.4]$     & $1.26_{-0.06}^{+0.07}$       & $1.27$   & $1.08_{-0.13}^{+0.16}$       & $0.83$ \\
$x_{\omega\Delta}$                   & $[0.6, 1.4]$     & $1.23_{-0.07}^{+0.08}$       & $1.23$   & $1.00_{-0.15}^{+0.19}$       & $0.74$ \\
$\gap\,(\mathrm{MeV})$             & $[0.0, 200.0]$   & ---                          & ---      & $125.96_{-66.48}^{+51.13}$   & $157.08$ \\
$a_4$                                & $[0.6, 1.0]$     & ---                          & ---      & $0.82_{-0.13}^{+0.12}$       & $0.92$ \\
$B^{1/4}\,(\mathrm{MeV})$            & $[130.0, 180.0]$ & ---                          & ---      & $149.67_{-12.62}^{+11.12}$   & $158.21$ \\
\hline
\hline
\end{tabular*}
\caption{Prior ranges, marginalized posterior estimates (median and $1\sigma$ credible intervals) detailed in Figs.~\ref{fig:corner1F} and \ref{fig:corner2F}, and highest-likelihood values for EOS parameters in the 1F and 2F scenarios. Due to the use of uniform priors, the highest-likelihood estimates match the maximum a posteriori values.}
\label{tab:priors_vs_posteriors_wide}
\end{table*}
In the 1F scenario, the posterior prefers comparatively small values of $\nsat$, $K_{\rm sat}$, $E_{\rm sym}$, and $L_{\rm sym}$, together with an intermediate value of the Dirac effective mass, whereas $E_{\rm sat}$ remains essentially uninformed by the data. The three hyperon potentials remain broad over their prior support, indicating that the present dataset carries little direct information on the strange sector. By contrast, the delta resonance sector shows a preference for large values of both $x_{\sigma\Delta}$ and $x_{\omega\Delta}$. Their visible correlation is partly inherited from the prior strip enforced by Eqs.~(\ref{eq:deltaprior1},~\ref{eq:deltaprior2}), but the data reinforce it further by pinning the joint distribution toward the upper part of the allowed strip. 

A sequence of intermediate 1F runs, in which individual likelihood contributions are added one at a time, clarifies the origin of these trends. With the $\chi$EFT likelihood alone, the dominant effect is on the isovector sector: the low-density PNM band efficiently constrains $E_{\rm sym}$ and $L_{\rm sym}$ and shows a preference for comparatively large values of $m^\ast/m$. Adding the HIC information without astrophysical constraints leaves the low-density isovector structure largely intact, while further selecting the isoscalar sector: $m^\ast/m$ peaks around $0.8$ and $K_{\rm sat}$ settles in a region around $240\pm20\,\mathrm{MeV}$. Including instead the astrophysical dataset without HIC yields a comparable constraint on $K_{\rm sat}$, but pushes $m^\ast/m$ in the opposite direction, toward significantly smaller values, and drives the $\Delta$-resonances couplings to the large values found in the final 1F posterior. 

The clearest tension between the HIC and astrophysical data therefore concerns $m^\ast/m$: HIC prefers a comparatively large effective mass, whereas the astrophysical constraints favor a substantially smaller one. The final 1F posterior can thus be read as the outcome of a nontrivial compromise between these two tendencies. The effective mass settles at an intermediate value between the HIC-only and astro-only cases; $E_{\rm sym}$ and $L_{\rm sym}$ shift slightly downward with respect to the astro-only case while remaining broadly compatible with the $\chi$EFT-only result at the $1\sigma$ level; and $K_{\rm sat}$ sits near the lower edge of the $\sim 240\pm20\,\mathrm{MeV}$ region preferred by the partial runs.

At the same time, the $\Delta$ resonances sector is driven to large couplings, so that the model exploits the early appearance of $\Delta$ resonances to soften the EOS at intermediate densities, while relying on the associated repulsion to preserve sufficient high-density support for the heaviest pulsars. This structural behavior is qualitatively reminiscent of recent Bayesian hybrid-star analyses~\cite{Ayriyan:2025rub}, in which the favored solutions feature an early onset of quark matter, softening the EOS at low and intermediate densities, together with a sufficient high density stiffening mediated by vectorial repulsive interaction to sustain large maximum masses. In the present 1F posterior the same structural pattern emerges, but the mechanism remains entirely hadronic and is mediated by the $\Delta$ resonances sector rather than by deconfinement. 

To be more precise, we can identify two different contributions from astrophysical data to the parameter posterior. Astrophysical data suggesting intermediate mass and relatively low radii, such as PSR J0614-3329 and GW170817, provide a posterior very similar to the one obtained from HIC (namely, high $m^*/m$). On the other hand, massive PSR J0740+6620 and PSR J0952-0607 are the contributions requiring a small $m^*/m$, which provide the main tension, and higher values for the delta couplings. Thus, again, the tension is not actually between astrophysical and HIC data, but between HIC plus astrophysical data preferring small radii for $\sim 1.4\Msol$ objects and high-mass astrophysical data.

The 2F posterior differs from the 1F one in precisely the way expected if the support of the heaviest observed objects is shifted to the QS branch. The hadronic sector no longer needs to account for masses above $\sim 2\,\Msol$ and thus it is no longer forced into the finely balanced compromise between intermediate-density compactness and high-density support that characterizes the 1F posterior. This is reflected in the posterior by a preference for larger $m^\ast/m$ and a smaller and less constrained $\Delta$ resonance couplings than in the 1F case. On the other hand, the symmetry-sector parameters remain compatible with the 1F case. 

The quark-sector parameters are only moderately constrained. Very small values of $a_4$ are disfavored, consistently with the fact that they tend to promote rehadronization, which is excluded at the prior level through the condition Eq.~\eqref{eq:no_rehadronization}. The bag constant $B^{1/4}$ and the CFL pairing gap $\gap$ are not sharply pinned down individually, but they display a clear positive correlation, reflecting the fact that both contribute to setting the compactness and maximum mass of the QS branch: larger pairing gaps can compensate, at least partially, for larger values of $B$, provided that the resulting EOS still supports the heaviest observed objects within the Bodmer--Witten stability window.

\subsection{Equation of state and composition}
\label{sec:results_eos_composition}

\begin{figure*}[t]
    \centering
    \begin{subfigure}{0.49\textwidth}
        \centering
        \begin{overpic}[width=\linewidth]{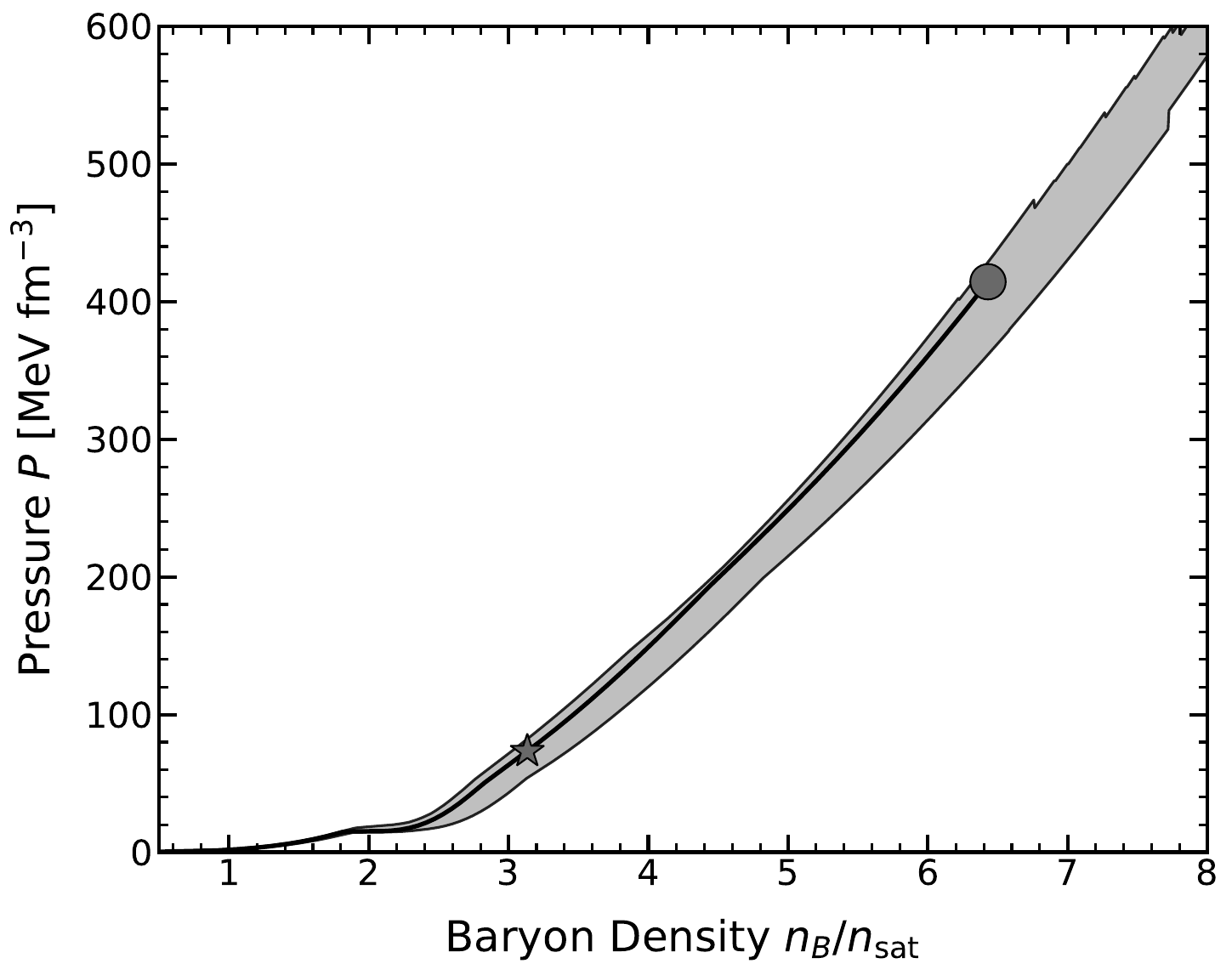}
            \put(43, 178){\textbf{(a)}}
        \end{overpic}
        \label{fig:thermo_1f}
    \end{subfigure}
    \hfill
    \begin{subfigure}{0.49\textwidth}
        \centering
        \begin{overpic}[width=\linewidth]{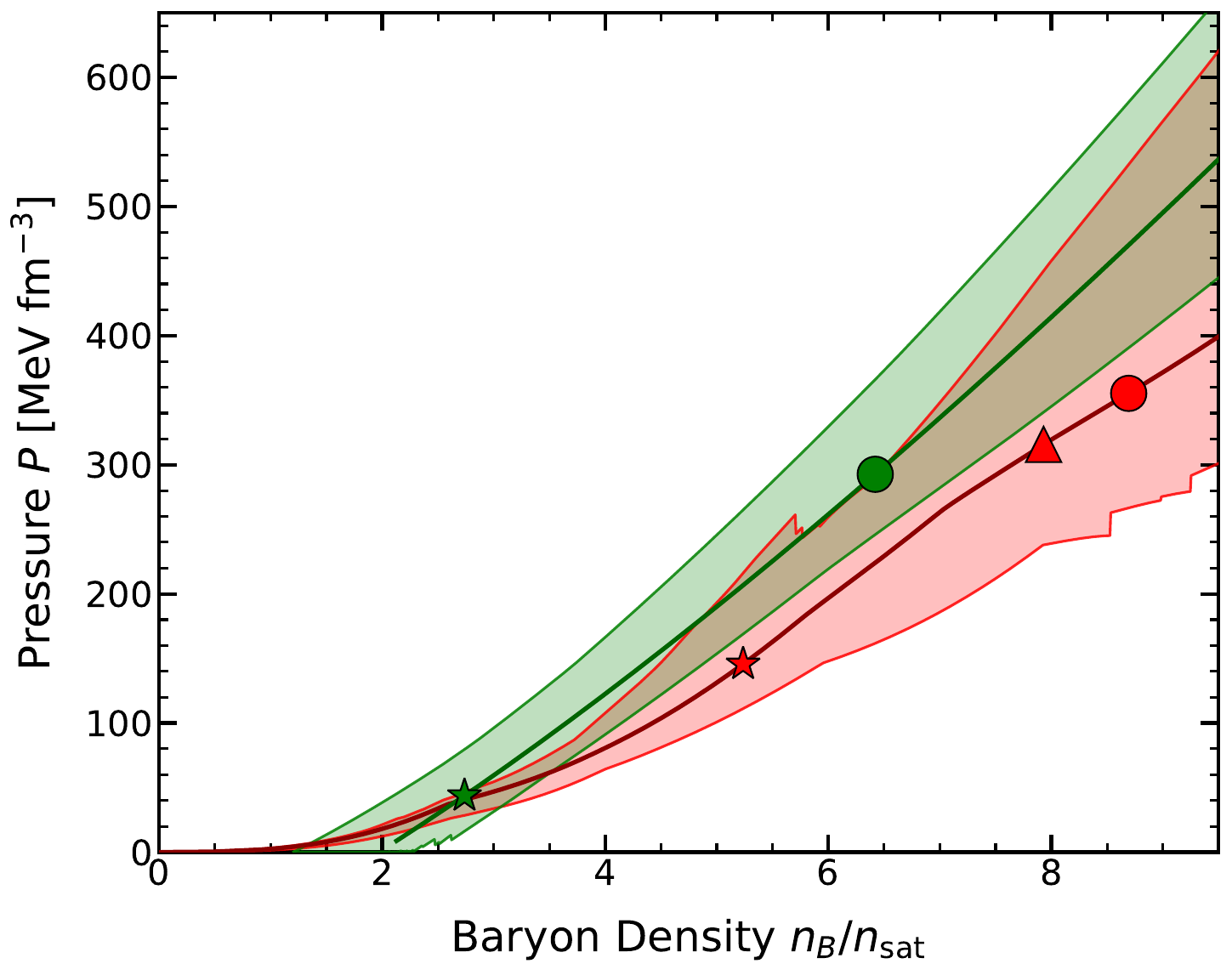}
            \put(43, 178){\textbf{(b)}}
        \end{overpic}
        \label{fig:thermo_2f}
    \end{subfigure}
    \caption{Pressure $P$ as a function of baryon density $n_B$ for the (a) 1F and (b) 2F scenarios. Shaded envelopes represent the 68\% highest-likelihood intervals, while the thick central lines denote the highest-likelihood models. Hadronic matter in the 1F scenario is shown in black, whereas in the 2F scenario, hadronic matter is shown in red and SQM in green. The star and circle markers indicate the central baryon density of respectively a $1.4\Msol$ and a $M_{f,\rm TOV}^{\rm max}$ compact object for the highest-likelihood parametrization, while the triangle marks the point where the strangeness density equals the saturation density for the HS in the 2F scenario, which sets $M_{\rm conv}$.}
    \label{fig:thermo_comparison}
\end{figure*}

\begin{figure*}[t]
    \centering
    
    \begin{subfigure}{0.49\textwidth}
        \centering
        \includegraphics[width=\linewidth]{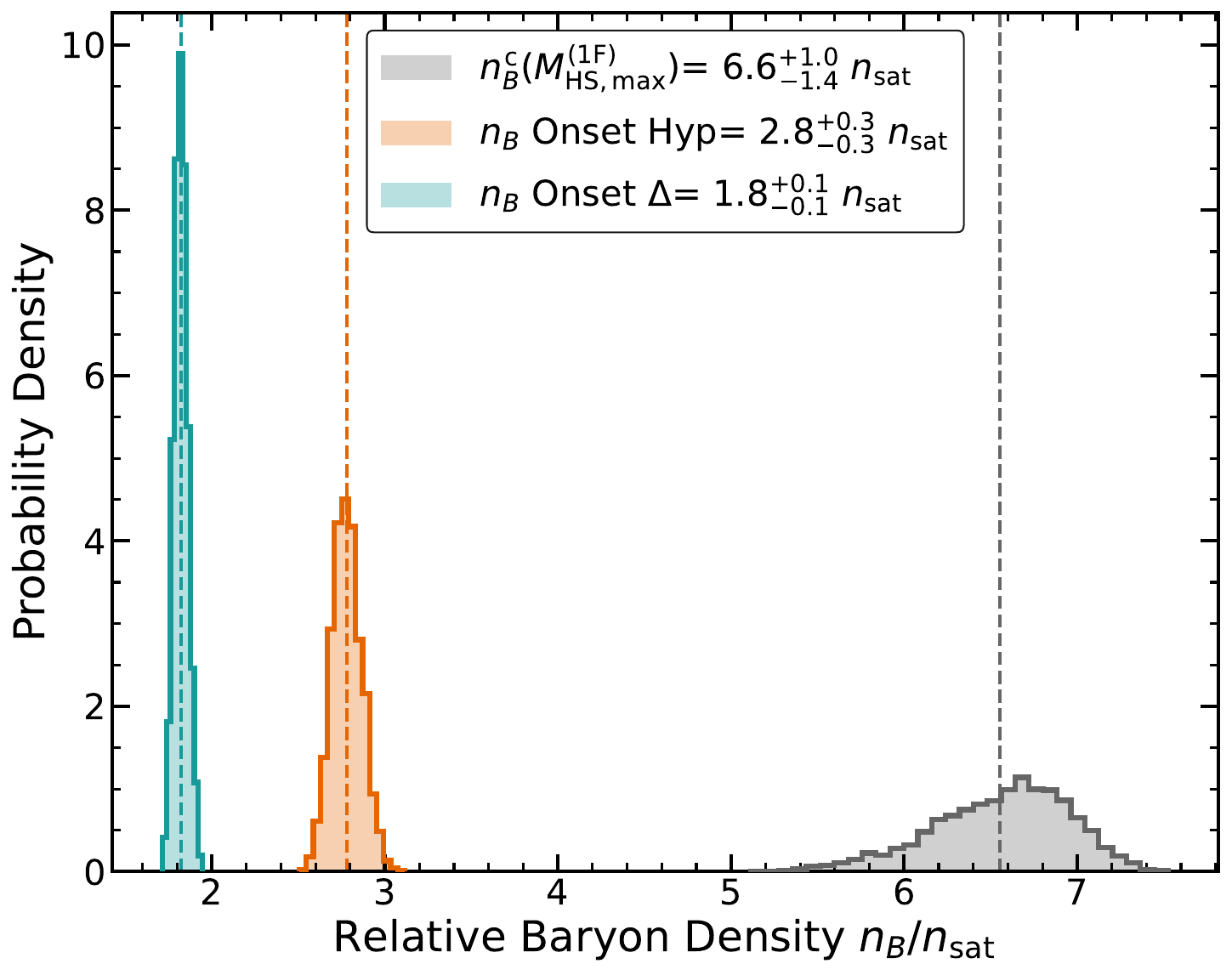}
        \caption{}
        \label{}
    \end{subfigure}
    \hfill
    \begin{subfigure}{0.49\textwidth}
        \centering
        \includegraphics[width=\linewidth]{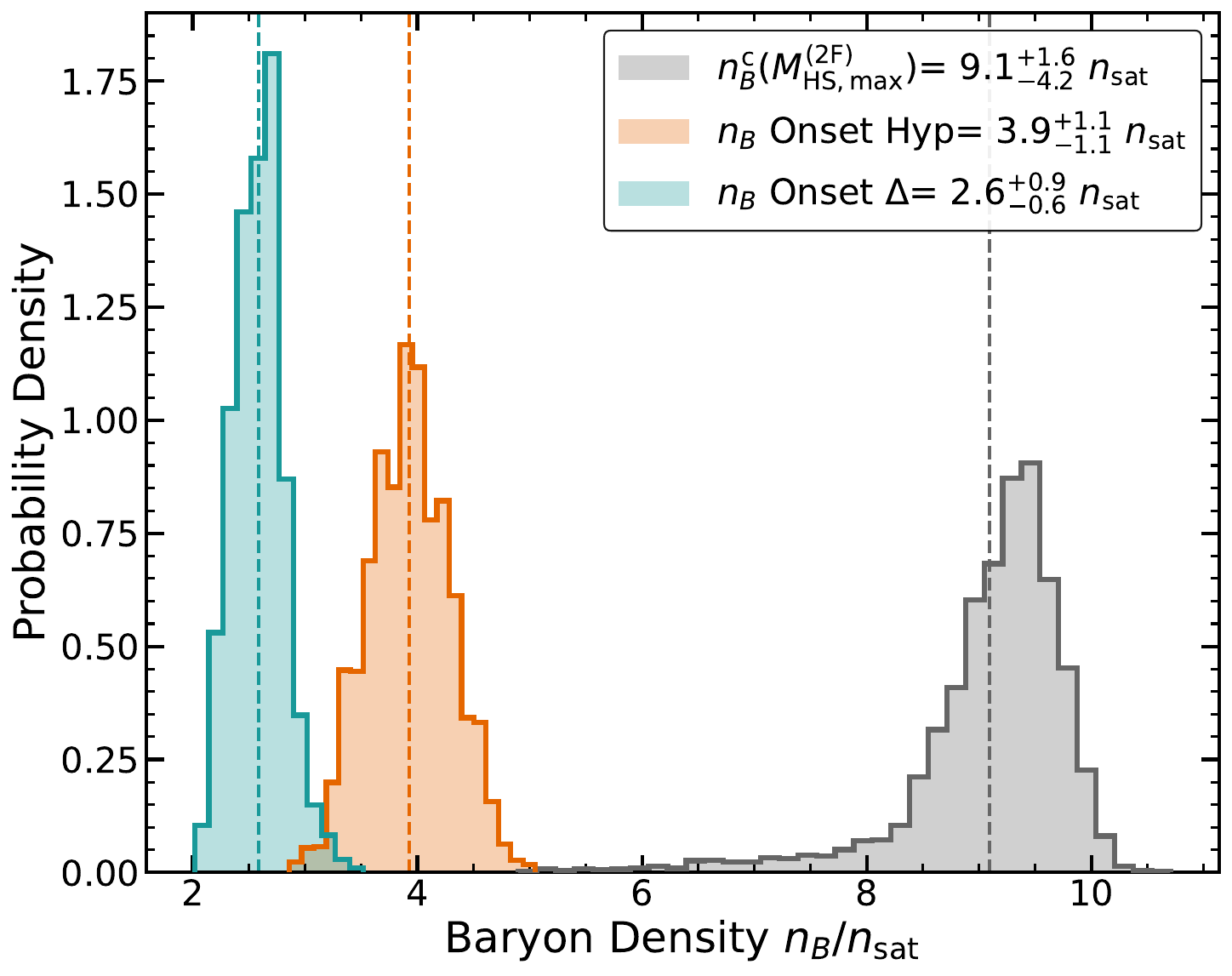}
        \caption{}
        \label{}
    \end{subfigure}
\caption{Probability density distributions of the central baryon density of the maximum TOV mass configuration (gray), the hyperon onset density (orange), and the $\Delta$-resonance onset density (cyan). Panel (a) shows the results for the 1F scenario, while panel (b) corresponds to the 2F scenario. These distributions are evaluated using the parametrizations that fall within the $68\%$ credible region (i.e., those producing the envelopes in Fig.~\ref{fig:thermo_comparison}). Dashed vertical lines denote the posterior means for each distribution.}
    \label{fig:dens_hist}
\end{figure*}

Figure~\ref{fig:thermo_comparison} shows the pressure as a function of the baryon density for the EOS parametrizations with the highest likelihoods in both the 1F and 2F scenarios. The solid lines represent the maximum-likelihood configurations, while the shaded areas indicate the $68\%$ credible regions. Additionally, the posterior distributions for the onset densities of $\Delta$ resonances and hyperons, as well as the central density of the maximum-mass configuration for these parametrizations, are summarized in Fig.~\ref{fig:dens_hist}.

The hadronic EOS is well constrained in the low-density regime (up to $\nsat$) by $\chi$EFT. At intermediate baryon densities, HIC data and astrophysical observations of relatively compact stars drive the preference for a soft EOS. Conversely, a stiff EOS is required at high baryon densities to support the most massive pulsars.

In the 1F scenario, the posterior EOS is forced into a compromise to satisfy both of these contrasting requirements. This balance is achieved through an early onset of $\Delta$ resonances, which appear at $n_B^{(\rm onset\, \Delta)} \simeq 1.8\nsat$, well before hyperons, which start to be produced at $n_B^{(\rm onset\, Hyp.)} \simeq 2.8\nsat$. Consequently, the first major rearrangement in the matter composition is driven by the $\Delta$ resonance sector rather than by strangeness. The appearance of $\Delta$ resonances induces an initial strong softening of the EOS, leading to a small region of quasi-constant pressure (and a quasi-inflection visible in the stellar sequence, see Sec.~\ref{sec:results_macroscopic}), followed by a strong stiffening of the $\Delta$ resonance channel (as discussed in Sec.~\ref{sec:results_posteriors}). Although the EOS exhibits this region of quasi-constant pressure, it is not a true first-order phase transition: the pressure remains strictly continuous without a Maxwell plateau. Notably, our present analysis explicitly excludes mechanically unstable EOS segments at the prior level, precluding the investigation of true first-order phase transitions triggered by the $\Delta$ resonances onset. A more general treatment would allow for parametrizations that lead to mechanical instabilities upon the appearance of new degrees of freedom, which could then be resolved through an appropriate phase transition construction.

On the other hand, in the 2F scenario, the metastable hadronic EOS easily accommodates the soft constraints at intermediate densities, while the required stiff behavior at high densities is naturally provided by the SQM branch. In the 2F posterior, the $\Delta$ onset is shifted to $n_B^{(\rm onset\, \Delta)} \simeq 2.6\nsat$, and the hyperon onset occurs at $n_B^{(\rm onset\, Hyp.)} \simeq 3.9\nsat$. Although the hierarchy between the $\Delta$ and hyperon onsets is maintained, both thresholds are pushed to higher densities compared to the 1F posterior.

Finally, the relative ordering of the $\Delta$ and hyperon thresholds has an additional implication for the 2F scenario. The early appearance of $\Delta$ resonances delays the buildup of strangeness in the hadronic phase, as already discussed in~\cite{Drago:2014oja}. This is relevant because hyperons provide the strange degrees of freedom from which an SQM seed may nucleate.
Thus, the $\Delta$ sector plays a dual role: it softens the hadronic EOS and, by postponing the appearance of the strangeness fraction needed for conversion to the QS branch, allows the HS branch to reach larger masses with respect to the case in which only hyperons are considered.


\begin{figure*}[t]
   \centering
    
   \begin{subfigure}{0.49\textwidth}
       \centering
       \includegraphics[width=\linewidth]{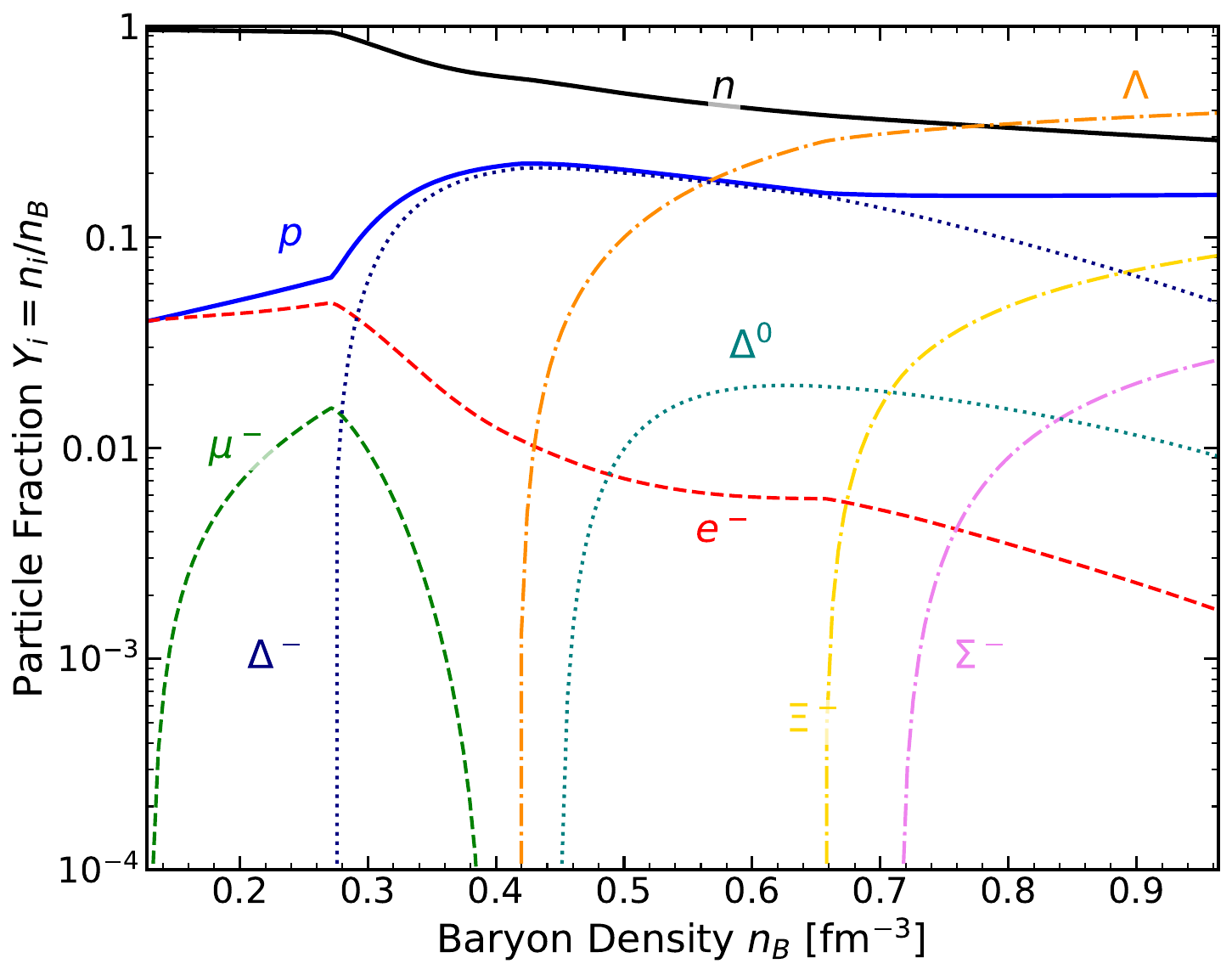}
       \caption{}
       \label{}
   \end{subfigure}
   \hfill
   \begin{subfigure}{0.49\textwidth}
       \centering
       \includegraphics[width=\linewidth]{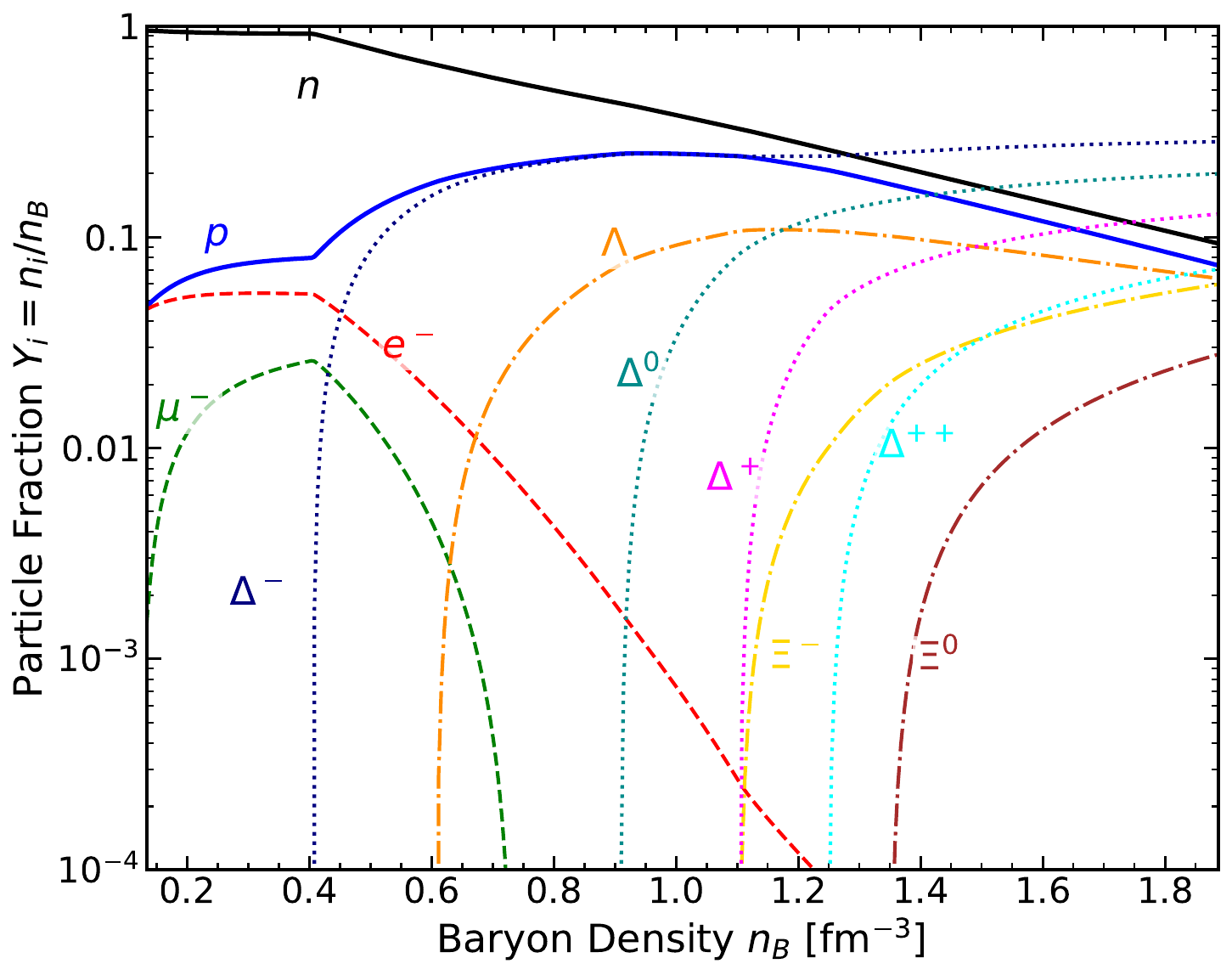}
       \caption{}
       \label{}
   \end{subfigure}
   \caption{Particle fractions $Y_i = n_i/n_B$ of hadronic matter as a function of the baryon density $n_B$ for the highest-likelihood models in the 1F (a) and 2F (b) scenarios. Both panels illustrate the composition of charge-neutral matter in $\beta-$equilibrium.}
   \label{fig:parti_frac}
\end{figure*}

As shown in Fig.~\ref{fig:parti_frac}, HSs in the 1F and 2F scenarios exhibit distinct average compositions regarding $\Delta$ and hyperon populations. Consequently, analyzing the thermal evolution of these compact stars provides a complementary method to discriminate between the two frameworks. Moreover, transport properties such as bulk and shear viscosity are highly sensitive to particle composition; therefore, phenomena like r-mode instability windows could be rather different between the two scenarios. Extending this analysis to incorporate cooling data will be taken into account for future research.

\subsection{Compact star properties}
\label{sec:results_macroscopic}

\begin{figure*}[t]
    \centering
    \begin{subfigure}{0.49\textwidth}
        \centering
        \begin{overpic}[width=\linewidth]{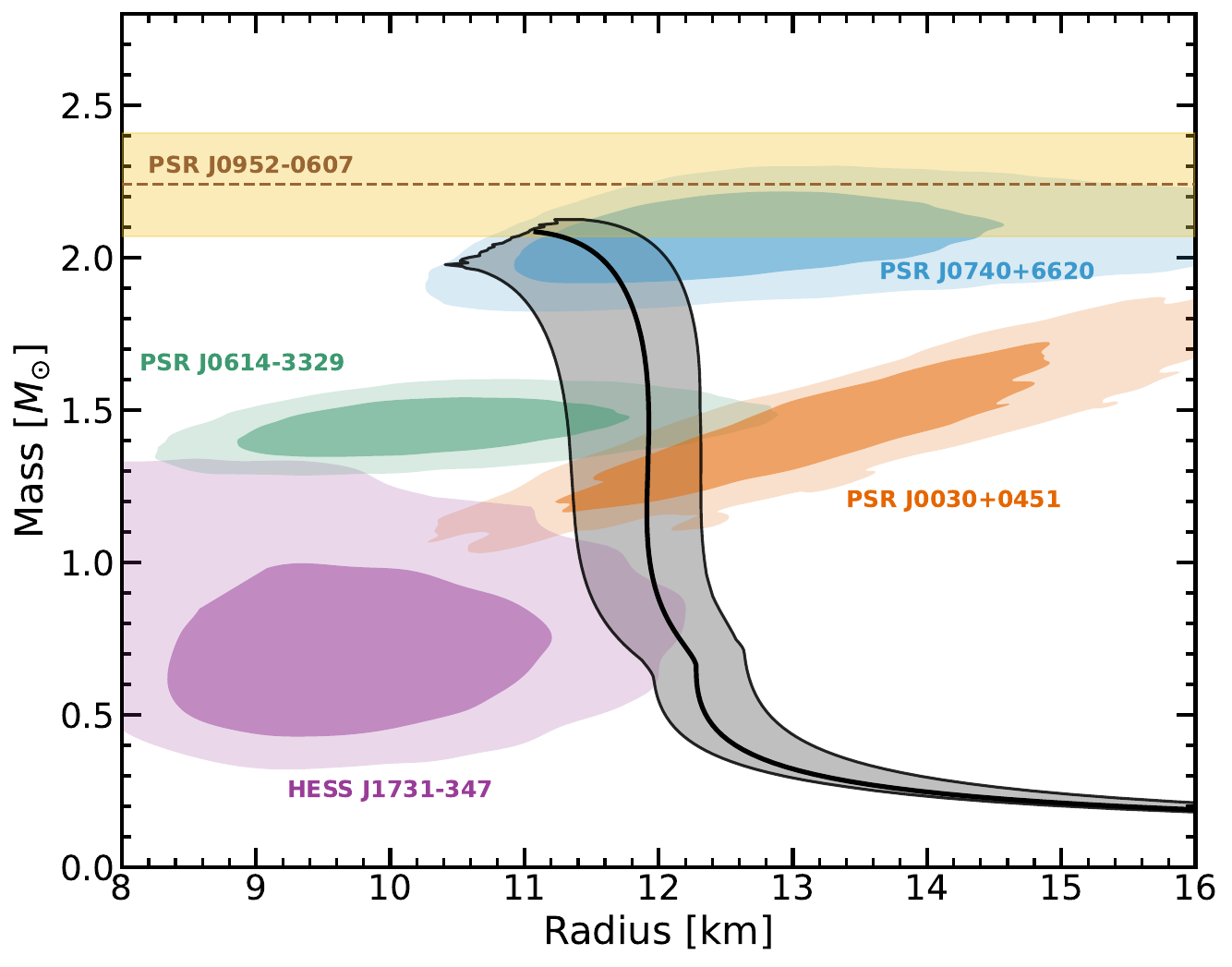}
            \put(33, 178){\textbf{(a)}} 
        \end{overpic}
        \label{fig:mr_1f}
    \end{subfigure}
    \hfill
    \begin{subfigure}{0.49\textwidth}
        \centering
        \begin{overpic}[width=\linewidth]{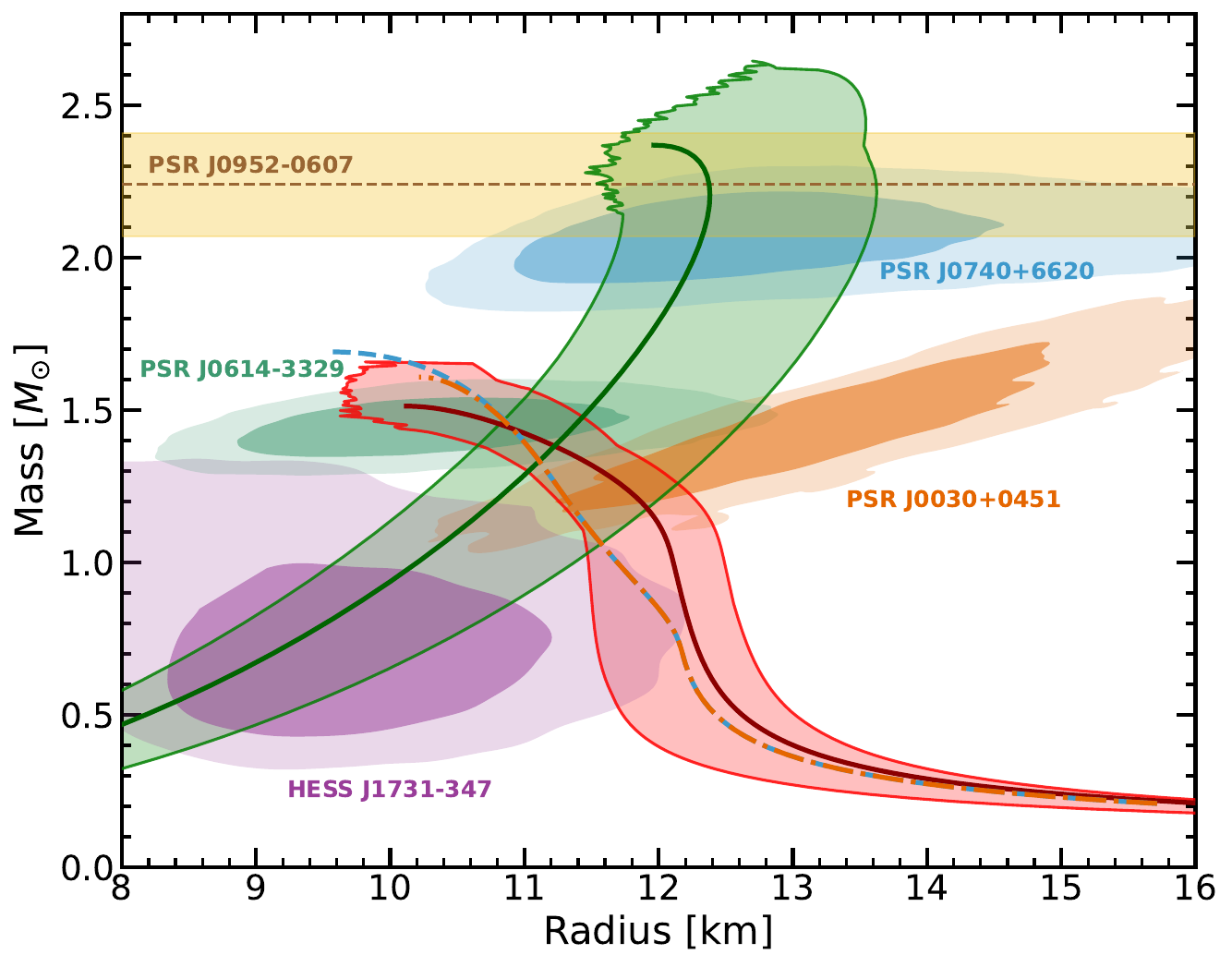}
            \put(33, 178){\textbf{(b)}}
        \end{overpic}
        \label{fig:mr_2f}
    \end{subfigure}
    \caption{Mass--radius relations for the 1F and 2F scenarios obtained from the TOV equations. Shaded envelopes and color schemes follow the conventions described in Fig.~\ref{fig:thermo_comparison}. Panel (a) shows the results for the 1F scenario, while panel (b) corresponds to the 2F scenario. Overlaid probability contours represent the 68\% and 95\% credible intervals for astrophysical constraints from NICER, HESS J1731--347, and GW170817. The blue and orange dashed/dot-dashed curves show the standard SFHo-HD reference models~\cite{Becerra:2025ryy,Bombaci:2020vgw,Drago:2013fsa} with $x_{\sigma\Delta}=1.15$ and $x_{\omega\Delta}=x_{\rho\Delta}=1$, respectively with and without the $\phi$ field.}
    \label{fig:mr_comparison}
\end{figure*}

\begin{figure*}[t]
    \centering
    \begin{subfigure}{0.49\textwidth}
        \centering
        \begin{overpic}[width=\linewidth]{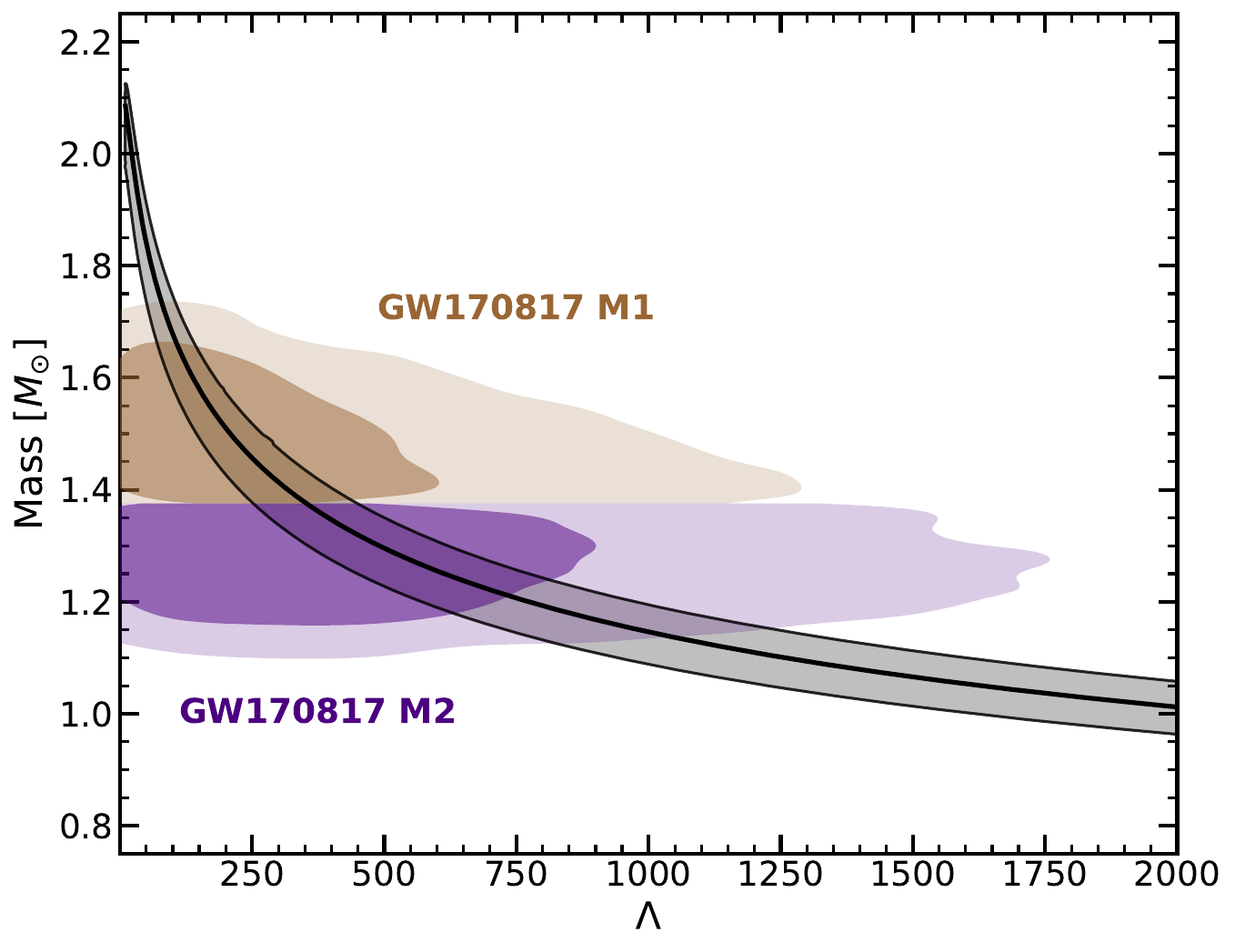}
            \put(33, 178){\textbf{(a)}}
        \end{overpic}
        \label{fig:lambda_1f}
    \end{subfigure}
    \hfill
    \begin{subfigure}{0.49\textwidth}
        \centering
        \begin{overpic}[width=\linewidth]{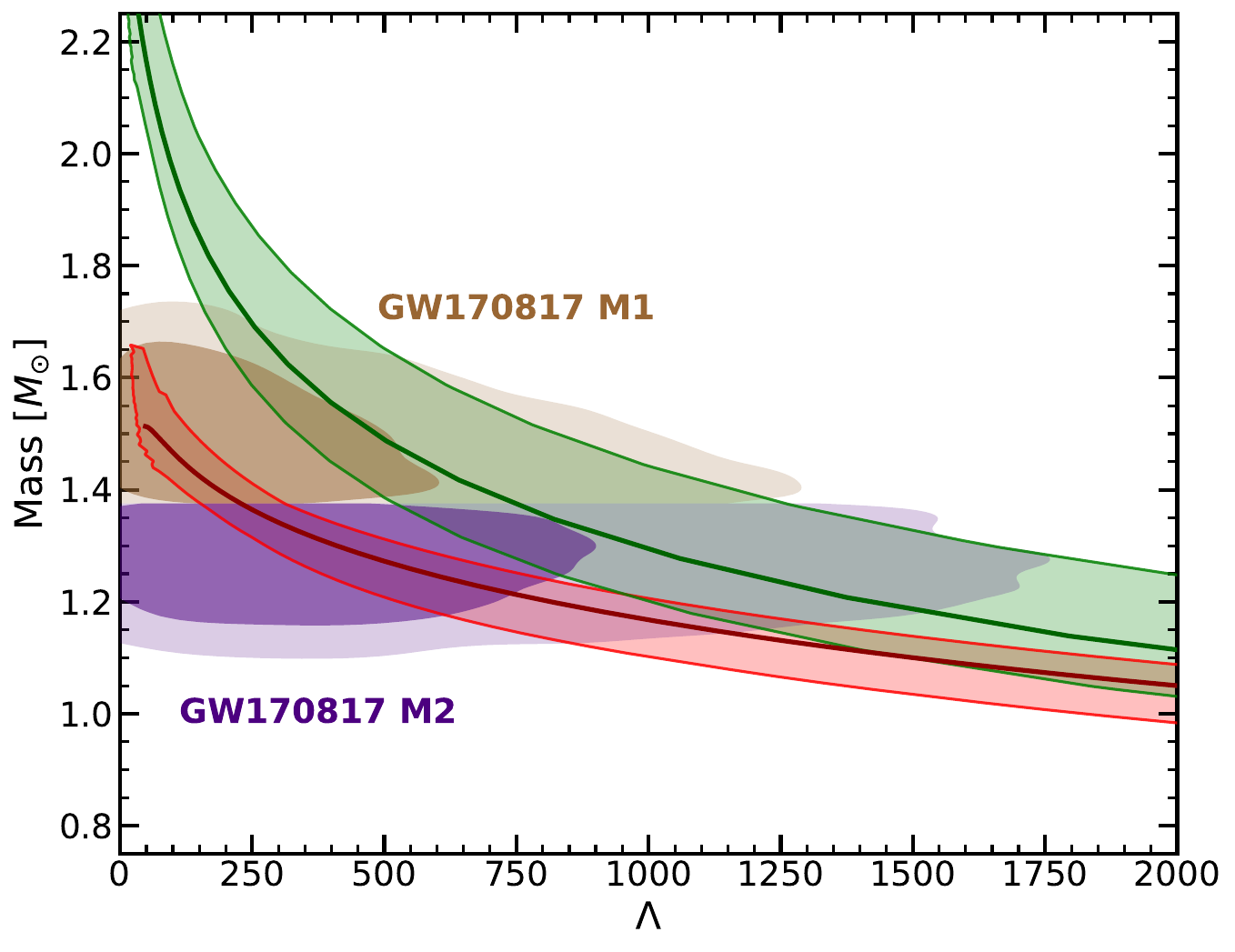}
            \put(33, 178){\textbf{(b)}}
        \end{overpic}
        \label{}
    \end{subfigure}
    \caption{Mass-tidal deformability $\Lambda$ relations for the 1F \textbf{a} and 2F \textbf{b} scenarios. Shaded envelopes and color schemes follow the conventions described in Fig.~\ref{fig:thermo_comparison}.}
    \label{fig:mlambda_comparison}
\end{figure*}

We now discuss the macroscopic stellar properties and how the results for the EOSs presented in Secs.~\ref{sec:results_posteriors} and~\ref{sec:results_eos_composition} reflect into the mass--radius ($M\text{-}R$) and mass--tidal-deformability ($M\text{-}\Lambda$) relations. The $68\%$ credible envelopes are shown in Fig.~\ref{fig:mr_comparison} and Fig.~\ref{fig:mlambda_comparison}.

In the 1F scenario, the $M\text{-}R$ envelope is shaped by a single-sequence compromise between compactness at intermediate masses and support for the heaviest pulsars. The strongest pull toward compact configurations comes from the HIC likelihood, the tidal information from GW170817, and especially the small radius inferred for PSR~J0614--3329. Since PSR~J0614--3329 lies close to the canonical HS mass, its small inferred radius directly constrains the intermediate-density part of the EOS. At the same time, the same hadronic sequence must remain sufficiently stiff at higher densities to support PSR~J0740$+$6620 and PSR~J0952--0607. In order to fulfil those constraints, the resulting $M\text{-}R$ relation develops the quasi-inflection visible in Fig.~\ref{fig:mr_comparison}(a). This feature is the macroscopic counterpart of the rapid $\Delta$-driven softening followed by stiffening discussed in Sec.~\ref{sec:results_eos_composition}. 

The same compromise is visible in the $M\text{-}\Lambda$ relation shown in Fig.~\ref{fig:mlambda_comparison}(a). In the 1F case, the posterior envelope follows a relatively narrow hadronic band: the EOS must be soft enough to keep $\Lambda$ small for intermediate values of the mass, as required by GW170817, but not too soft to violate the 
two-solar-mass limit. It is important, however, not to interpret the two GW170817 shaded regions in Fig.~\ref{fig:mlambda_comparison} as two independent measurements of the two component stars. They are two-dimensional projections of the full four-dimensional posterior density
$q_{\rm GW170817}(M_1,M_2,\Lambda_1,\Lambda_2)$ used in the likelihood. The component masses and tidal deformabilities are correlated through the accurately measured chirp mass, the mass ratio, and the effective tidal deformability $\tilde{\Lambda}$. Thus, Fig.~\ref{fig:mlambda_comparison} should be read only as a visualization of the marginal support of the GW posterior, while the actual constraint entering the inference is the joint four-dimensional one.


\begin{figure}
    \centering
    \includegraphics[width=1\linewidth]{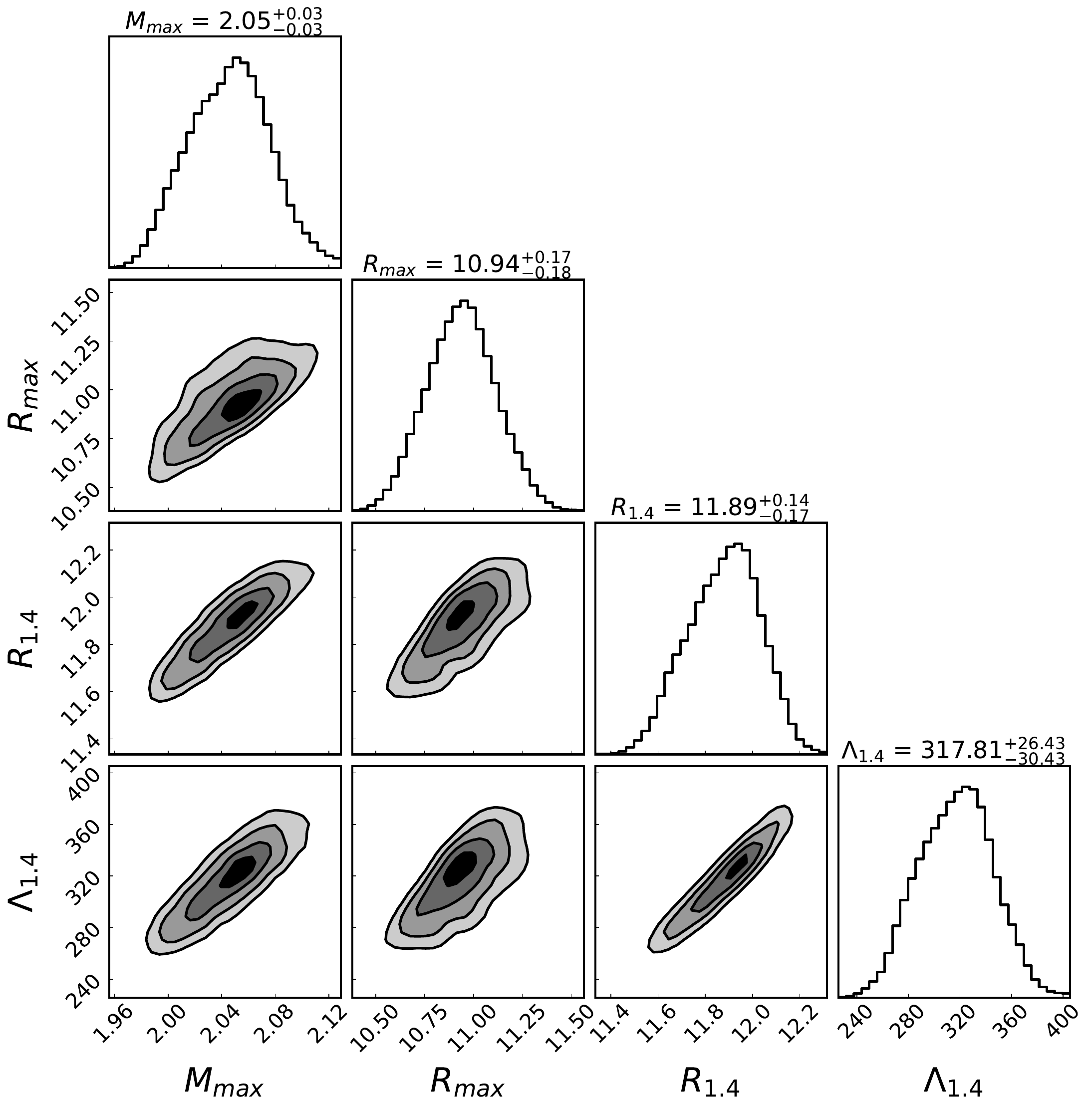}
    \caption{Macroscopic observable posteriors for the HSs in the 1F scenario.}
    \label{fig:macro_1f}
\end{figure}

The posterior distributions of the main macroscopic quantities in the 1F scenario are summarized in Fig.~\ref{fig:macro_1f}. The maximum mass is tightly concentrated around
$M_{\rm HS,\max}^{\rm(1F)}=2.05^{+0.03}_{-0.03}\,\Msol$, close to the lower edge allowed by the heaviest precisely measured pulsars. This value satisfies the robust $\sim2\,\Msol$ constraint, but it is marginally consistent with 
the PSR~J0952--0607 constraint. It is also compatible with the upper bounds inferred from the standard one-family interpretations of GW170817, which typically suggest $M_{\rm HS,\max}^{\rm(1F)}\lesssim2.2$--$2.3\,\Msol$, \cite{Rezzolla:2017aly,Shibata:2019ctb}. 
The corresponding radii and tidal deformability,
$R_{\rm HS,\max}^{\rm(1F)}=10.94^{+0.17}_{-0.18}\,\mathrm{km}$,
$R_{\rm HS,1.4}^{\rm(1F)}=11.89^{+0.14}_{-0.17}\,\mathrm{km}$, and
$\Lambda_{\rm HS,1.4}^{\rm(1F)}=317^{+26}_{-30}$,
place the 1F posterior in a relatively compact region of the standard HS parameter space. The value of $R_{\rm HS,1.4}^{\rm(1F)}$ is close to the lower side of many multimessenger EOS reconstructions, consistently with the small-radius pull from PSR~J0614--3329 and the tidal constraints from GW170817, while $\Lambda_{\rm HS,1.4}^{\rm(1F)}$ is sufficiently small to avoid the large-tidal-deformability region disfavored by the inspiral data.

\begin{figure*}[t]
    \centering
    \begin{subfigure}{0.49\textwidth}
        \centering
        \begin{overpic}[width=\linewidth]{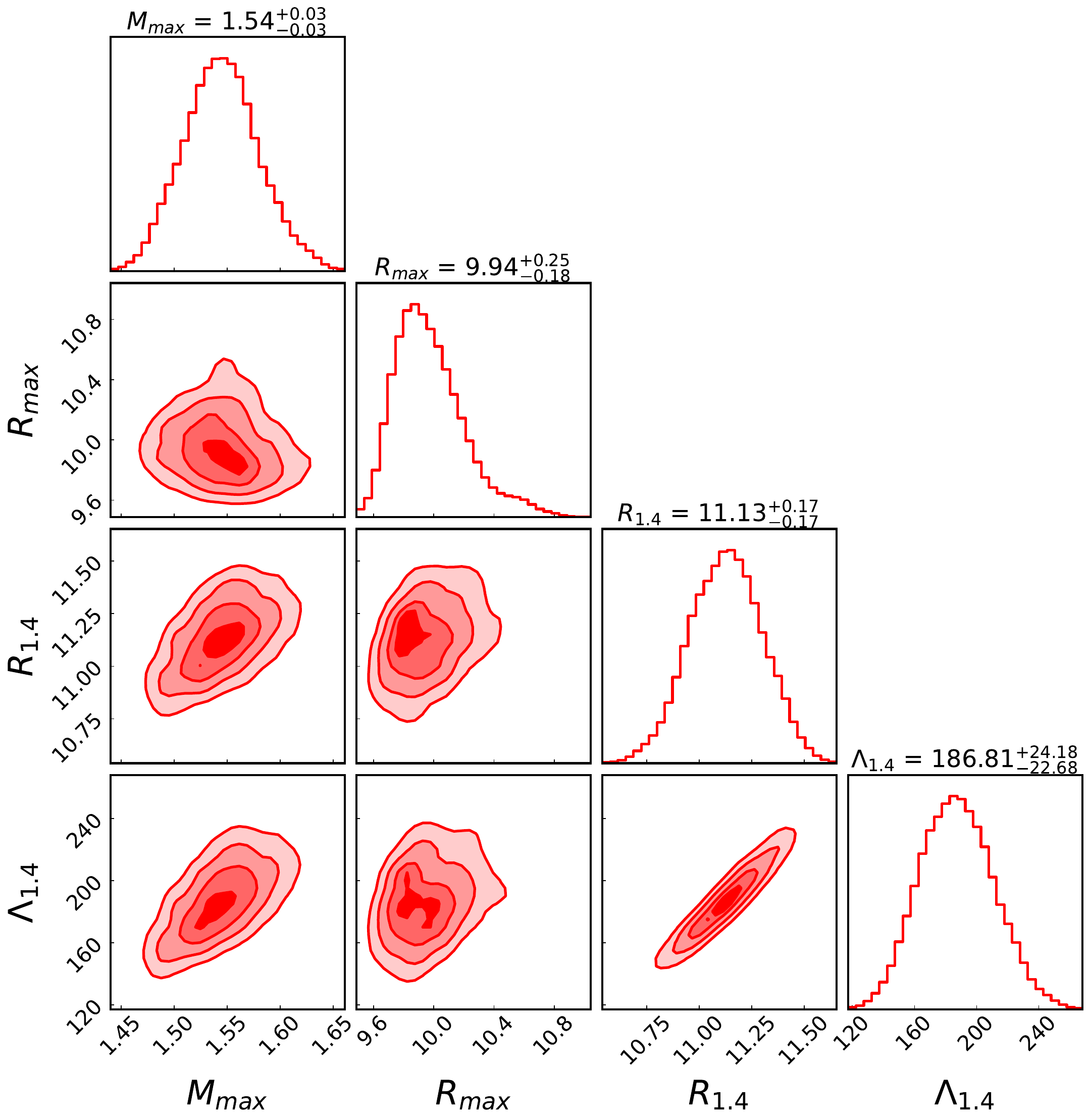}
            \put(223, 238){\textbf{(a)}}
        \end{overpic}
        \label{fig:corner_macro_hadron_2F}
    \end{subfigure}
    \hfill
    \begin{subfigure}{0.49\textwidth}
        \centering
        \begin{overpic}[width=\linewidth]{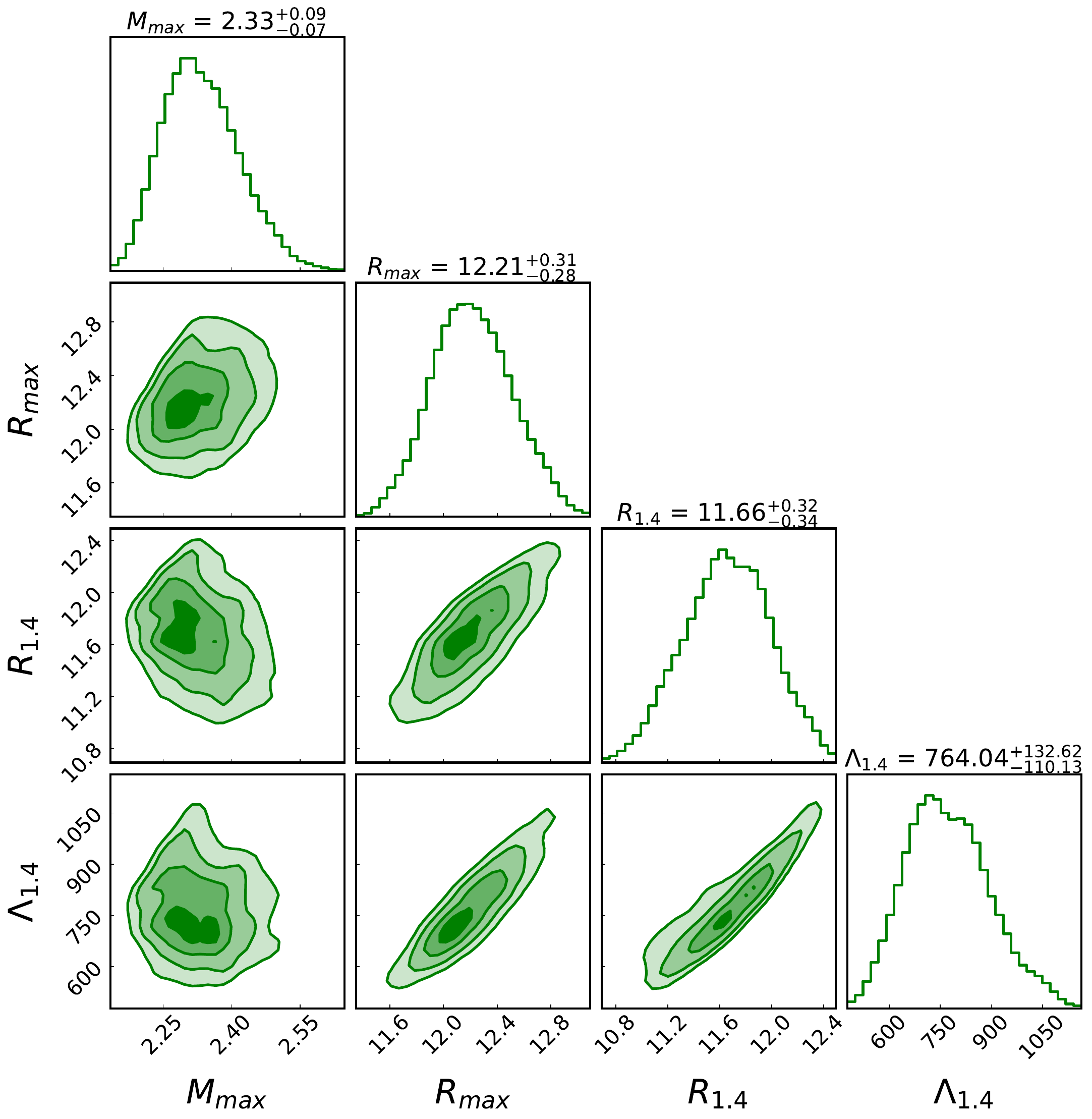}
            \put(223, 238){\textbf{(b)}}
        \end{overpic}
        \label{}
    \end{subfigure}
    \caption{Macroscopic observable posteriors for the HSs (panel a) and QSs (panel b) in the 2F scenario.}
    \label{fig:corner_macro_2F}
\end{figure*}

In the 2F scenario, the same observational requirements are accommodated in a structurally different way. The posterior support separates into a compact hadronic branch and a stiffer self-bound QS branch, as shown in Fig.~\ref{fig:mr_comparison}(b). The hadronic branch occupies the low-radius, low- and intermediate-mass region. The high-mass constraints, instead, are supported by the QS branch. 

This separation is also evident in Fig.~\ref{fig:mlambda_comparison}(b). The hadronic branch predicts smaller tidal deformabilities at intermediate masses, while the QS branch extends to larger masses and follows a distinct $M$--$\Lambda$ trajectory. 

The macroscopic posteriors for the 2F scenario, displayed in Fig.~\ref{fig:corner_macro_2F}, make the division of roles between the two branches explicit. The hadronic branch has
$M_{\rm HS,max}^{\rm(2F)}=1.54^{+0.03}_{-0.03}\,\Msol$,
$R_{\rm HS,max}^{\rm(2F)}=9.94^{+0.25}_{-0.18}\,\mathrm{km}$,
$R_{\rm HS,1.4}^{\rm(2F)}=11.13^{+0.17}_{-0.17}\,\mathrm{km}$, and
$\Lambda_{\rm HS,1.4}^{\rm(2F)}=186^{+24}_{-23}$.
These values would be excluded in a one-family interpretation, because such a branch cannot support the heaviest pulsars.

The QS branch, by contrast, reaches
$M_{\rm QS,max}^{\rm(2F)}=2.33^{+0.09}_{-0.07}\,\Msol$,
with
$R_{\rm QS,max}^{\rm(2F)}=12.21^{+0.31}_{-0.28}\,\mathrm{km}$,
$R_{\rm QS,1.4}^{\rm(2F)}=11.66^{+0.32}_{-0.34}\,\mathrm{km}$, and
$\Lambda_{\rm QS,1.4}^{\rm(2F)}=764^{+132}_{-110}$.
Thus, the QS branch comfortably supports the largest masses used in the present analysis, including the PSR~J0952--0607 constraint, while remaining inside the Bodmer--Witten stability window imposed in the prior. The relatively large value of $\Lambda_{1.4}^{\rm QS}$ compared with the hadronic branch, despite comparable radii, reflects the different internal structure and surface properties of self-bound stars. 

As a useful benchmark, Fig.~\ref{fig:mr_comparison} also shows two reference SFHo-HD sequences commonly adopted in two-families scenario  studies~\cite{Bombaci:2020vgw,Becerra:2025ryy,Guerrini:2026_inpreparation}, for which $x_{\sigma\Delta}=1.15$ and $x_{\omega\Delta}=x_{\rho\Delta}=1$, with and without the $\phi$ field. These curves are broadly compatible with the inferred 2F hadronic band, but they lie on its more compact side: relative to our maximum-likelihood EOS, the standard SFHo-HD setup produces a stronger softening associated with the onset of $\Delta$ resonances.

A possibly interesting complementary analysis concerns the population of compact stars. The observed mass distribution of compact stars, in particular of recycled millisecond pulsars, has been argued to be bimodal, with a low-mass component near $\sim 1.4\Msol$ and a high-mass component near $\sim 1.8\Msol$~\cite{Antoniadis:2016hxz}. While such a feature is naturally ascribed to distinct formation channels and birth masses, we note that in the 2F scenario a bimodal-looking population could also arise as the superposition of two individually unimodal distributions associated with the HS and the QS branches, which populate partially offset mass ranges. Assessing this possibility would require modeling the formation and production scenarios of the two families and their relative abundance as a function of mass, i.e.\ the weights $\eta_f^{(2F)}(M)$ introduced in Sec.~\ref{sec:priors_mass_distributions}; such a population-level analysis is beyond the scope of the present work but could provide additional means of discriminating between the 1F and 2F scenarios.

As in the 1F case, the GW170817 contours shown in the figure are only two-dimensional projections of the full four-dimensional posterior. This point is especially important in the 2F scenario, because each point in the GW170817 posterior can be associated with different latent family assignments for the two components: HS--HS, HS--QS, QS--HS, or QS--QS. The probability of a given channel is therefore not obtained by independently matching the two components to projected $M$--$\Lambda$ contours, but by evaluating the joint density $q_{\rm GW170817}(M_1,M_2,\Lambda_1,\Lambda_2)$ along the corresponding pair of theoretical branches, as in Eq.~\eqref{eq:L_GW170817_2F_compact}.

To illustrate this point, we consider the EOS with the highest likelihood, shown by the solid red and green curves in Fig.~\ref{fig:mlambda_comparison}(b). For this EOS we evaluate the four branch-conditioned contributions to the GW170817 likelihood by imposing the theoretical $\Lambda_f^{\rm TOV}(M)$ relation using the highest likelihood parametrization and by marginalizing over the component masses. Here the first label refers to the more massive component, $M_1\geq M_2$. The corresponding EOS-conditioned effective tidal deformabilities are
\begin{align}
\tilde{\Lambda}_{\rm HSHS}^{\rm GW170817}=239\pm33,\qquad
\tilde{\Lambda}_{\rm HSQS}^{\rm GW170817}=550\pm42,\\
\tilde{\Lambda}_{\rm QSHS}^{\rm GW170817}=553\pm48,\qquad
\tilde{\Lambda}_{\rm QSQS}^{\rm GW170817}=850\pm60.\notag
\end{align}
Replacing one HS with a QS moves the binary toward larger tidal deformability, and the QS--QS interpretation gives the largest $\tilde{\Lambda}$ among the four possibilities. 
For comparison, the same estimation in the 1F scenario gives $\tilde{\Lambda}_{\rm HSHS}^{\rm GW170817}=367\pm31$, namely between the 2F HS-HS and 2F mixed channels.
Again for the maximum-likelihood EOS, the branch-conditioned reconstruction of GW170817 gives essentially the same mass sector in all 2F channels, with $M_1\simeq 1.47\,\Msol$, $M_2\simeq 1.27\,\Msol$, and $q\simeq 0.87$. The relevant channel dependence appears instead in the tidal deformabilities, since the same mass distribution is projected onto different HS or QS $M$--$\Lambda$ branches.

Normalizing the four channel likelihoods for the same maximum-likelihood EOS gives
\begin{align}
\prob_{\rm HSHS}^{\rm GW170817}\simeq0.61,\qquad
\prob_{\rm HSQS}^{\rm GW170817}\simeq0.16,\\
\prob_{\rm QSHS}^{\rm GW170817}\simeq0.19,\qquad
\prob_{\rm QSQS}^{\rm GW170817}\simeq0.04.\notag
\end{align}

Thus, for this particular EOS\footnote{
These numbers should not be interpreted as final posterior probabilities for the four possible GW170817 compositions. They are conditional on a single representative EOS, namely the maximum-likelihood model, and on the adopted branch priors. A complete probability assignment would require computing the normalized channel evidences, i.e. marginalizing each channel likelihood over the full EOS-parameter posterior rather than evaluating it only at the best-likelihood point. The present calculation should therefore be regarded as a diagnostic of the best-fit EOS rather than as a population-level inference of the GW170817 family assignment.} 
and for the adopted branch priors, the HS--HS interpretation is the most probable one. However, the mixed channels are not negligible: taken together, HS--QS and QS--HS account for about one third of the normalized likelihood. This shows that, with the data included in the present analysis alone, the family assignment of the two compact objects in GW170817 cannot be uniquely inferred. 
The QS--QS channel is instead disfavored by the GW170817 posterior for this EOS, mainly because the corresponding tidal deformability lies in the high-$\tilde{\Lambda}$ tail of the distribution.

Additional multimessenger information may help to further discriminate among the possible interpretations. In particular, the kilonova AT2017gfo associated with GW170817 constrains the ejecta properties and the post-merger evolution, and may therefore provide complementary information on the nature of the binary; see the discussion in Sec.~\ref{sec:gw170817}. 

\begin{figure}
    \centering
    \includegraphics[width=1\linewidth]{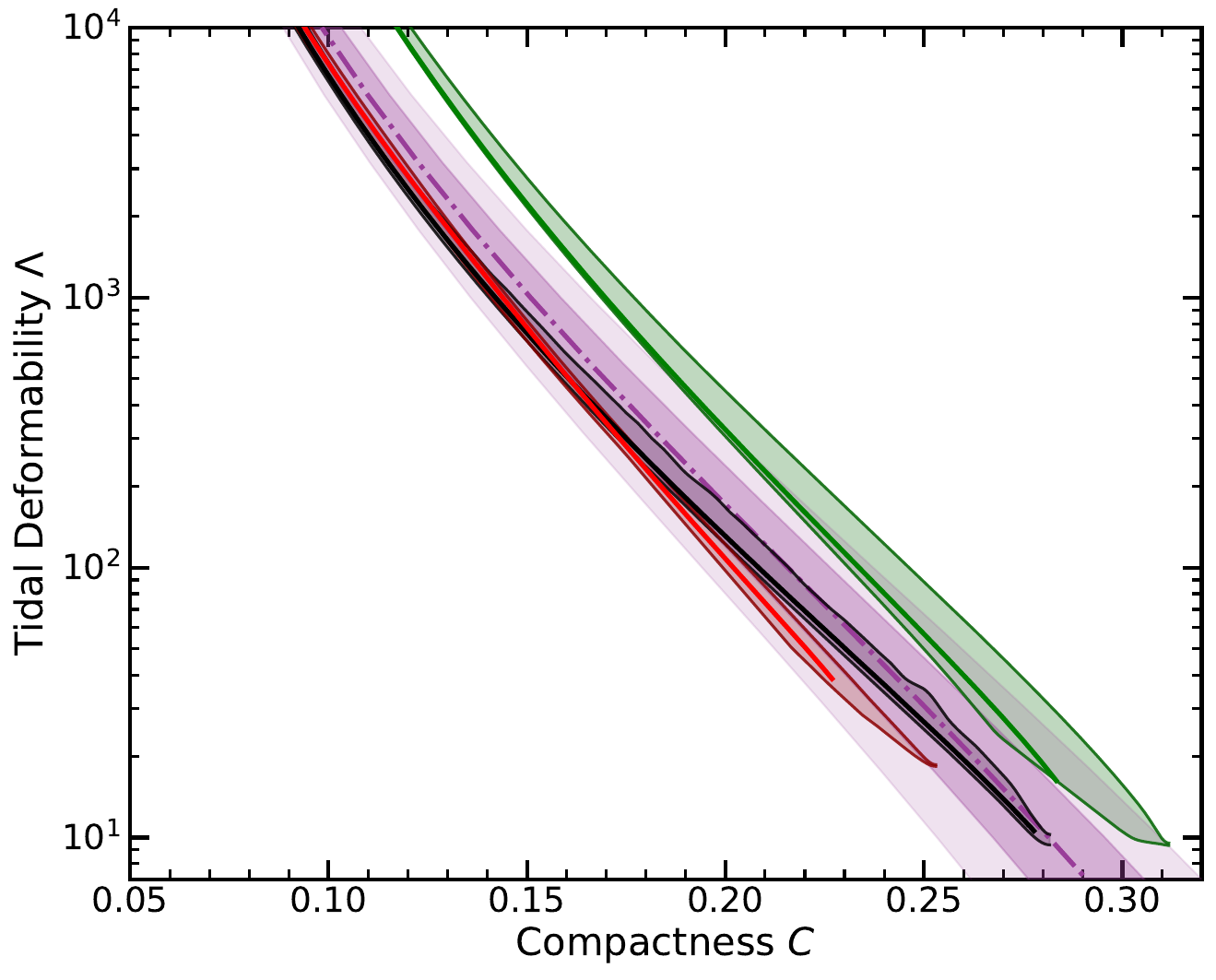}
    \caption{Compactness-tidal deformability $\Lambda$ relations for the 1F and 2F scenarios. Shaded envelopes and color schemes for the hadronic and strange quark branches follow the conventions described in Fig.~\ref{fig:mr_comparison}. The purple curve shows the $\Lambda$--$C$ quasi-universal relation introduced in~\cite{Maselli:2013mva} and plotted using the updated fit of~\cite{Godzieba:2020bbz}. The purple shaded bands are indicative $\pm 5\%$ and $\pm 10\%$ relative-error envelopes motivated by the residuals quoted in~\cite{Godzieba:2020bbz}, and should not be interpreted as Bayesian credible intervals.}
    \label{fig:compatness_compare}
\end{figure}

Finally, Fig.~\ref{fig:compatness_compare} shows the tidal deformability as a function of the compactness $C=M/R$ of the 1F posterior and of the two branches of the 2F posterior, and compare them with the quasi-universal $C$--$\Lambda$  relation~\cite{Maselli:2013mva,Yagi:2016bkt,Godzieba:2020bbz,Lowrey:2024anh}, shown as the purple band. 
The 1F posterior and the HS branch of the 2F posterior follow the hadronic band, while the QS branch departs from it. A direct test based on Fig.~\ref{fig:compatness_compare} alone would require an independent measurement of the compactness of the inspiral components, which the gravitational-wave inspiral signal cannot provide. 

\subsection{Model comparison}
\label{sec:results_bayesianfactors}

Table~\ref{Tab:1F_comparisonhesse_HIC} reports the Bayesian evidence for the 1F and 2F hypotheses under different combinations of astrophysical and laboratory data. Since the 2F hypothesis contains three additional parameters for the SQM EOS, the comparison cannot be based on the highest-likelihood alone. The relevant quantity is the Bayesian evidence, which averages the likelihood over the prior volume and therefore accounts for the larger parameter space of the 2F model. All rows include the $\chi$EFT constraint, while the last three rows additionally include the HIC flow data.

\begin{table}[htbp]
    \centering
    \begin{tabular}{l|c|c|c}
        \hline
        \hline
        \textbf{Data} ($\chi$EFT +)&
        \textbf{$\ln \mathcal{Z}_{\rm 1F}$} &
        \textbf{$\ln \mathcal{Z}_{\rm 2F}$} &
        \textbf{$\ln \mathcal{B}_{\rm 2F,1F}$} \\
        \hline
 Astro. (no HESS, no J0614)  & $-28.5$ & $-25.5$ & $3.0$ \\
        \hline
        Astro. (no HESS)  & $-32.2$ & $-26.5$ & $5.7$ \\
        \hline
        All astro. & $-37.6$ & $-28.4$ & $9.2$ \\
        \hline
        HIC + astro. (no HESS, no J0614)  & $-32.1$ & $-25.5$ & $6.5$ \\
        \hline
        HIC + astro. (no HESS)  & $-34.8$ & $-26.3$ & $8.5$ \\
        \hline
        HIC + all astro. & $-40.5$ & $-28.6$ & $11.9$ \\
        \hline
        \hline
    \end{tabular}
    \caption{Bayesian log-evidence $\ln \mathcal{Z}$ evaluated for different combinations of observational and laboratory constraints. The last column reports the log-Bayes factor $\ln \mathcal{B}_{\rm 2F,1F}=\ln \mathcal{Z}_{\rm 2F}-\ln \mathcal{Z}_{\rm 1F}$.}
    \label{Tab:1F_comparisonhesse_HIC}
\end{table}

The 2F scenario is favoured for all dataset combinations considered. In the most conservative astrophysical setup, where both HESS~J1731--347 and PSR~J0614--3329 are removed, we already find $\ln \mathcal{B}_{\rm 2F,1F}=3.0$. This indicates that, within the explored priors, the additional QS branch is not penalized by the evidence and already helps to accommodate the combination of heavy-pulsar and tidal constraints without forcing all objects to lie on a single hadronic sequence. Including PSR~J0614--3329 while still excluding HESS raises the preference to $5.7$. PSR~J0614--3329 contributes in the expected direction because its small inferred radius for an intermediate value of the mass enhances, in a 1F interpretation, the tension between intermediate-density softness and the high-density stiffness required to support the massive pulsars. Adding HESS~J1731--347 further raises the preference to $9.2$. The role of HESS~J1731--347 is qualitatively different: a very light and compact object is difficult to accommodate on a single ordinary hadronic sequence, while in 2F it can be naturally associated with the low-mass part of the QS branch.

The inclusion of the HIC flow constraint further increases the preference for 2F across all astrophysical sub-datasets. Without the HESS~J1731--347 and PSR~J0614--3329 data, $\ln \mathcal{B}_{\rm 2F,1F}$ increases from $3.0$ to $6.5$; with PSR~J0614--3329 included, from $5.7$ to $8.5$; and for the full dataset, from $9.2$ to $11.9$. As discussed above, the HIC constraint restricts the allowed values of the pressure at intermediate densities. This makes the 1F compromise more difficult: the same hadronic branch must remain sufficiently soft to satisfy the low-radius and HIC information, but also stiff enough at higher densities to support PSR~J0740$+$6620 and PSR~J0952--0607. In the 2F scenario this tension is relaxed, because the hadronic branch can remain relatively soft while the heaviest compact stars can be described by the QS branch.

Overall, the Bayes factors indicate that the current dataset is more naturally organized by two disconnected compact-star families than by a single purely hadronic sequence. For the complete dataset we find $\ln \mathcal{B}_{\rm 2F,1F}=11.9$, corresponding to very strong, or decisive, support for the 2F scenario on conventional Bayes-factor scales. This qualitative label should not be overinterpreted, since the numerical evidence depends on the adopted EOS models, parametrization, prior ranges, source-mass priors, and dataset selection. The robustness of this conclusion with respect to specific modeling assumptions and analysis choices is examined in Sec.~\ref{sec:consistency}.

\subsection{Consistency checks}
\label{sec:consistency}

The model comparison in Sec.~\ref{sec:results_bayesianfactors} shows a preference for the 2F scenario. We now test how stable this conclusion is under a few changes in the analysis setup.
The purpose of these checks is not to define new baseline models, but to verify that the preference for 2F is not driven by a single technical choice. Unless otherwise stated, all tests are performed for the HIC+all-astrophysical dataset, for which the baseline result is $\ln\mathcal{Z}_{\rm 1F}=-40.5$, $\ln\mathcal{Z}_{\rm 2F}=-28.6$, and $\ln\mathcal{B}_{\rm 2F,1F}=11.9$.

First, fixing the hyperon potentials to the central values of their priors gives $\ln\mathcal{Z}_{\rm 1F}=-40.0$ and $\ln\mathcal{Z}_{\rm 2F}=-28.4$. The corresponding log-Bayes factor remains essentially unchanged, $\ln\mathcal{B}_{\rm 2F,1F}\simeq 11.6$. The result is also stable with respect to a broadening of the SQM prior support. Using the enlarged ranges $B^{1/4}\in[130,220]\,\mathrm{MeV}$, $a_4\in[0.1,1.0]$, and $\gap\in[0,300]\,\mathrm{MeV}$, we find $\ln\mathcal{Z}_{\rm 2F}=-28.1$, very close to the baseline value. Thus, the preference for 2F is not driven by the precise boundaries of the adopted SQM prior nor by the marginalization over the hyperon-potential ranges.

We also tested the treatment of the GW170817 source-family channels. Motivated by the discussion in Sec.~\ref{sec:gw170817} on the kilonova interpretation, removing the QS--QS channel gives $\ln\mathcal{Z}_{\rm 2F}=-28.1$. Removing both the QS--QS and HS--HS channels gives instead $\ln\mathcal{Z}_{\rm 2F}=-29.8$. The latter choice has a larger impact, but even in this case the 2F evidence remains well above the 1F value.

The most relevant shifts are associated with the source-mass prior. For the 1F hypothesis, raising the lower mass bound from $M_{\min}=0$ to $M_{\min}=1.17\Msol$ increases the evidence from $-40.46$ to $-38.9$. Although this may look counterintuitive because it removes most of the low-mass region preferred by HESS~J1731--347, it is a natural prior-volume effect: the HESS~J1731--347 likelihood is reduced, but the mass interval over which the likelihood is averaged is also reduced for all mass--radius sources. The net result is therefore an increase in the 1F evidence. Even with this more favorable choice for 1F, the Bayes factor remains large,
$\ln\mathcal{B}_{\rm 2F,1F}\simeq 10.3$. A similar conclusion holds for the 2F source-mass prior. Allowing the hadronic branch to extend over the broader interval $M_{\min}^{\rm HS}=0$ to $M_{\max}^{\rm HS}=M_{\max}^{\rm TOV}$ lowers the evidence to
$\ln\mathcal{Z}_{\rm 2F}=-32.6$. This decrease is expected: the enlarged HS mass support adds prior volume and additional source-family configurations without a corresponding gain in likelihood. Nevertheless, the resulting log-Bayes factor remains positive and sizeable.

As an additional strong test, we considered the possibility that the secondary component of GW190814 is a compact star rather than a black hole. Including this object as a compact-star constraint gives $\ln\mathcal{Z}_{\rm 2F}=-30.6$, while the 1F evidence drops to $\ln\mathcal{Z}_{\rm 1F}=-87.5$. This behavior is expected, since such a large compact-star mass can be accommodated by the QS branch in the 2F scenario, whereas it is essentially inaccessible to the purely hadronic 1F sequences considered here.

Overall, these checks show that the numerical value of the Bayes factor is sensitive to source-mass-prior choices and to the assumed GW170817 channel content, yielding maximum variations in $\ln\mathcal{B}$ of up to some units. However, none of the tested variations reverses the model comparison. The preference for the 2F scenario therefore appears robust when considering the full dataset, which includes both the HIC and all astrophysical constraints. In contrast, given the systematic spread highlighted by these tests, the statistical preference for the 2F scenario in the minimal dataset (excluding HESS J1731-347, PSR J0614-3329, and HIC data) is less solid.

\subsection{Limitations and comparison with previous works}
\label{sec:limitations}

Although we have structured this analysis to be as data-driven as possible within a physics-informed framework, certain model dependencies and methodological limitations remain.

First, compared with agnostic parametrizations of the pressure or sound speed, our framework is less flexible.
Its advantage is instead its direct physical interpretability: the inferred EOS behaviour can be related to microscopic quantities such as nuclear empirical parameters, hyperon potentials, $\Delta$-resonance couplings, and SQM bag-model parameters. Consequently, our results remain tied to the specific choice of the RMF Lagrangian. Adding further non-linear meson interactions, such as $\omega$--$\rho$ cross couplings or $\omega$ self-interaction terms, considering additional density-dependent couplings, see e.g. \cite{Fortin:2017dsj}, or relaxing the assumption $x_{\rho\Delta}=1$, could provide extra freedom to tune the density dependence of the EOS and, in particular, to partially disentangle its low- and high-density behavior.
Such extensions could therefore affect the quantitative strength of the Bayes factors comparing the 1F and 2F scenarios.

Our results for the 1F scenario are qualitatively consistent with the Bayesian analysis of hadronic stars considering hyperons and $\Delta$-resonances by Parmar \textit{et al.} \cite{Parmar:2025csx}, in which it has been found that $\Delta$-resonances can efficiently soften the EOS at low and intermediate densities while preserving enough high-density stiffness to satisfy the $2\Msol$ constraint.
A direct quantitative comparison is however not straightforward, because the two analyses differ both in the adopted dataset and in the underlying Lagrangian.
On the observational side, Parmar \textit{et al.} constrain their models using nuclear saturation properties, $\chi$EFT information for symmetric and pure neutron matter, NICER constraints on PSR~J0030$+$0451 and PSR~J0740$+$6620, and tidal information from GW170817. In our case, the 1F posterior is further constrained by the small-radius inference for PSR~J0614--3329, the exceptionally light and compact object HESS~J1731--347, the high mass of PSR~J0952--0607, and HIC collective-flow information. These additional constraints sharpen the tension between compact configurations at low and intermediate masses and the requirement of supporting the heaviest observed pulsars.
The two analyses also differ at the Lagrangian level. Parmar \textit{et al.} adopt a density-dependent relativistic hadron framework in which the $\sigma$, $\omega$, and $\rho$ meson couplings are explicit functions of density, while the mesonic sector contains only the standard quadratic terms.
By contrast, our 1F model is based on a non-linear RMF Lagrangian with $\sigma$ self-interactions and a density-dependent $\rho$ coupling, but density-independent isoscalar $\sigma$ and $\omega$ couplings.
Moreover, Parmar \textit{et al.} vary $x_{\sigma\Delta}$ and $x_{\omega\Delta}$ independently and keep the hyperon optical potentials fixed, whereas  here we vary the hyperon potentials within phenomenological ranges and restrict the allowed $\Delta$ couplings through laboratory-motivated constraint.
The quantitative comparison of macroscopic stellar properties yields compatible conclusions. The main difference is the significantly narrower posterior obtained in our 1F inference. Intermediate runs, in which different datasets are incorporated one at a time, indicate that this narrowing is not solely a consequence of the larger dataset, but is also tied to the more restricted density dependence of our hadronic functional.
Similarly, the quasi-first-order behavior at the $\Delta$ onset is not primarily induced by the additional data. Instead, it is most likely connected to this same restricted density dependence within our hadronic Lagrangian.

A related structural limitation is that hybrid stars, namely hadronic stars with deconfined quark cores, are not included as a third competing hypothesis in the present Bayesian comparison. Therefore, our Bayes factor explicitly compares a purely hadronic 1F scenario with a disconnected 2F scenario, but it does not decide whether the same data would instead favor hybrid-star configurations.
Recent physics-informed Bayesian analyses have found that hybrid EOSs with a hadron--quark phase transition remain compatible with current multimessenger constraints, and may even be statistically preferred over purely hadronic baselines for specific quark-matter models~\cite{Ayriyan:2025rub,Albino:2025puc}.

A complete model-selection analysis should include a third hybrid-star hypothesis. We will leave this analysis for a future work.

Finally, our observational dataset is deliberately conservative. We have not included radius and mass--radius constraints derived from thermonuclear X-ray bursts or from quiescent low-mass X-ray binaries, such as the compilation of~\cite{Ozel:2016oaf} and the analysis of 4U~1702--429 of~\cite{Nattila:2017wtj}; see also~\cite{Drago:2015cea,Bombaci:2020vgw} and references therein.

Nevertheless, these data point in a direction that is naturally accommodated by the 2F scenario and that exacerbates the tension in the 1F one, for the same reason as PSR~J0614--3329: a small radius at a canonical stellar mass is difficult to reconcile, within a single hadronic sequence, with the simultaneous requirement of supporting the heaviest pulsars, whereas in 2F it is naturally associated with the compact hadronic branch.
Concretely, the $(M,R)$ region inferred for 4U~1702--429~\cite{Nattila:2017wtj} is broadly compatible both with the QS branch of our 2F posterior and with the HS configurations of our 1F posterior, while the $\lesssim 11\,\mathrm{km}$ canonical-mass radii suggested by some qLMXB and burst analyses~\cite{Ozel:2016oaf} would increase the pressure on the 1F sequence and thus further strengthen the statistical preference for 2F.

\section{Summary and conclusions}
\label{sec:conclusions}

Determining the EOS and the QCD phases of cold dense matter remains one of the central open problems in nuclear physics and astrophysics. The two-families scenario, in which HSs and QSs coexist on two disconnected branches, offers a natural way to reconcile observations of very compact ordinary-mass objects with the existence of compact stars exceeding $2\Msol$, while at the same time relieving the hyperon puzzle. Although the scenario has been extensively explored at the model level, it had not yet been confronted with a Bayesian analysis of the available astrophysical and laboratory data.

In this work we have performed such an analysis. We compared two physically distinct hypotheses on the same footing: a one-family (1F) scenario in which all observed compact stars are HSs described by a single RMF EOS including nucleons, hyperons, and $\Delta$ resonances, and a two-families (2F) scenario in which the same hadronic sector (but metastable) coexists with an SQM branch, described by a CFL bag-model EOS satisfying the Bodmer--Witten absolute-stability condition. Both hypotheses were explored within a common Bayesian framework, using up to 14 EOS parameters, $\chi$EFT constraints on cold pure neutron matter, heavy-ion-collision (HIC) flow data on symmetric matter, and astrophysical inputs from PSR~J0030$+$0451, PSR~J0740$+$6620, PSR~J0614--3329, HESS~J1731--347, PSR~J0952--0607, and GW170817. In the 2F case the family label of each source was treated as a latent variable and marginalized over, and the binary nature of GW170817 was implemented through the full joint four-dimensional posterior in $(M_1,M_2,\Lambda_1,\Lambda_2)$.

The parameter posteriors, the inferred EOS, and the corresponding stellar properties show that the 1F and 2F scenarios realize the same observational requirements through structurally different EOS-level mechanisms. In the 1F case, a single hadronic sequence must combine intermediate-density softening, required by the HIC band and by the small radii and tidal deformability inferred for ordinary-mass objects, with sufficient high-density stiffness to support the heaviest pulsars. The 1F posterior accommodates this compromise through an early onset of $\Delta$ resonances, at $n_B^{( \rm onset \, \Delta)}\simeq 1.8\nsat$, followed by a strong high-density stiffening mediated again by the $\Delta$ couplings. The macroscopic counterpart of this compromise is a characteristic quasi-inflection of the mass--radius relation and a maximum mass of $M_{\rm HS, max}^{(1\rm F)}=2.05^{+0.03}_{-0.03}\Msol$. In the 2F case, these roles are separated. Non-nucleonic degrees of freedom make the metastable hadronic soft, leading to a compact HS branch ($R_{\rm HS,1.4}^{\rm (2F)}=11.13^{+0.17}_{-0.17}\,\mathrm{km}$, with $M^{\rm (2F)}_{\rm HS,max}=1.54^{+0.03}_{-0.03}\Msol$), while absolutely stable SQM supports the most massive compact stars on the QS branch ($M_{\rm QS,max}^{\rm (2F)}=2.33^{+0.09}_{-0.07}\Msol$). Correspondingly, the 2F hadronic posterior prefers a larger effective mass and a less extreme tuning of the $\Delta$ sector than its 1F counterpart. The 1F tension visible in the posterior is therefore not between HIC and astrophysical information taken separately, but between the combination of HIC plus the small-radius astrophysical observations on one side, and the heavy-pulsar constraints on the other.

The Bayesian model comparison reflects this picture quantitatively. The 2F  scenario is favored for all dataset combinations considered, and the log-Bayes factor grows as observational constraints sensitive to small radii and to large masses are added. PSR~J0614--3329, HESS~J1731--347, and the HIC flow band each contribute in distinct ways: PSR~J0614--3329 enhances the 1F tension between intermediate-density softness and high-density stiffness at an ordinary stellar mass; HESS~J1731--347 introduces a low-mass and compact object that the 2F scenario can naturally place on the QS branch; HIC restricts the stiffness at intermediate densities and therefore makes the 1F hadronic compromise more difficult. For the full dataset we find $\ln\mathcal{B}_{\rm 2F,1F}=11.9$, which corresponds to decisive evidence in favor of the 2F scenario. The consistency checks performed in 
Sec.~\ref{sec:consistency}, varying hyperon-potential priors, the SQM prior support, the GW170817 source-family channels, and the source-mass prior, do not reverse this conclusion. However, while this preference remains robust when analyzing the complete dataset, the systematic variations highlighted by these checks indicate that the statistical evidence in favor of the 2F scenario is significantly less definitive for restricted datasets that omit PSR~J0614--3329, HESS~J1731--347, and the HIC flow band constraints.

Several limitations should be kept in mind. Our results are conditional on the adopted RMF framework for the hadronic sector, and could be refined by enlarging the model space, for instance by including additional density-dependent or non-linear meson couplings, or by relaxing the assumption $x_{\rho\Delta}=1$. The most important structural extension is the inclusion of hybrid stars in the 1F scenario. These extensions are left to future works. 
Ultimately, achieving a more definitive conclusion will require a larger and more precise body of data. In general, future observations providing stronger indications of small radii or tidal deformability at canonical stellar masses (or, equivalently, a soft EOS at intermediate densities from HIC data), the detection of even more massive compact objects, or, crucially, the precise measurement of two distinct masses with the same radius would substantially increase the statistical weight in favor of the 2F scenario.

\bibliography{biblio}

@article{Logoteta:2019utx,
    author = "Logoteta, Domenico and Vidana, Isaac and Bombaci, Ignazio",
    title = "{Impact of chiral hyperonic three-body forces on neutron stars}",
    eprint = "1906.11722",
    archivePrefix = "arXiv",
    primaryClass = "nucl-th",
    doi = "10.1140/epja/i2019-12909-9",
    journal = "Eur. Phys. J. A",
    volume = "55",
    number = "11",
    pages = "207",
    year = "2019"
}

@article{Traversi:2021fad,
    author = "Traversi, Silvia and Char, Prasanta and Pagliara, Giuseppe and Drago, Alessandro",
    title = "{Speed of sound in dense matter and two families of compact stars}",
    eprint = "2102.02357",
    archivePrefix = "arXiv",
    primaryClass = "astro-ph.HE",
    doi = "10.1051/0004-6361/202141544",
    journal = "Astron. Astrophys.",
    volume = "660",
    pages = "A62",
    year = "2022"
}

@article{Albino:2024ymc,
    author = "Albino, Milena and Malik, Tuhin and Ferreira, M{\'a}rcio and Provid{\^e}ncia, Constan{\c{c}}a",
    title = "{Hybrid star properties with the NJL and mean field approximation of QCD models: A Bayesian approach}",
    eprint = "2406.15337",
    archivePrefix = "arXiv",
    primaryClass = "nucl-th",
    doi = "10.1103/PhysRevD.110.083037",
    journal = "Phys. Rev. D",
    volume = "110",
    number = "8",
    pages = "083037",
    year = "2024"
}

@article{PREX:2021umo,
    author = "Adhikari, D. and others",
    collaboration = "PREX",
    title = "{Accurate Determination of the Neutron Skin Thickness of $^{208}$Pb through Parity-Violation in Electron Scattering}",
    eprint = "2102.10767",
    archivePrefix = "arXiv",
    primaryClass = "nucl-ex",
    reportNumber = "JLAB-PHY-21-3420",
    doi = "10.1103/PhysRevLett.126.172502",
    journal = "Phys. Rev. Lett.",
    volume = "126",
    number = "17",
    pages = "172502",
    year = "2021"
}

@article{CREX:2022kgg,
    author = "Adhikari, D. and others",
    collaboration = "CREX",
    title = "{Precision Determination of the Neutral Weak Form Factor of Ca48}",
    eprint = "2205.11593",
    archivePrefix = "arXiv",
    primaryClass = "nucl-ex",
    doi = "10.1103/PhysRevLett.129.042501",
    journal = "Phys. Rev. Lett.",
    volume = "129",
    number = "4",
    pages = "042501",
    year = "2022"
}

@article{Reed:2021nqk,
    author = "Reed, Brendan T. and Fattoyev, F. J. and Horowitz, C. J. and Piekarewicz, J.",
    title = "{Implications of PREX-2 on the Equation of State of Neutron-Rich Matter}",
    eprint = "2101.03193",
    archivePrefix = "arXiv",
    primaryClass = "nucl-th",
    doi = "10.1103/PhysRevLett.126.172503",
    journal = "Phys. Rev. Lett.",
    volume = "126",
    number = "17",
    pages = "172503",
    year = "2021"
}

@article{Reinhard:2021utv,
    author = "Reinhard, Paul-Gerhard and Roca-Maza, Xavier and Nazarewicz, Witold",
    title = "{Information Content of the Parity-Violating Asymmetry in Pb208}",
    eprint = "2105.15050",
    archivePrefix = "arXiv",
    primaryClass = "nucl-th",
    doi = "10.1103/PhysRevLett.127.232501",
    journal = "Phys. Rev. Lett.",
    volume = "127",
    number = "23",
    pages = "232501",
    year = "2021"
}

@article{Zhang:2022bni,
    author = "Zhang, Zhen and Chen, Lie-Wen",
    title = "{Bayesian inference of the symmetry energy and the neutron skin in Ca48 and Pb208 from CREX and PREX-2}",
    eprint = "2207.03328",
    archivePrefix = "arXiv",
    primaryClass = "nucl-th",
    doi = "10.1103/PhysRevC.108.024317",
    journal = "Phys. Rev. C",
    volume = "108",
    number = "2",
    pages = "024317",
    year = "2023"
}

@article{LIGOScientific:2020zkf,
    author = "Abbott, R. and others",
    collaboration = "LIGO Scientific, Virgo",
    title = "{GW190814: Gravitational Waves from the Coalescence of a 23 Solar Mass Black Hole with a 2.6 Solar Mass Compact Object}",
    eprint = "2006.12611",
    archivePrefix = "arXiv",
    primaryClass = "astro-ph.HE",
    reportNumber = "LIGO-P190814",
    doi = "10.3847/2041-8213/ab960f",
    journal = "Astrophys. J. Lett.",
    volume = "896",
    number = "2",
    pages = "L44",
    year = "2020"
}

@article{Romani:2022jhd,
    author = "Romani, Roger W. and Kandel, D. and Filippenko, Alexei V. and Brink, Thomas G. and Zheng, WeiKang",
    title = "{PSR J0952\ensuremath{-}0607: The Fastest and Heaviest Known Galactic Neutron Star}",
    eprint = "2207.05124",
    archivePrefix = "arXiv",
    primaryClass = "astro-ph.HE",
    doi = "10.3847/2041-8213/ac8007",
    journal = "Astrophys. J. Lett.",
    volume = "934",
    number = "2",
    pages = "L17",
    year = "2022"
}

@article{Parmar:2025csx,
    author = "Parmar, Vishal and Thapa, Vivek Baruah and Sinha, Monika and Bombaci, Ignazio",
    title = "{Exploring $\Delta$-resonance in neutron stars: implications from astrophysical and nuclear observations}",
    eprint = "2503.07256",
    archivePrefix = "arXiv",
    primaryClass = "astro-ph.HE",
    month = "3",
    year = "2025"
}

@article{Miller:2021qha,
    author = "Miller, M. C. and others",
    title = "{The Radius of PSR J0740+6620 from NICER and XMM-Newton Data}",
    eprint = "2105.06979",
    archivePrefix = "arXiv",
    primaryClass = "astro-ph.HE",
    doi = "10.3847/2041-8213/ac089b",
    journal = "Astrophys. J. Lett.",
    volume = "918",
    number = "2",
    pages = "L28",
    year = "2021"
}

@article{Miller:2019cac,
    author = "Miller, M. C. and others",
    title = "{PSR J0030+0451 Mass and Radius from $NICER$ Data and Implications for the Properties of Neutron Star Matter}",
    eprint = "1912.05705",
    archivePrefix = "arXiv",
    primaryClass = "astro-ph.HE",
    doi = "10.3847/2041-8213/ab50c5",
    journal = "Astrophys. J. Lett.",
    volume = "887",
    number = "1",
    pages = "L24",
    year = "2019"
}

@article{Salmi:2024bss,
    author = "Salmi, Tuomo and others",
    title = "{A NICER View of PSR J1231{\ensuremath{-}}1411: A Complex Case}",
    eprint = "2409.14923",
    archivePrefix = "arXiv",
    primaryClass = "astro-ph.HE",
    doi = "10.3847/1538-4357/ad81d2",
    journal = "Astrophys. J.",
    volume = "976",
    number = "1",
    pages = "58",
    year = "2024"
}

@article{Choudhury:2024xbk,
    author = "Choudhury, Devarshi and others",
    title = "{A NICER View of the Nearest and Brightest Millisecond Pulsar: PSR J0437{\textendash}4715}",
    eprint = "2407.06789",
    archivePrefix = "arXiv",
    primaryClass = "astro-ph.HE",
    doi = "10.3847/2041-8213/ad5a6f",
    journal = "Astrophys. J. Lett.",
    volume = "971",
    number = "1",
    pages = "L20",
    year = "2024"
}

@article{Miller:2025qfq,
    author = "Miller, M. C. and others",
    title = "{The Radius of PSR J0437-4715 from NICER Data}",
    eprint = "2512.08790",
    archivePrefix = "arXiv",
    primaryClass = "astro-ph.HE",
    month = "12",
    year = "2025"
}

@article{Qi:2025mpn,
    author = "Qi, Liqiang and others",
    title = "{PSR J1231{\textendash}1411 Revisited: Pulse Profile Analysis of X-Ray Observation}",
    eprint = "2502.09147",
    archivePrefix = "arXiv",
    primaryClass = "astro-ph.HE",
    doi = "10.3847/1538-4357/adb42f",
    journal = "Astrophys. J.",
    volume = "981",
    number = "2",
    pages = "99",
    year = "2025"
}

@article{Doroshenko:2022nwp,
    author = {Doroshenko, Victor and Suleimanov, Valery and P{\"u}hlhofer, Gerd and Santangelo, Andrea},
    title = "{A strangely light neutron star within a supernova remnant}",
    doi = "10.1038/s41550-022-01800-1",
    journal = "Nature Astron.",
    volume = "6",
    number = "12",
    pages = "1444--1451",
    year = "2022"
}

@article{Mauviard:2025dmd,
    author = "Mauviard, Lucien and others",
    title = "{A NICER View of the 1.4 M$_{⊙}$ Edge-on Pulsar PSR J0614-3329}",
    eprint = "2506.14883",
    archivePrefix = "arXiv",
    primaryClass = "astro-ph.HE",
    doi = "10.3847/1538-4357/ae145d",
    journal = "Astrophys. J.",
    volume = "995",
    number = "1",
    pages = "60",
    year = "2025"
}

@article{Vinciguerra:2023qxq,
    author = "Vinciguerra, Serena and others",
    title = "{An Updated Mass{\textendash}Radius Analysis of the 2017{\textendash}2018 NICER Data Set of PSR J0030+0451}",
    eprint = "2308.09469",
    archivePrefix = "arXiv",
    primaryClass = "astro-ph.HE",
    doi = "10.3847/1538-4357/acfb83",
    journal = "Astrophys. J.",
    volume = "961",
    number = "1",
    pages = "62",
    year = "2024"
}

@article{Lonardoni:2014bwa,
    author = "Lonardoni, Diego and Lovato, Alessandro and Gandolfi, Stefano and Pederiva, Francesco",
    title = "{Hyperon Puzzle: Hints from Quantum Monte Carlo Calculations}",
    eprint = "1407.4448",
    archivePrefix = "arXiv",
    primaryClass = "nucl-th",
    reportNumber = "LA-UR-14-25265",
    doi = "10.1103/PhysRevLett.114.092301",
    journal = "Phys. Rev. Lett.",
    volume = "114",
    number = "9",
    pages = "092301",
    year = "2015"
}

@article{Tews:2015ufa,
    author = "Tews, I. and Gandolfi, S. and Gezerlis, A. and Schwenk, A.",
    title = "{Quantum Monte Carlo calculations of neutron matter with chiral three-body forces}",
    eprint = "1507.05561",
    archivePrefix = "arXiv",
    primaryClass = "nucl-th",
    doi = "10.1103/PhysRevC.93.024305",
    journal = "Phys. Rev. C",
    volume = "93",
    number = "2",
    pages = "024305",
    year = "2016"
}

@article{Huang:2023grj,
    author = "Huang, Chun and Raaijmakers, Geert and Watts, Anna L. and Tolos, Laura and Provid\^encia, Constan\c{c}a",
    title = "{Constraining a relativistic mean field model using neutron star mass\textendash{}radius measurements I: nucleonic models}",
    eprint = "2303.17518",
    archivePrefix = "arXiv",
    primaryClass = "astro-ph.HE",
    doi = "10.1093/mnras/stae844",
    journal = "Mon. Not. Roy. Astron. Soc.",
    volume = "529",
    number = "4",
    pages = "4650--4665",
    year = "2024"
}

@article{Huang:2024rvj,
    author = "Huang, Chun and Tolos, Laura and Provid\^encia, Constan\c{c}a and Watts, Anna",
    title = "{Constraining a relativistic mean field model using neutron star mass-radius measurements II: Hyperonic models}",
    eprint = "2410.14572",
    archivePrefix = "arXiv",
    primaryClass = "astro-ph.HE",
    doi = "10.1093/mnras/stae2792",
    month = "10",
    year = "2024"
}

@article{Drago:2014oja,
    author = "Drago, Alessandro and Lavagno, Andrea and Pagliara, Giuseppe and Pigato, Daniele",
    title = "{Early appearance of \ensuremath{\Delta} isobars in neutron stars}",
    eprint = "1407.2843",
    archivePrefix = "arXiv",
    primaryClass = "astro-ph.SR",
    doi = "10.1103/PhysRevC.90.065809",
    journal = "Phys. Rev. C",
    volume = "90",
    number = "6",
    pages = "065809",
    year = "2014"
}

@article{Baym:1971pw,
    author = "Baym, Gordon and Pethick, Christopher and Sutherland, Peter",
    title = "{The Ground state of matter at high densities: Equation of state and stellar models}",
    doi = "10.1086/151216",
    journal = "Astrophys. J.",
    volume = "170",
    pages = "299--317",
    year = "1971"
}

@article{Malik:2022jqc,
    author = "Malik, Tuhin and Provid\^encia, Constan\c{c}a",
    title = "{Bayesian inference of signatures of hyperons inside neutron stars}",
    eprint = "2205.15843",
    archivePrefix = "arXiv",
    primaryClass = "nucl-th",
    doi = "10.1103/PhysRevD.106.063024",
    journal = "Phys. Rev. D",
    volume = "106",
    number = "6",
    pages = "063024",
    year = "2022"
}

@article{Raaijmakers:2019qny,
    author = "Raaijmakers, G. and others",
    title = "{A $NICER$ view of PSR J0030+0451: Implications for the dense matter equation of state}",
    eprint = "1912.05703",
    archivePrefix = "arXiv",
    primaryClass = "astro-ph.HE",
    doi = "10.3847/2041-8213/ab451a",
    journal = "Astrophys. J. Lett.",
    volume = "887",
    number = "1",
    pages = "L22",
    year = "2019"
}

@article{Takatsy:2023xzf,
    author = {Takatsy, Janos and Kovacs, Peter and Wolf, Gy{\"o}rgy and Schaffner-Bielich, J{\"u}rgen},
    title = "{What neutron stars tell about the hadron-quark phase transition: A Bayesian study}",
    eprint = "2303.00013",
    archivePrefix = "arXiv",
    primaryClass = "astro-ph.HE",
    doi = "10.1103/PhysRevD.108.043002",
    journal = "Phys. Rev. D",
    volume = "108",
    number = "4",
    pages = "043002",
    year = "2023"
}

@article{LeFevre:2016vpp,
    author = "Le F{\`e}vre, A. and Leifels, Y. and Hartnack, C. and Aichelin, J.",
    title = "{Origin of elliptic flow and its dependence on the equation of state in heavy ion reactions at intermediate energies}",
    eprint = "1611.07500",
    archivePrefix = "arXiv",
    primaryClass = "nucl-th",
    doi = "10.1103/PhysRevC.98.034901",
    journal = "Phys. Rev. C",
    volume = "98",
    number = "3",
    pages = "034901",
    year = "2018"
}

@article{Russotto:2016ucm,
    author = "Russotto, P. and others",
    title = "{Results of the ASY-EOS experiment at GSI: The symmetry energy at suprasaturation density}",
    eprint = "1608.04332",
    archivePrefix = "arXiv",
    primaryClass = "nucl-ex",
    doi = "10.1103/PhysRevC.94.034608",
    journal = "Phys. Rev. C",
    volume = "94",
    number = "3",
    pages = "034608",
    year = "2016"
}

@article{Miao:2021nuq,
    author = "Miao, Zhiqiang and Jiang, Jin-Liang and Li, Ang and Chen, Lie-Wen",
    title = "{Bayesian Inference of Strange Star Equation of State Using the GW170817 and GW190425 Data}",
    eprint = "2107.13997",
    archivePrefix = "arXiv",
    primaryClass = "astro-ph.HE",
    doi = "10.3847/2041-8213/ac194d",
    journal = "Astrophys. J. Lett.",
    volume = "917",
    number = "2",
    pages = "L22",
    year = "2021"
}

@article{Grundler:2025mcz,
    author = "Grundler, Xavier and Li, Bao-An",
    title = "{Bayesian quantification of observability and equation of state of twin stars}",
    eprint = "2506.13677",
    archivePrefix = "arXiv",
    primaryClass = "astro-ph.HE",
    doi = "10.1103/hsd4-j54y",
    journal = "Phys. Rev. D",
    volume = "112",
    number = "10",
    pages = "103012",
    year = "2025"
}

@article{Annala:2023cwx,
    author = {Annala, Eemeli and Gorda, Tyler and Hirvonen, Joonas and Komoltsev, Oleg and Kurkela, Aleksi and N{\"a}ttil{\"a}, Joonas and Vuorinen, Aleksi},
    title = "{Strongly interacting matter exhibits deconfined behavior in massive neutron stars}",
    eprint = "2303.11356",
    archivePrefix = "arXiv",
    primaryClass = "astro-ph.HE",
    reportNumber = "HIP-2023-5/TH, HIP-2023-5/TH",
    doi = "10.1038/s41467-023-44051-y",
    journal = "Nature Commun.",
    volume = "14",
    number = "1",
    pages = "8451",
    year = "2023"
}

@article{Pang:2023dqj,
    author = "Pang, Peter T. H. and Sivertsen, Lars and Somasundaram, Rahul and Dietrich, Tim and Sen, Srimoyee and Tews, Ingo and Coughlin, Michael W. and Van Den Broeck, Chris",
    title = "{Probing quarkyonic matter in neutron stars with the Bayesian nuclear-physics multimessenger astrophysics framework}",
    eprint = "2308.15067",
    archivePrefix = "arXiv",
    primaryClass = "nucl-th",
    reportNumber = "LA-UR-23-29118, NP3M-P2300017",
    doi = "10.1103/PhysRevC.109.025807",
    journal = "Phys. Rev. C",
    volume = "109",
    number = "2",
    pages = "025807",
    year = "2024"
}

@article{Godzieba:2020bbz,
    author = "Godzieba, Daniel A. and Gamba, Rossella and Radice, David and Bernuzzi, Sebastiano",
    title = "{Updated universal relations for tidal deformabilities of neutron stars from phenomenological equations of state}",
    eprint = "2012.12151",
    archivePrefix = "arXiv",
    primaryClass = "astro-ph.HE",
    doi = "10.1103/PhysRevD.103.063036",
    journal = "Phys. Rev. D",
    volume = "103",
    number = "6",
    pages = "063036",
    year = "2021"
}

@article{Kopp:2025aez,
    author = {K{\"o}pp, F{\'a}bio and Lenzi, C{\'e}sar H. and Flores, C{\'e}sar V. and Menezes and D{\'e}bora P.},
    title = "{A Bayesian Approach Study of Hybrid Neutron Stars}",
    eprint = "2512.08911",
    archivePrefix = "arXiv",
    primaryClass = "nucl-th",
    month = "12",
    year = "2025"
}

@article{Shahrbaf:2021cjz,
    author = "Shahrbaf, Mahboubeh and Anti{\'c}, Sofija and Ayriyan, A. and Blaschke, David and Grunfeld, Ana Gabriela",
    title = "{Constraining free parameters of a color superconducting nonlocal Nambu{\textendash}Jona-Lasinio model using Bayesian analysis of neutron stars mass and radius measurements}",
    eprint = "2105.00029",
    archivePrefix = "arXiv",
    primaryClass = "nucl-th",
    doi = "10.1103/PhysRevD.107.054011",
    journal = "Phys. Rev. D",
    volume = "107",
    number = "5",
    pages = "054011",
    year = "2023"
}

@article{Li:2020wbw,
    author = "Li, Ang and Miao, Zhi-Qiang and Jiang, Jin-Liang and Tang, Shao-Peng and Xu, Ren-Xin",
    title = "{Bayesian inference of quark star equation of state using the NICER PSR J0030+0451 data}",
    eprint = "2009.12571",
    archivePrefix = "arXiv",
    primaryClass = "astro-ph.HE",
    doi = "10.1093/mnras/stab2029",
    journal = "Mon. Not. Roy. Astron. Soc.",
    volume = "506",
    number = "4",
    pages = "5916--5922",
    year = "2021"
}

@article{Gorda:2022jvk,
    author = "Gorda, Tyler and Komoltsev, Oleg and Kurkela, Aleksi",
    title = "{Ab-initio QCD Calculations Impact the Inference of the Neutron-star-matter Equation of State}",
    eprint = "2204.11877",
    archivePrefix = "arXiv",
    primaryClass = "nucl-th",
    doi = "10.3847/1538-4357/acce3a",
    journal = "Astrophys. J.",
    volume = "950",
    number = "2",
    pages = "107",
    year = "2023"
}

@article{Char:2025nli,
    author = "Char, Prasanta and Mondal, Chiranjib and Alezraa, Timoth{\'e} and Gulminelli, Francesca and Oertel, Micaela",
    title = "{Properties of Neutron Stars with Hyperons within a Relativistic Metamodel}",
    eprint = "2510.00997",
    archivePrefix = "arXiv",
    primaryClass = "nucl-th",
    month = "10",
    year = "2025"
}

@article{Montefusco:2026jlq,
    author = "Montefusco, Gabriele and Antonelli, Marco and Gulminelli, Francesca",
    title = "{An Asymptotically Causal Metamodel for Neutron Star Equations of State}",
    eprint = "2604.00196",
    archivePrefix = "arXiv",
    primaryClass = "nucl-th",
    month = "3",
    year = "2026"
}

@article{Margueron:2017eqc,
    author = "Margueron, J\'er\^ome and Hoffmann Casali, Rudiney and Gulminelli, Francesca",
    title = "{Equation of state for dense nucleonic matter from metamodeling. I. Foundational aspects}",
    eprint = "1708.06894",
    archivePrefix = "arXiv",
    primaryClass = "nucl-th",
    reportNumber = "INT-PUB-17-029",
    doi = "10.1103/PhysRevC.97.025805",
    journal = "Phys. Rev. C",
    volume = "97",
    number = "2",
    pages = "025805",
    year = "2018"
}

@article{Malik:2022zol,
    author = "Malik, Tuhin and Ferreira, M\'arcio and Agrawal, B. K. and Provid\^encia, Constan\c{c}a",
    title = "{Relativistic Description of Dense Matter Equation of State and Compatibility with Neutron Star Observables: A Bayesian Approach}",
    eprint = "2201.12552",
    archivePrefix = "arXiv",
    primaryClass = "nucl-th",
    doi = "10.3847/1538-4357/ac5d3c",
    journal = "Astrophys. J.",
    volume = "930",
    number = "1",
    pages = "17",
    year = "2022"
}

@article{Traversi:2020aaa,
    author = "Traversi, Silvia and Char, Prasanta and Pagliara, Giuseppe",
    title = "{Bayesian Inference of Dense Matter Equation of State within Relativistic Mean Field Models using Astrophysical Measurements}",
    eprint = "2002.08951",
    archivePrefix = "arXiv",
    primaryClass = "astro-ph.HE",
    doi = "10.3847/1538-4357/ab99c1",
    journal = "Astrophys. J.",
    volume = "897",
    pages = "165",
    year = "2020"
}

@article{Drischler:2020hwi,
    author = "Drischler, C. and Furnstahl, R. J. and Melendez, J. A. and Phillips, D. R.",
    title = "{How Well Do We Know the Neutron-Matter Equation of State at the Densities Inside Neutron Stars? A Bayesian Approach with Correlated Uncertainties}",
    eprint = "2004.07232",
    archivePrefix = "arXiv",
    primaryClass = "nucl-th",
    doi = "10.1103/PhysRevLett.125.202702",
    journal = "Phys. Rev. Lett.",
    volume = "125",
    number = "20",
    pages = "202702",
    year = "2020"
}

@article{Drischler:2017wtt,
    author = "Drischler, C. and Hebeler, K. and Schwenk, A.",
    title = "{Chiral interactions up to next-to-next-to-next-to-leading order and nuclear saturation}",
    eprint = "1710.08220",
    archivePrefix = "arXiv",
    primaryClass = "nucl-th",
    doi = "10.1103/PhysRevLett.122.042501",
    journal = "Phys. Rev. Lett.",
    volume = "122",
    number = "4",
    pages = "042501",
    year = "2019"
}

@article{Tolos:2020aln,
    author = "Tolos, Laura and Fabbietti, Laura",
    title = "{Strangeness in Nuclei and Neutron Stars}",
    eprint = "2002.09223",
    archivePrefix = "arXiv",
    primaryClass = "nucl-ex",
    doi = "10.1016/j.ppnp.2020.103770",
    journal = "Prog. Part. Nucl. Phys.",
    volume = "112",
    pages = "103770",
    year = "2020"
}

@article{Drischler:2021kxf,
    author = "Drischler, C. and Holt, J. W. and Wellenhofer, C.",
    title = "{Chiral Effective Field Theory and the High-Density Nuclear Equation of State}",
    eprint = "2101.01709",
    archivePrefix = "arXiv",
    primaryClass = "nucl-th",
    doi = "10.1146/annurev-nucl-102419-041903",
    journal = "Ann. Rev. Nucl. Part. Sci.",
    volume = "71",
    pages = "403--432",
    year = "2021"
}

@article{MUSES:2023hyz,
    author = "Kumar, Rajesh and others",
    collaboration = "MUSES",
    title = "{Theoretical and experimental constraints for the equation of state of dense and hot matter}",
    eprint = "2303.17021",
    archivePrefix = "arXiv",
    primaryClass = "nucl-th",
    doi = "10.1007/s41114-024-00049-6",
    journal = "Living Rev. Rel.",
    volume = "27",
    number = "1",
    pages = "3",
    year = "2024"
}

@article{Drischler:2024ebw,
    author = "Drischler, C. and Giuliani, P. G. and Bezoui, S. and Piekarewicz, J. and Viens, F.",
    title = "{Bayesian mixture model approach to quantifying the empirical nuclear saturation point}",
    eprint = "2405.02748",
    archivePrefix = "arXiv",
    primaryClass = "nucl-th",
    doi = "10.1103/PhysRevC.110.044320",
    journal = "Phys. Rev. C",
    volume = "110",
    number = "4",
    pages = "044320",
    year = "2024"
}

@article{Gal:2016boi,
    author = "Gal, A. and Hungerford, E. V. and Millener, D. J.",
    title = "{Strangeness in nuclear physics}",
    eprint = "1605.00557",
    archivePrefix = "arXiv",
    primaryClass = "nucl-th",
    doi = "10.1103/RevModPhys.88.035004",
    journal = "Rev. Mod. Phys.",
    volume = "88",
    number = "3",
    pages = "035004",
    year = "2016"
}

@article{Lattimer:2023rpe,
    author = "Lattimer, James M.",
    title = "{Constraints on Nuclear Symmetry Energy Parameters}",
    eprint = "2301.03666",
    archivePrefix = "arXiv",
    primaryClass = "nucl-th",
    doi = "10.3390/particles6010003",
    journal = "Particles",
    volume = "6",
    number = "1",
    pages = "30--56",
    year = "2023"
}

@article{Anglani:2013gfu,
    author = "Anglani, Roberto and Casalbuoni, Roberto and Ciminale, Marco and Ippolito, Nicola and Gatto, Raoul and Mannarelli, Massimo and Ruggieri, Marco",
    title = "{Crystalline color superconductors}",
    eprint = "1302.4264",
    archivePrefix = "arXiv",
    primaryClass = "hep-ph",
    doi = "10.1103/RevModPhys.86.509",
    journal = "Rev. Mod. Phys.",
    volume = "86",
    pages = "509--561",
    year = "2014"
}

@article{Typel:2009sy,
    author = "Typel, S. and Ropke, G. and Klahn, T. and Blaschke, D. and Wolter, H. H.",
    title = "{Composition and thermodynamics of nuclear matter with light clusters}",
    eprint = "0908.2344",
    archivePrefix = "arXiv",
    primaryClass = "nucl-th",
    doi = "10.1103/PhysRevC.81.015803",
    journal = "Phys. Rev. C",
    volume = "81",
    pages = "015803",
    year = "2010"
}

@article{Drago:2015cea,
    author = "Drago, Alessandro and Lavagno, Andrea and Pagliara, Giuseppe and Pigato, Daniele",
    title = "{The scenario of two families of compact stars}: {1. Equations of state, mass-radius relations and binary systems}",
    eprint = "1509.02131",
    archivePrefix = "arXiv",
    primaryClass = "astro-ph.SR",
    doi = "10.1140/epja/i2016-16040-3",
    journal = "Eur. Phys. J. A",
    volume = "52",
    number = "2",
    pages = "40",
    year = "2016"
}

@article{Drago:2015dea,
    author = "Drago, Alessandro and Pagliara, Giuseppe",
    title = "{The scenario of two families of compact stars}: {2. Transition from hadronic to quark matter and explosive phenomena}",
    eprint = "1509.02134",
    archivePrefix = "arXiv",
    primaryClass = "astro-ph.SR",
    doi = "10.1140/epja/i2016-16041-2",
    journal = "Eur. Phys. J. A",
    volume = "52",
    number = "2",
    pages = "41",
    year = "2016"
}

@article{Typel:2005ba,
    author = "Typel, S.",
    title = "{Relativistic model for nuclear matter and atomic nuclei with momentum-dependent self-energies}",
    eprint = "nucl-th/0501056",
    archivePrefix = "arXiv",
    doi = "10.1103/PhysRevC.71.064301",
    journal = "Phys. Rev. C",
    volume = "71",
    pages = "064301",
    year = "2005"
}

@article{Typel:1999yq,
    author = "Typel, S. and Wolter, H. H.",
    title = "{Relativistic mean field calculations with density dependent meson nucleon coupling}",
    doi = "10.1016/S0375-9474(99)00310-3",
    journal = "Nucl. Phys. A",
    volume = "656",
    pages = "331--364",
    year = "1999"
}

@article{Glendenning:1991es,
    author = "Glendenning, N. K. and Moszkowski, S. A.",
    title = "{Reconciliation of neutron star masses and binding of the lambda in hypernuclei}",
    reportNumber = "LBL-30645",
    doi = "10.1103/PhysRevLett.67.2414",
    journal = "Phys. Rev. Lett.",
    volume = "67",
    pages = "2414--2417",
    year = "1991"
}

@article{Drago:2013fsa,
    author = "Drago, Alessandro and Lavagno, Andrea and Pagliara, Giuseppe",
    title = "{Can very compact and very massive neutron stars both exist?}",
    eprint = "1309.7263",
    archivePrefix = "arXiv",
    primaryClass = "nucl-th",
    doi = "10.1103/PhysRevD.89.043014",
    journal = "Phys. Rev. D",
    volume = "89",
    number = "4",
    pages = "043014",
    year = "2014"
}

@article{LIGOScientific:2018hze,
    author = "Abbott, B. P. and others",
    collaboration = "LIGO Scientific, Virgo",
    title = "{Properties of the binary neutron star merger GW170817}",
    eprint = "1805.11579",
    archivePrefix = "arXiv",
    primaryClass = "gr-qc",
    doi = "10.1103/PhysRevX.9.011001",
    journal = "Phys. Rev. X",
    volume = "9",
    number = "1",
    pages = "011001",
    year = "2019"
}

@article{Lattimer:2012nd,
    author = "Lattimer, James M.",
    title = "{The nuclear equation of state and neutron star masses}",
    eprint = "1305.3510",
    archivePrefix = "arXiv",
    primaryClass = "nucl-th",
    doi = "10.1146/annurev-nucl-102711-095018",
    journal = "Ann. Rev. Nucl. Part. Sci.",
    volume = "62",
    pages = "485--515",
    year = "2012"
}

@article{Pagliara:2013tza,
    author = {Pagliara, Giuseppe and Herzog, Matthias and R{\"o}pke, Friedrich K.},
    title = "{Combustion of a neutron star into a strange quark star: The neutrino signal}",
    eprint = "1304.6884",
    archivePrefix = "arXiv",
    primaryClass = "astro-ph.HE",
    doi = "10.1103/PhysRevD.87.103007",
    journal = "Phys. Rev. D",
    volume = "87",
    number = "10",
    pages = "103007",
    year = "2013"
}

@article{Alford:2007xm,
    author = {Alford, Mark G. and Schmitt, Andreas and Rajagopal, Krishna and Sch{\"a}fer, Thomas},
    title = "{Color superconductivity in dense quark matter}",
    eprint = "0709.4635",
    archivePrefix = "arXiv",
    primaryClass = "hep-ph",
    reportNumber = "MIT-CTP-3861",
    doi = "10.1103/RevModPhys.80.1455",
    journal = "Rev. Mod. Phys.",
    volume = "80",
    pages = "1455--1515",
    year = "2008"
}

@article{Alford:2001zr,
    author = "Alford, Mark G. and Rajagopal, Krishna and Reddy, Sanjay and Wilczek, Frank",
    title = "{The Minimal CFL nuclear interface}",
    eprint = "hep-ph/0105009",
    archivePrefix = "arXiv",
    reportNumber = "GUTPA-01-04-03, MIT-CTP-3123, DOE-ER-41132-110-INT01",
    doi = "10.1103/PhysRevD.64.074017",
    journal = "Phys. Rev. D",
    volume = "64",
    pages = "074017",
    year = "2001"
}

@article{Weissenborn:2011qu,
    author = {Weissenborn, Simon and Sagert, Irina and Pagliara, Giuseppe and Hempel, Matthias and Schaffner-Bielich, J{\"u}rgen},
    title = "{Quark Matter In Massive Neutron Stars}",
    eprint = "1102.2869",
    archivePrefix = "arXiv",
    primaryClass = "astro-ph.HE",
    doi = "10.1088/2041-8205/740/1/L14",
    journal = "Astrophys. J. Lett.",
    volume = "740",
    pages = "L14",
    year = "2011"
}

@article{Borsanyi:2020fev,
    author = "Borsanyi, Szabolcs and Fodor, Zoltan and Guenther, Jana N. and Kara, Ruben and Katz, Sandor D. and Parotto, Paolo and Pasztor, Attila and Ratti, Claudia and Szabo, Kalman K.",
    title = "{QCD Crossover at Finite Chemical Potential from Lattice Simulations}",
    eprint = "2002.02821",
    archivePrefix = "arXiv",
    primaryClass = "hep-lat",
    doi = "10.1103/PhysRevLett.125.052001",
    journal = "Phys. Rev. Lett.",
    volume = "125",
    number = "5",
    pages = "052001",
    year = "2020"
}

@article{Kurkela:2014vha,
    author = {Kurkela, Aleksi and Fraga, Eduardo S. and Schaffner-Bielich, J{\"u}rgen and Vuorinen, Aleksi},
    title = "{Constraining neutron star matter with Quantum Chromodynamics}",
    eprint = "1402.6618",
    archivePrefix = "arXiv",
    primaryClass = "astro-ph.HE",
    reportNumber = "CERN-PH-TH-2014-032, HIP-2014-02-TH",
    doi = "10.1088/0004-637X/789/2/127",
    journal = "Astrophys. J.",
    volume = "789",
    pages = "127",
    year = "2014"
}

@article{Baym:2017whm,
    author = "Baym, Gordon and Hatsuda, Tetsuo and Kojo, Toru and Powell, Philip D. and Song, Yifan and Takatsuka, Tatsuyuki",
    title = "{From hadrons to quarks in neutron stars: a review}",
    eprint = "1707.04966",
    archivePrefix = "arXiv",
    primaryClass = "astro-ph.HE",
    reportNumber = "RIKEN-ITHEMS-REPORT-17, RIKEN-QHP-316, RIKEN-iTHEMS-Report-17",
    doi = "10.1088/1361-6633/aaae14",
    journal = "Rept. Prog. Phys.",
    volume = "81",
    number = "5",
    pages = "056902",
    year = "2018"
}

@article{Borsanyi:2021sxv,
    author = "Bors{\'a}nyi, S. and Fodor, Z. and Guenther, J. N. and Kara, R. and Katz, S. D. and Parotto, P. and P{\'a}sztor, A. and Ratti, C. and Szab{\'o}, K. K.",
    title = "{Lattice QCD equation of state at finite chemical potential from an alternative expansion scheme}",
    eprint = "2102.06660",
    archivePrefix = "arXiv",
    primaryClass = "hep-lat",
    doi = "10.1103/PhysRevLett.126.232001",
    journal = "Phys. Rev. Lett.",
    volume = "126",
    number = "23",
    pages = "232001",
    year = "2021"
}

@article{Burgio:2021vgk,
    author = "Burgio, G. F. and Schulze, H. -J. and Vidana, I. and Wei, J. -B.",
    title = "{Neutron stars and the nuclear equation of state}",
    eprint = "2105.03747",
    archivePrefix = "arXiv",
    primaryClass = "nucl-th",
    doi = "10.1016/j.ppnp.2021.103879",
    journal = "Prog. Part. Nucl. Phys.",
    volume = "120",
    pages = "103879",
    year = "2021"
}

@article{Alford:2013aca,
    author = "Alford, Mark G. and Han, Sophia and Prakash, Madappa",
    title = "{Generic conditions for stable hybrid stars}",
    eprint = "1302.4732",
    archivePrefix = "arXiv",
    primaryClass = "astro-ph.SR",
    doi = "10.1103/PhysRevD.88.083013",
    journal = "Phys. Rev. D",
    volume = "88",
    number = "8",
    pages = "083013",
    year = "2013"
}

@article{Constantinou:2025wxj,
    author = "Constantinou, Constantinos and Guerrini, Mirco and Zhao, Tianqi and Han, Sophia and Prakash, Madappa",
    title = "{Framework for phase transitions between the Maxwell and Gibbs constructions at finite temperature}",
    eprint = "2506.20418",
    archivePrefix = "arXiv",
    primaryClass = "nucl-th",
    reportNumber = "N3AS-25-003, INT-PUB-25-003",
    doi = "10.1103/8l3m-tdlc",
    journal = "Phys. Rev. D",
    volume = "112",
    number = "9",
    pages = "094014",
    year = "2025"
}

@article{Weber:2004kj,
    author = "Weber, Fridolin",
    title = "{Strange quark matter and compact stars}",
    eprint = "astro-ph/0407155",
    archivePrefix = "arXiv",
    doi = "10.1016/j.ppnp.2004.07.001",
    journal = "Prog. Part. Nucl. Phys.",
    volume = "54",
    pages = "193--288",
    year = "2005"
}

@article{Drago:2015fpa,
    author = "Drago, Alessandro and Pagliara, Giuseppe",
    title = "{Combustion of a hadronic star into a quark star: the turbulent and the diffusive regimes}",
    eprint = "1506.08337",
    archivePrefix = "arXiv",
    primaryClass = "nucl-th",
    doi = "10.1103/PhysRevC.92.045801",
    journal = "Phys. Rev. C",
    volume = "92",
    number = "4",
    pages = "045801",
    year = "2015"
}

@article{Berezhiani:2002ks,
    author = "Berezhiani, Z. and Bombaci, I. and Drago, Alessandro and Frontera, F. and Lavagno, A.",
    title = "{Gamma-ray bursts from delayed collapse of neutron stars to quark matter stars}",
    eprint = "astro-ph/0209257",
    archivePrefix = "arXiv",
    doi = "10.1086/367756",
    journal = "Astrophys. J.",
    volume = "586",
    pages = "1250--1253",
    year = "2003"
}

@article{Guerrini:2025mxx,
    author = "Guerrini, Mirco and Pagliara, Giuseppe and Lavagno, Andrea and Drago, Alessandro",
    title = "{Role of Thermal Fluctuations in Nucleation of Three-Flavor Quark Matter}",
    eprint = "2506.00139",
    archivePrefix = "arXiv",
    primaryClass = "nucl-th",
    doi = "10.3390/universe11080258",
    journal = "Universe",
    volume = "11",
    number = "8",
    pages = "258",
    year = "2025"
}

@article{Guerrini:2026_inpreparation,
    author = "Guerrini, Mirco and Pagliara, Giuseppe and Lavagno, Andrea and Drago, Alessandro",
    title = "{Deconfinement phase transition in proto-neutron stars:\\
testing the coexistence of strange quark stars and neutron stars}",
    eprint = "",
    archivePrefix = "",
    primaryClass = "nucl-th",
    doi = "",
    journal = "in preparation",
    volume = "",
    number = "",
    pages = "",
    year = "2026"
}

@article{Drago:2018jds,
    author = "Drago, Alessandro and Pagliara, Giuseppe and Traversi, Silvia",
    editor = "Amati, L. and Bozzo, E. and Della Valle, M. and Gotz, D. and O'Brien, P.",
    title = "{A multi-messenger analysis of neutron star mergers}",
    eprint = "1802.01696",
    archivePrefix = "arXiv",
    primaryClass = "astro-ph.HE",
    journal = "Mem. Soc. Ast. It.",
    volume = "89",
    number = "2",
    pages = "236--244",
    year = "2018"
}

@article{Fonseca:2021wxt,
    author = "Fonseca, E. and others",
    title = "{Refined Mass and Geometric Measurements of the High-mass PSR J0740+6620}",
    eprint = "2104.00880",
    archivePrefix = "arXiv",
    primaryClass = "astro-ph.HE",
    doi = "10.3847/2041-8213/ac03b8",
    journal = "Astrophys. J. Lett.",
    volume = "915",
    number = "1",
    pages = "L12",
    year = "2021"
}

@misc{gw170817data,
  author       = {{Gravitational Wave Open Science Center}},
  title        = {{GW170817-v3 event page and posterior samples from GWTC-1-confident}},
  year         = {2020},
  doi          = {10.7935/82H3-HH23},
  url          = {https://gwosc.org/eventapi/html/GWTC-1-confident/GW170817/v3/},
  note         = {Accessed 2026-04-21}
}

@article{Lowrey:2024anh,
    author = "Lowrey, Tristen and Yagi, Kent and Yunes, Nicol{\'a}s",
    title = "{Improved analytic Love-C relations for neutron stars}",
    eprint = "2410.06358",
    archivePrefix = "arXiv",
    primaryClass = "gr-qc",
    doi = "10.1103/PhysRevD.111.024075",
    journal = "Phys. Rev. D",
    volume = "111",
    number = "2",
    pages = "024075",
    year = "2025"
}

@article{Albino:2025puc,
    author = "Albino, Milena and Malik, Tuhin and Ferreira, M{\'a}rcio and Provid{\^e}ncia, Constan{\c{c}}a",
    title = "{Bayesian inference of hybrid stars with large quark cores}",
    eprint = "2511.02653",
    archivePrefix = "arXiv",
    primaryClass = "nucl-th",
    doi = "10.1103/jrz4-zjq1",
    journal = "Phys. Rev. D",
    volume = "113",
    number = "8",
    pages = "083019",
    year = "2026"
}

@article{Maselli:2013mva,
    author = "Maselli, Andrea and Cardoso, Vitor and Ferrari, Valeria and Gualtieri, Leonardo and Pani, Paolo",
    title = "{Equation-of-state-independent relations in neutron stars}",
    eprint = "1304.2052",
    archivePrefix = "arXiv",
    primaryClass = "gr-qc",
    doi = "10.1103/PhysRevD.88.023007",
    journal = "Phys. Rev. D",
    volume = "88",
    number = "2",
    pages = "023007",
    year = "2013"
}

@article{Yagi:2016bkt,
    author = "Yagi, Kent and Yunes, Nicol{\'a}s",
    title = "{Approximate Universal Relations for Neutron Stars and Quark Stars}",
    eprint = "1608.02582",
    archivePrefix = "arXiv",
    primaryClass = "gr-qc",
    doi = "10.1016/j.physrep.2017.03.002",
    journal = "Phys. Rept.",
    volume = "681",
    pages = "1--72",
    year = "2017"
}

@article{Huth:2021bsp,
    author = "Huth, S. and others",
    title = "{Constraining Neutron-Star Matter with Microscopic and Macroscopic Collisions}",
    eprint = "2107.06229",
    archivePrefix = "arXiv",
    primaryClass = "nucl-th",
    reportNumber = "LA-UR-21-22072",
    doi = "10.1038/s41586-022-04750-w",
    journal = "Nature",
    volume = "606",
    pages = "276--280",
    year = "2022"
}

@article{LIGOScientific:2018cki,
    author = "Abbott, B. P. and others",
    collaboration = "LIGO Scientific, Virgo",
    title = "{GW170817: Measurements of neutron star radii and equation of state}",
    eprint = "1805.11581",
    archivePrefix = "arXiv",
    primaryClass = "gr-qc",
    reportNumber = "LIGO-P1800115",
    doi = "10.1103/PhysRevLett.121.161101",
    journal = "Phys. Rev. Lett.",
    volume = "121",
    number = "16",
    pages = "161101",
    year = "2018"
}

@article{Haensel:1986qb,
    author = "Haensel, P. and Zdunik, J. L. and Schaeffer, R.",
    title = "{Strange quark stars}",
    journal = "Astron. Astrophys.",
    volume = "160",
    pages = "121--128",
    year = "1986"
}

@article{Nattila:2017wtj,
    author = {N{\"a}ttil{\"a}, J. and Miller, M. C. and Steiner, A. W. and Kajava, J. J. E. and Suleimanov, V. F. and Poutanen, J.},
    title = "{Neutron star mass and radius measurements from atmospheric model fits to X-ray burst cooling tail spectra}",
    eprint = "1709.09120",
    archivePrefix = "arXiv",
    primaryClass = "astro-ph.HE",
    doi = "10.1051/0004-6361/201731082",
    journal = "Astron. Astrophys.",
    volume = "608",
    pages = "A31",
    year = "2017"
}

@article{Antoniadis:2013pzd,
    author = "Antoniadis, John and others",
    title = "{A Massive Pulsar in a Compact Relativistic Binary}",
    eprint = "1304.6875",
    archivePrefix = "arXiv",
    primaryClass = "astro-ph.HE",
    doi = "10.1126/science.1233232",
    journal = "Science",
    volume = "340",
    pages = "6131",
    year = "2013"
}

@article{Sagun:2023rzp,
    author = "Sagun, Violetta and Giangrandi, Edoardo and Dietrich, Tim and Ivanytskyi, Oleksii and Negreiros, Rodrigo and Provid{\^e}ncia, Constan{\c{c}}a",
    title = "{What Is the Nature of the HESS J1731-347 Compact Object?}",
    eprint = "2306.12326",
    archivePrefix = "arXiv",
    primaryClass = "astro-ph.HE",
    doi = "10.3847/1538-4357/acfc9e",
    journal = "Astrophys. J.",
    volume = "958",
    number = "1",
    pages = "49",
    year = "2023"
}

@article{Bombaci:2016xuj,
    author = "Bombaci, Ignazio and Logoteta, Domenico and Vida{\~n}a, Isaac and Provid{\^e}ncia, Constan{\c{c}}a",
    title = "{Quark matter nucleation in neutron stars and astrophysical implications}",
    eprint = "1601.04559",
    archivePrefix = "arXiv",
    primaryClass = "astro-ph.HE",
    doi = "10.1140/epja/i2016-16058-5",
    journal = "Eur. Phys. J. A",
    volume = "52",
    number = "3",
    pages = "58",
    year = "2016"
}

@article{Madsen:1998uh,
    author = "Madsen, Jes",
    editor = "Cleymans, J. and Geyer, H. B. and Scholtz, F. G.",
    title = "{Physics and astrophysics of strange quark matter}",
    eprint = "astro-ph/9809032",
    archivePrefix = "arXiv",
    doi = "10.1007/BFb0107314",
    journal = "Lect. Notes Phys.",
    volume = "516",
    pages = "162--203",
    year = "1999"
}

@article{Alcock:1986hz,
    author = "Alcock, Charles and Farhi, Edward and Olinto, Angela",
    title = "{Strange stars}",
    doi = "10.1086/164679",
    journal = "Astrophys. J.",
    volume = "310",
    pages = "261--272",
    year = "1986"
}

@article{Bodmer:1971we,
    author = "Bodmer, A. R.",
    title = "{Collapsed nuclei}",
    doi = "10.1103/PhysRevD.4.1601",
    journal = "Phys. Rev. D",
    volume = "4",
    pages = "1601--1606",
    year = "1971"
}

@article{Witten:1984rs,
    author = "Witten, Edward",
    title = "{Cosmic Separation of Phases}",
    reportNumber = "PRINT-84-0400 (IAS,PRINCETON)",
    doi = "10.1103/PhysRevD.30.272",
    journal = "Phys. Rev. D",
    volume = "30",
    pages = "272--285",
    year = "1984"
}

@article{Constantinou:2023ged,
    author = "Constantinou, Constantinos and Zhao, Tianqi and Han, Sophia and Prakash, Madappa",
    title = "{Framework for phase transitions between the Maxwell and Gibbs constructions}",
    eprint = "2302.04289",
    archivePrefix = "arXiv",
    primaryClass = "nucl-th",
    reportNumber = "N3AS-23-004",
    doi = "10.1103/PhysRevD.107.074013",
    journal = "Phys. Rev. D",
    volume = "107",
    number = "7",
    pages = "074013",
    year = "2023"
}

@article{Han:2019bub,
    author = "Han, Sophia and Mamun, M. A. A. and Lalit, S. and Constantinou, C. and Prakash, M.",
    title = "{Treating quarks within neutron stars}",
    eprint = "1906.04095",
    archivePrefix = "arXiv",
    primaryClass = "astro-ph.HE",
    doi = "10.1103/PhysRevD.100.103022",
    journal = "Phys. Rev. D",
    volume = "100",
    number = "10",
    pages = "103022",
    year = "2019"
}

@article{Annala:2019puf,
    author = {Annala, Eemeli and Gorda, Tyler and Kurkela, Aleksi and N{\"a}ttil{\"a}, Joonas and Vuorinen, Aleksi},
    title = "{Evidence for quark-matter cores in massive neutron stars}",
    eprint = "1903.09121",
    archivePrefix = "arXiv",
    primaryClass = "astro-ph.HE",
    reportNumber = "CERN-TH-2019-031, HIP-2019-7/TH",
    doi = "10.1038/s41567-020-0914-9",
    journal = "Nature Phys.",
    volume = "16",
    number = "9",
    pages = "907--910",
    year = "2020"
}

@article{Chatterjee:2015pua,
    author = "Chatterjee, Debarati and Vida{\~n}a, Isaac",
    title = "{Do hyperons exist in the interior of neutron stars?}",
    eprint = "1510.06306",
    archivePrefix = "arXiv",
    primaryClass = "nucl-th",
    doi = "10.1140/epja/i2016-16029-x",
    journal = "Eur. Phys. J. A",
    volume = "52",
    number = "2",
    pages = "29",
    year = "2016"
}

@article{Vidana:2018bdi,
    author = "Vida{\~n}a, Isaac",
    title = "{Hyperons: the strange ingredients of the nuclear equation of state}",
    eprint = "1803.00504",
    archivePrefix = "arXiv",
    primaryClass = "nucl-th",
    doi = "10.1098/rspa.2018.0145",
    journal = "Proc. Roy. Soc. Lond. A",
    volume = "474",
    pages = "0145",
    year = "2018"
}

@article{Riley:2021pdl,
    author = "Riley, Thomas E. and others",
    title = "{A NICER View of the Massive Pulsar PSR J0740+6620 Informed by Radio Timing and XMM-Newton Spectroscopy}",
    eprint = "2105.06980",
    archivePrefix = "arXiv",
    primaryClass = "astro-ph.HE",
    doi = "10.3847/2041-8213/ac0a81",
    journal = "Astrophys. J. Lett.",
    volume = "918",
    number = "2",
    pages = "L27",
    year = "2021"
}

@article{Oertel:2016bki,
    author = {Oertel, M. and Hempel, M. and Kl{\"a}hn, T. and Typel, S.},
    title = "{Equations of state for supernovae and compact stars}",
    eprint = "1610.03361",
    archivePrefix = "arXiv",
    primaryClass = "astro-ph.HE",
    doi = "10.1103/RevModPhys.89.015007",
    journal = "Rev. Mod. Phys.",
    volume = "89",
    number = "1",
    pages = "015007",
    year = "2017"
}

@article{Sorensen:2023zkk,
    author = "Sorensen, Agnieszka and others",
    title = "{Dense nuclear matter equation of state from heavy-ion collisions}",
    eprint = "2301.13253",
    archivePrefix = "arXiv",
    primaryClass = "nucl-th",
    reportNumber = "INT-PUB-23-001, LA-UR-23-20514, LLNL-TR-844629",
    doi = "10.1016/j.ppnp.2023.104080",
    journal = "Prog. Part. Nucl. Phys.",
    volume = "134",
    pages = "104080",
    year = "2024"
}

@article{Komoltsev:2021jzg,
    author = "Komoltsev, Oleg and Kurkela, Aleksi",
    title = "{How Perturbative QCD Constrains the Equation of State at Neutron-Star Densities}",
    eprint = "2111.05350",
    archivePrefix = "arXiv",
    primaryClass = "nucl-th",
    doi = "10.1103/PhysRevLett.128.202701",
    journal = "Phys. Rev. Lett.",
    volume = "128",
    number = "20",
    pages = "202701",
    year = "2022"
}

@article{Burgio:2018yix,
    author = "Burgio, G. F. and Drago, A. and Pagliara, G. and Schulze, H. -J. and Wei, J. -B.",
    title = "{Are Small Radii of Compact Stars Ruled out by GW170817/AT2017gfo?}",
    eprint = "1803.09696",
    archivePrefix = "arXiv",
    primaryClass = "astro-ph.HE",
    doi = "10.3847/1538-4357/aac6ee",
    journal = "Astrophys. J.",
    volume = "860",
    number = "2",
    pages = "139",
    year = "2018"
}

@article{Drago:2017bnf,
    author = "Drago, Alessandro and Pagliara, Giuseppe",
    title = "{Merger of two neutron stars: predictions from the two-families scenario}",
    eprint = "1710.02003",
    archivePrefix = "arXiv",
    primaryClass = "astro-ph.HE",
    doi = "10.3847/2041-8213/aaa40a",
    journal = "Astrophys. J. Lett.",
    volume = "852",
    number = "2",
    pages = "L32",
    year = "2018"
}

@article{Li:2018qaw,
    author = "Li, Jia Jie and Sedrakian, Armen and Weber, Fridolin",
    title = "{Competition between delta isobars and hyperons and properties of compact stars}",
    eprint = "1803.03661",
    archivePrefix = "arXiv",
    primaryClass = "nucl-th",
    doi = "10.1016/j.physletb.2018.06.051",
    journal = "Phys. Lett. B",
    volume = "783",
    pages = "234--240",
    year = "2018"
}

@article{Shibata:2019ctb,
    author = "Shibata, Masaru and Zhou, Enping and Kiuchi, Kenta and Fujibayashi, Sho",
    title = "{Constraint on the maximum mass of neutron stars using GW170817 event}",
    eprint = "1905.03656",
    archivePrefix = "arXiv",
    primaryClass = "astro-ph.HE",
    doi = "10.1103/PhysRevD.100.023015",
    journal = "Phys. Rev. D",
    volume = "100",
    number = "2",
    pages = "023015",
    year = "2019"
}

@article{Rezzolla:2017aly,
    author = "Rezzolla, Luciano and Most, Elias R. and Weih, Lukas R.",
    title = "{Using gravitational-wave observations and quasi-universal relations to constrain the maximum mass of neutron stars}",
    eprint = "1711.00314",
    archivePrefix = "arXiv",
    primaryClass = "astro-ph.HE",
    doi = "10.3847/2041-8213/aaa401",
    journal = "Astrophys. J. Lett.",
    volume = "852",
    number = "2",
    pages = "L25",
    year = "2018"
}

@article{Ribes:2019kno,
    author = "Ribes, Patricia and Ramos, Angels and Tolos, Laura and Gonzalez-Boquera, Claudia and Centelles, Mario",
    title = "{Interplay between $\Delta$ Particles and Hyperons in Neutron Stars}",
    eprint = "1907.08583",
    archivePrefix = "arXiv",
    primaryClass = "astro-ph.HE",
    doi = "10.3847/1538-4357/ab3a93",
    journal = "Astrophys. J.",
    volume = "883",
    pages = "168",
    year = "2019"
}

@article{Dietrich:2020efo,
    author = "Dietrich, Tim and Coughlin, Michael W. and Pang, Peter T. H. and Bulla, Mattia and Heinzel, Jack and Issa, Lina and Tews, Ingo and Antier, Sarah",
    title = "{Multimessenger constraints on the neutron-star equation of state and the Hubble constant}",
    eprint = "2002.11355",
    archivePrefix = "arXiv",
    primaryClass = "astro-ph.HE",
    reportNumber = "LA-UR-20-21470",
    doi = "10.1126/science.abb4317",
    journal = "Science",
    volume = "370",
    number = "6523",
    pages = "1450--1453",
    year = "2020"
}

@article{Bombaci:2020vgw,
    author = "Bombaci, I. and Drago, A. and Logoteta, D. and Pagliara, G. and Vida\~na, I.",
    title = "{Was GW190814 a Black Hole\textendash{}Strange Quark Star System?}",
    eprint = "2010.01509",
    archivePrefix = "arXiv",
    primaryClass = "nucl-th",
    doi = "10.1103/PhysRevLett.126.162702",
    journal = "Phys. Rev. Lett.",
    volume = "126",
    number = "16",
    pages = "162702",
    year = "2021"
}

@article{Pang:2022rzc,
    author = "Pang, Peter T. H. and others",
    title = "{NMMA: A nuclear-physics and multi-messenger astrophysics framework to analyze binary neutron star mergers}",
    eprint = "2205.08513",
    archivePrefix = "arXiv",
    primaryClass = "astro-ph.HE",
    reportNumber = "LA-UR-22-23872, LIGO-P2200150",
    month = "5",
    year = "2022"
}

@article{Landry:2018prl,
    author = "Landry, Philippe and Essick, Reed",
    title = "{Nonparametric inference of the neutron star equation of state from gravitational wave observations}",
    eprint = "1811.12529",
    archivePrefix = "arXiv",
    primaryClass = "gr-qc",
    doi = "10.1103/PhysRevD.99.084049",
    journal = "Phys. Rev. D",
    volume = "99",
    number = "8",
    pages = "084049",
    year = "2019"
}

@article{Beznogov:2024vcv,
    author = "Beznogov, Mikhail V. and Raduta, Adriana R.",
    title = "{Bayesian inference of the dense matter equation~of state built upon extended Skyrme interactions}",
    eprint = "2403.19325",
    archivePrefix = "arXiv",
    primaryClass = "nucl-th",
    doi = "10.1103/PhysRevC.110.035805",
    journal = "Phys. Rev. C",
    volume = "110",
    number = "3",
    pages = "035805",
    year = "2024"
}

@article{Char:2020utj,
    author = "Char, Prasanta and Traversi, Silvia and Pagliara, Giuseppe",
    editor = "Sedrakian, Armen",
    title = "{A Bayesian Analysis on Neutron Stars within Relativistic Mean Field Models}",
    doi = "10.3390/particles3030040",
    journal = "Particles",
    volume = "3",
    number = "3",
    pages = "621--629",
    year = "2020"
}

@article{Legred:2021hdx,
    author = "Legred, Isaac and Chatziioannou, Katerina and Essick, Reed and Han, Sophia and Landry, Philippe",
    title = "{Impact of the PSR J0740+6620 radius constraint on the properties of high-density matter}",
    eprint = "2106.05313",
    archivePrefix = "arXiv",
    primaryClass = "astro-ph.HE",
    reportNumber = "N3AS-21-010, INT-PUB-21-014",
    doi = "10.1103/PhysRevD.104.063003",
    journal = "Phys. Rev. D",
    volume = "104",
    number = "6",
    pages = "063003",
    year = "2021"
}

@article{Ayriyan:2025rub,
    author = "Ayriyan, Alexander and Ivanytskyi, Oleksii and Blaschke, David",
    title = "{Bayesian inference favors quark matter in neutron star interiors}",
    eprint = "2509.02554",
    archivePrefix = "arXiv",
    primaryClass = "nucl-th",
    month = "9",
    year = "2025"
}

@article{Speagle:2019ivv,
    author = "Speagle, Joshua S.",
    title = "{dynesty: a dynamic nested sampling package for estimating Bayesian posteriors and evidences}",
    eprint = "1904.02180",
    archivePrefix = "arXiv",
    primaryClass = "astro-ph.IM",
    doi = "10.1093/mnras/staa278",
    journal = "Mon. Not. Roy. Astron. Soc.",
    volume = "493",
    number = "3",
    pages = "3132--3158",
    year = "2020"
}

@software{sergey_koposov_2025_17268284,
  author       = {Sergey Koposov and
                  Josh Speagle and
                  Kyle Barbary and
                  Gregory Ashton and
                  Ed Bennett and
                  Johannes Buchner and
                  Carl Scheffler and
                  Colm Talbot and
                  Ben Cook and
                  James Guillochon and
                  Patricio Cubillos and
                  Andrés Asensio Ramos and
                  Matthieu Dartiailh and
                  Ilya and
                  Erik Tollerud and
                  Dustin Lang and
                  Ben Johnson and
                  jtmendel and
                  Edward Higson and
                  Thomas Vandal and
                  Tansu Daylan and
                  Ruth Angus and
                  patelR and
                  Phillip Cargile and
                  Patrick Sheehan and
                  Matt Pitkin and
                  Matthew Kirk and
                  Lu Xu and
                  Joel Leja and
                  joezuntz},
  title        = {joshspeagle/dynesty: v3.0.0},
  month        = oct,
  year         = 2025,
  publisher    = {Zenodo},
  version      = {v3.0.0},
  doi          = {10.5281/zenodo.17268284},
  url          = {https://doi.org/10.5281/zenodo.17268284},
  swhid        = {swh:1:dir:b3340f6aaa931bae6c6bcff6e7ddeb89ca508a15
                   ;origin=https://doi.org/10.5281/zenodo.3348367;vis
                   it=swh:1:snp:fa92422c27d5611b107216e9b766a871bb43b
                   bee;anchor=swh:1:rel:94af3a8ef6e50d997258e152c3833
                   3550bd7463f;path=joshspeagle-dynesty-217bc94
                  },
}

@article{Danielewicz:2002pu,
    author = "Danielewicz, Pawel and Lacey, Roy and Lynch, William G.",
    title = "{Determination of the equation of state of dense matter}",
    eprint = "nucl-th/0208016",
    archivePrefix = "arXiv",
    doi = "10.1126/science.1078070",
    journal = "Science",
    volume = "298",
    pages = "1592--1596",
    year = "2002"
}

@article{Suwa:2018uni,
    author = "Suwa, Yudai and Yoshida, Takashi and Shibata, Masaru and Umeda, Hideyuki and Takahashi, Koh",
    title = "{On the minimum mass of neutron stars}",
    eprint = "1808.02328",
    archivePrefix = "arXiv",
    primaryClass = "astro-ph.HE",
    reportNumber = "YITP-18-98",
    doi = "10.1093/mnras/sty2460",
    journal = "Mon. Not. Roy. Astron. Soc.",
    volume = "481",
    number = "3",
    pages = "3305--3312",
    year = "2018"
}

@article{Muller:2024aod,
    author = {M{\"u}ller, Bernhard and Heger, Alexander and Powell, Jade},
    title = "{Minimum Neutron Star Mass in Neutrino-Driven Supernova Explosions}",
    eprint = "2407.08407",
    archivePrefix = "arXiv",
    primaryClass = "astro-ph.HE",
    doi = "10.1103/PhysRevLett.134.071403",
    journal = "Phys. Rev. Lett.",
    volume = "134",
    number = "7",
    pages = "071403",
    year = "2025"
}

@article{DePietri:2019khb,
    author = "De Pietri, Roberto and Drago, Alessandro and Feo, Alessandra and Pagliara, Giuseppe and Pasquali, Michele and Traversi, Silvia and Wiktorowicz, Grzegorz",
    title = "{Merger of compact stars in the two-families scenario}",
    eprint = "1904.01545",
    archivePrefix = "arXiv",
    primaryClass = "astro-ph.HE",
    doi = "10.3847/1538-4357/ab2fd0",
    journal = "Astrophys. J.",
    volume = "881",
    number = "2",
    pages = "122",
    year = "2019"
}

@article{Fortin:2017dsj,
    author = "Fortin, M. and Oertel, M. and Provid{\^e}ncia, C.",
    title = "{Hyperons in hot dense matter: what do the constraints tell us for equation of state?}",
    eprint = "1711.09427",
    archivePrefix = "arXiv",
    primaryClass = "astro-ph.HE",
    doi = "10.1017/pasa.2018.32",
    journal = "Publ. Astron. Soc. Austral.",
    volume = "35",
    pages = "44",
    year = "2018"
}

@article{Lattimer:2012xj,
    author = "Lattimer, James M. and Lim, Yeunhwan",
    title = "{Constraining the Symmetry Parameters of the Nuclear Interaction}",
    eprint = "1203.4286",
    archivePrefix = "arXiv",
    primaryClass = "nucl-th",
    doi = "10.1088/0004-637X/771/1/51",
    journal = "Astrophys. J.",
    volume = "771",
    pages = "51",
    year = "2013"
}

@article{Passarella:2025zqb,
    author = "Passarella, Luca and Margueron, Jerome and Pagliara, Giuseppe",
    title = "{Relativistic mean-field predictions for the dense-matter equation~of state and application to neutron stars}",
    eprint = "2503.23028",
    archivePrefix = "arXiv",
    primaryClass = "nucl-th",
    doi = "10.1103/7qs4-wb95",
    journal = "Phys. Rev. C",
    volume = "112",
    number = "3",
    pages = "035805",
    year = "2025"
}

@article{Han:2018mtj,
    author = "Han, Sophia and Steiner, Andrew W.",
    title = "{Tidal deformability with sharp phase transitions in (binary) neutron stars}",
    eprint = "1810.10967",
    archivePrefix = "arXiv",
    primaryClass = "nucl-th",
    doi = "10.1103/PhysRevD.99.083014",
    journal = "Phys. Rev. D",
    volume = "99",
    number = "8",
    pages = "083014",
    year = "2019"
}

@dataset{doroshenko_2022_8232233,
  author       = {Doroshenko, Victor and
                  Suleimanov, Valery F. and
                  Pühlhofer, Gerd and
                  Santangelo, Andrea},
  title        = {MCMC samples for X-ray spectra fits summarised in
                   the paper "A strangely light neutron star"
                  },
  month        = jun,
  year         = 2022,
  publisher    = {Zenodo},
  version      = {V2},
  doi          = {10.5281/zenodo.8232233},
  url          = {https://doi.org/10.5281/zenodo.8232233},
}

@dataset{mauviard_2025_17380576,
  author       = {Mauviard, Lucien and
                  Guillot, Sebastien and
                  Salmi, Tuomo and
                  Choudhury, Devarshi and
                  Dorsman, Bas and
                  González-Caniulef, Denis and
                  Hoogkamer, Mariska and
                  Huppenkothen, Daniela and
                  Kazantsev, Christine and
                  Kini, Yves and
                  Olive, Jean-François and
                  Stammler, Pierre and
                  Watts, Anna and
                  Mendes, Melissa and
                  Rutherford, Nathan and
                  Schwenk, Achim and
                  Svensson, Isak and
                  Bogdanov, Slavko and
                  Kerr, Matthew and
                  Ray, Paul and
                  Guillemot, Lucas and
                  Cognard, Ismaël and
                  Theureau, Gilles},
  title        = {Data and Reproduction package for: "A NICER view
                   of the 1.4 solar-mass edge-on pulsar PSR
                   J0614-3329"
                  },
  month        = oct,
  year         = 2025,
  publisher    = {Zenodo},
  doi          = {10.5281/zenodo.17380576},
  url          = {https://doi.org/10.5281/zenodo.17380576},
}

@article{Antoniadis:2016hxz,
    author = "Antoniadis, John and Tauris, Thomas M. and Ozel, Feryal and Barr, Ewan and Champion, David J. and Freire, Paulo C. C.",
    title = "{The millisecond pulsar mass distribution: Evidence for bimodality and constraints on the maximum neutron star mass}",
    eprint = "1605.01665",
    archivePrefix = "arXiv",
    primaryClass = "astro-ph.HE",
    month = "5",
    year = "2016"
}

@article{Takatsy:2020bnx,
    author = "Tak{\'a}tsy, J{\'a}nos and Kov{\'a}cs, P{\'e}ter",
    title = "{Comment on ''Tidal Love numbers of neutron and self-bound quark stars''}",
    eprint = "2007.01139",
    archivePrefix = "arXiv",
    primaryClass = "astro-ph.HE",
    doi = "10.1103/PhysRevD.102.028501",
    journal = "Phys. Rev. D",
    volume = "102",
    number = "2",
    pages = "028501",
    year = "2020"
}

@article{Miao:2024qik,
    author = "Miao, Zhiqiang and Zhu, Zhenyu and Lai, Dong",
    title = "{Equation of State of Decompressed Quark Matter, and Observational Signatures of Quark-Star Mergers}",
    eprint = "2411.09013",
    archivePrefix = "arXiv",
    primaryClass = "astro-ph.HE",
    doi = "10.1103/zklh-27mr",
    journal = "Phys. Rev. Lett.",
    volume = "135",
    number = "9",
    pages = "091402",
    year = "2025"
}

@article{Ozel:2016oaf,
    author = {{\"O}zel, Feryal and Freire, Paulo},
    title = "{Masses, Radii, and the Equation of State of Neutron Stars}",
    eprint = "1603.02698",
    archivePrefix = "arXiv",
    primaryClass = "astro-ph.HE",
    doi = "10.1146/annurev-astro-081915-023322",
    journal = "Ann. Rev. Astron. Astrophys.",
    volume = "54",
    pages = "401--440",
    year = "2016"
}

@article{Cartaxo:2025jpi,
    author = "Cartaxo, Jo{\~a}o and Huang, Chun and Malik, Tuhin and Sourav, Shashwat and Yuan, Wen-Li and Zhou, Tianzhe and Liu, Xuezhi and Provid{\^e}ncia, Constan{\c{c}}a",
    title = "{Covariant Energy Density Functionals for Modeling the Equation of State of Neutron Star Matter: Cross-comparison Analysis Using CompactObject}",
    eprint = "2506.03112",
    archivePrefix = "arXiv",
    primaryClass = "nucl-th",
    reportNumber = "ET-0331A-25",
    doi = "10.3847/1538-4365/ae2310",
    journal = "Astrophys. J. Suppl.",
    volume = "282",
    number = "2",
    pages = "33",
    year = "2026"
}

@dataset{miller_2021_4670689,
  author       = {Miller, M.C. and
                  Lamb, F. K. and
                  Dittmann, A. J. and
                  Bogdanov, S. and
                  Arzoumanian, Z. and
                  Gendreau, K. C. and
                  Guillot, S. and
                  Ho, W. C. G. and
                  Lattimer, J. M. and
                  Morsink, S. M. and
                  Ray, P. S. and
                  Wolff, M. T. and
                  Baker, C. L. and
                  Cazeau, T. and
                  Manthripragada, S. and
                  Markwardt, C. B. and
                  Okajima, T. and
                  Pollard, S. and
                  Cognard, I. and
                  Cromartie, H. T. and
                  Fonseca, E. and
                  Guillemot, L. and
                  Kerr, M. and
                  Parthasarathy, A. and
                  Pennucci, T. T. and
                  Ransom, S. and
                  Stairs, I. and
                  Loewenstein, M.},
  title        = {NICER PSR J0740+6620 Illinois-Maryland MCMC
                   Samples
                  },
  month        = apr,
  year         = 2021,
  publisher    = {Zenodo},
  doi          = {10.5281/zenodo.4670689},
  url          = {https://doi.org/10.5281/zenodo.4670689},
}

@article{Brandes:2023hma,
    author = "Brandes, Len and Weise, Wolfram and Kaiser, Norbert",
    title = "{Evidence against a strong first-order phase transition in neutron star cores: Impact of new data}",
    eprint = "2306.06218",
    archivePrefix = "arXiv",
    primaryClass = "nucl-th",
    doi = "10.1103/PhysRevD.108.094014",
    journal = "Phys. Rev. D",
    volume = "108",
    number = "9",
    pages = "094014",
    year = "2023"
}

@dataset{miller_2019_3473466,
  author       = {Miller, M. C. and
                  Lamb, F. K. and
                  Dittmann, A. J. and
                  Bogdanov, S. and
                  Arzoumanian, Z. and
                  Gendreau, K. C. and
                  Guillot, S. and
                  Harding, A. K. and
                  Ho, W. C. G. and
                  Lattimer, J. M. and
                  Ludlam, R. M. and
                  Mahmoodifar, S. and
                  Morsink, S. M. and
                  Ray, P. S. and
                  Strohmayer, T. E. and
                  Wood, K. S. and
                  Enoto, T. and
                  Foster, R. and
                  Okajima, T. and
                  Prigozhin, G. and
                  Soong, Y.},
  title        = {NICER PSR J0030+0451 Illinois-Maryland MCMC
                   Samples
                  },
  month        = dec,
  year         = 2019,
  publisher    = {Zenodo},
  version      = {1.0.0},
  doi          = {10.5281/zenodo.3473466},
  url          = {https://doi.org/10.5281/zenodo.3473466},
}

@article{DiClemente:2022wqp,
    author = "Di Clemente, Francesco and Drago, Alessandro and Pagliara, Giuseppe",
    title = "{Is the Compact Object Associated with HESS J1731-347 a Strange Quark Star? A Possible Astrophysical Scenario for Its Formation}",
    eprint = "2211.07485",
    archivePrefix = "arXiv",
    primaryClass = "astro-ph.HE",
    doi = "10.3847/1538-4357/ad445b",
    journal = "Astrophys. J.",
    volume = "967",
    number = "2",
    pages = "159",
    year = "2024"
}

@article{Becerra:2025ryy,
    author = "Becerra, L. M. and Cipolletta, F. and Drago, A. and Guerrini, M. and Lavagno, A. and Pagliara, G. and Rueda, J. A.",
    title = "{On the formation of strange quark stars from supernova in compact binaries}",
    eprint = "2507.22033",
    archivePrefix = "arXiv",
    primaryClass = "astro-ph.HE",
    doi = "10.1016/j.jheap.2025.100491",
    journal = "JHEAp",
    volume = "50",
    pages = "100491",
    year = "2026"
}

@article{Raaijmakers:2021uju,
    author = "Raaijmakers, G. and Greif, S. K. and Hebeler, K. and Hinderer, T. and Nissanke, S. and Schwenk, A. and Riley, T. E. and Watts, A. L. and Lattimer, J. M. and Ho, W. C. G.",
    title = "{Constraints on the Dense Matter Equation of State and Neutron Star Properties from NICER{\textquoteright}s Mass{\textendash}Radius Estimate of PSR J0740+6620 and Multimessenger Observations}",
    eprint = "2105.06981",
    archivePrefix = "arXiv",
    primaryClass = "astro-ph.HE",
    doi = "10.3847/2041-8213/ac089a",
    journal = "Astrophys. J. Lett.",
    volume = "918",
    number = "2",
    pages = "L29",
    year = "2021"
}

@article{Zhao:2024gjz,
    author = "Zhao, Tianqi and Lin, Zidu and Kumar, Bharat and Steiner, Andrew W. and Prakash, Madappa",
    title = "{Characterizing the nuclear models informed by PREX and CREX: A view from Bayesian inference}",
    eprint = "2406.05267",
    archivePrefix = "arXiv",
    primaryClass = "nucl-th",
    doi = "10.1103/472x-9cxj",
    journal = "Phys. Rev. Res.",
    volume = "7",
    number = "4",
    pages = "043335",
    year = "2025"
}

\newpage
\clearpage

\clearpage
\onecolumngrid
\appendix

\section{Details of the equations of state}
\label{app:EOS_details}

This appendix collects the mean-field equations and the thermodynamic quantities used to construct the hadronic and quark EOS. 

\subsection{Hadronic sector}
\label{app:had_fields}
In uniform matter at the mean-field level, spatial derivatives vanish and only the time-like components of the vector fields survive. For baryons belonging to the octet and to the $\Delta$ quartet, the Euler--Lagrange equations obtained from Eq.~\eqref{eq:Lagrangian} reduce to
\begin{subequations}
\label{eq:fields}
\begin{align}
\omega_0
&=
\sum_i \frac{g_{\omega i}}{m_\omega^2}\,n_i,
\\
\phi_0
&=
\sum_i \frac{g_{\phi i}}{m_\phi^2}\,n_i,
\\
\rho_{03}
&=
\sum_i \frac{g_{\rho i}(n_B)}{m_\rho^2}\,I_{3i}\,n_i,
\\
m_\sigma^2\sigma
&=
- b\,m_N\,g_{\sigma N}(g_{\sigma N}\sigma)^2
- c\,g_{\sigma N}(g_{\sigma N}\sigma)^3
+ \sum_i g_{\sigma i}\,n_{s,i}.
\end{align}
\end{subequations}
Here the index $i$ runs only over baryons.
Leptons do not couple to the meson fields and therefore do not contribute to Eqs.~\eqref{eq:fields}; they enter only the charge-neutrality condition and the total thermodynamic quantities below.
The third component of isospin is denoted by $I_{3i}$, with eigenvalues $\pm1/2$ for the nucleon doublet and $+3/2,+1/2,-1/2,-3/2$ for $\Delta^{++},\Delta^+,\Delta^0,\Delta^-$.
The hidden-strangeness coupling satisfies $g_{\phi i}=0$ for non-strange baryons.

At zero temperature, the number and scalar densities of baryon $i$ are
\begin{align}
n_i
&=
\frac{2J_i+1}{6\pi^2}\,k_{F_i}^3,
\\
n_{s,i}
&=
\frac{2J_i+1}{4\pi^2}\,
m_i^*
\left[
k_{F_i}E_{F_i}^*
-
m_i^{*\,2}
\ln\!\left(\frac{k_{F_i}+E_{F_i}^*}{m_i^*}\right)
\right],
\label{eq:nsi}
\end{align}
where
\begin{equation}
E_{F_i}^*=\sqrt{k_{F_i}^2+m_i^{*\,2}},
\qquad
m_i^*=m_i-g_{\sigma i}\sigma .
\end{equation}
The spin degeneracy is $2J_i+1=2$ for the baryon octet and $2J_i+1=4$ for the $\Delta$ quartet.
The total baryon density is
\begin{equation}
n_B=\sum_i n_i .
\end{equation}

The single-particle energy of baryon $i$ is
\begin{equation}
\epsilon_i(k)
=
g_{\omega i}\omega_0
+
g_{\phi i}\phi_0
+
g_{\rho i}(n_B) I_{3i}\rho_{03}
+
\Sigma_r
+
\sqrt{k^2+m_i^{*\,2}},
\label{eq:ei}
\end{equation}
and its chemical potential is $\mu_i=\epsilon_i(k_{F_i})$.
The rearrangement contribution induced by the density dependence of the $\rho$ coupling is
\begin{equation}
\Sigma_r
=
\frac{\partial g_{\rho N}(n_B)}{\partial n_B}
\rho_{03}
\sum_j x_{\rho j}I_{3j}n_j,
\label{eq:Sigmar}
\end{equation}
where $g_{\rho j}=x_{\rho j}g_{\rho N}$, $x_{\rho N}=1$, and $x_{\rho\Delta}=1$ in the models considered here.
The term $\Sigma_r$ is required for thermodynamic consistency in the presence of density-dependent couplings.

In $\beta$-equilibrated stellar matter, the baryon chemical potentials are written as
\begin{equation}
\mu_i=\mu_B+C_i\mu_C,
\end{equation}
where $C_i$ is the electric charge of species $i$.
For leptons,
\begin{equation}
\mu_\ell=C_\ell\mu_C,
\qquad
\ell=e,\mu,
\end{equation}
so that $\mu_e=\mu_\mu=-\mu_C$ when muons are present.
Charge neutrality reads
\begin{equation}
\sum_i C_i n_i - n_e - n_\mu =0,
\label{eq:charge_neutrality}
\end{equation}
with $n_\mu=0$ whenever $\mu_e<m_\mu$.
For symmetric nuclear matter and pure neutron matter, the same mean-field equations are solved with the appropriate fixed isospin composition, and with the lepton and strange sectors switched off.

The baryonic contribution to the energy density is
\begin{equation}
\begin{split}
\varepsilon_B =&\
\tfrac{1}{2}m_\sigma^2\sigma^2
+ \tfrac{1}{3}b\,m_N(g_{\sigma N}\sigma)^3
+ \tfrac{1}{4}c\,(g_{\sigma N}\sigma)^4
\\
&+
\tfrac{1}{2}m_\omega^2\omega_0^2
+
\tfrac{1}{2}m_\rho^2\rho_{03}^2
+
\tfrac{1}{2}m_\phi^2\phi_0^2
\\
&+
\sum_i
\frac{2J_i+1}{16\pi^2}
\left[
k_{F_i}E_{F_i}^*
\left(2k_{F_i}^2+m_i^{*\,2}\right)
-
m_i^{*\,4}
\ln\!\left(\frac{k_{F_i}+E_{F_i}^*}{m_i^*}\right)
\right].
\end{split}
\label{eq:epsB}
\end{equation}

The baryonic pressure is
\begin{equation}
\begin{split}
P_B =&\
-\tfrac{1}{2}m_\sigma^2\sigma^2
- \tfrac{1}{3}b\,m_N(g_{\sigma N}\sigma)^3
- \tfrac{1}{4}c\,(g_{\sigma N}\sigma)^4
\\
&+
\tfrac{1}{2}m_\omega^2\omega_0^2
+
\tfrac{1}{2}m_\rho^2\rho_{03}^2
+
\tfrac{1}{2}m_\phi^2\phi_0^2
+
n_B\Sigma_r
\\
&+
\sum_i
\frac{2J_i+1}{48\pi^2}
\left[
k_{F_i}E_{F_i}^*
\left(2k_{F_i}^2-3m_i^{*\,2}\right)
+
3m_i^{*\,4}
\ln\!\left(\frac{k_{F_i}+E_{F_i}^*}{m_i^*}\right)
\right].
\end{split}
\label{eq:PB}
\end{equation}
The rearrangement contribution appears explicitly in the pressure, but not in the energy density.
Equivalently, Eq.~\eqref{eq:PB} follows from
\begin{equation}
P_B=\sum_i \mu_i n_i-\varepsilon_B
\end{equation}
once the chemical potentials include $\Sigma_r$.

For stellar matter, the total energy density and pressure also include the free degenerate lepton gas,
\begin{equation}
\varepsilon_H=\varepsilon_B+\sum_{\ell=e,\mu}\varepsilon_\ell,
\qquad
P_H=P_B+\sum_{\ell=e,\mu}P_\ell .
\label{eq:epsP_total_had}
\end{equation}
At zero temperature,
\begin{align}
n_\ell
&=
\frac{k_{F_\ell}^3}{3\pi^2},
\\
\varepsilon_\ell
&=
\frac{1}{8\pi^2}
\left[
k_{F_\ell}E_{F_\ell}
\left(2k_{F_\ell}^2+m_\ell^2\right)
-
m_\ell^4
\ln\!\left(\frac{k_{F_\ell}+E_{F_\ell}}{m_\ell}\right)
\right],
\\
P_\ell
&=
\frac{1}{24\pi^2}
\left[
k_{F_\ell}E_{F_\ell}
\left(2k_{F_\ell}^2-3m_\ell^2\right)
+
3m_\ell^4
\ln\!\left(\frac{k_{F_\ell}+E_{F_\ell}}{m_\ell}\right)
\right],
\end{align}
with $E_{F_\ell}=\sqrt{k_{F_\ell}^2+m_\ell^2}$ and $\mu_\ell=E_{F_\ell}$.
The Gibbs energy per baryon is
\begin{equation}
\mu_H\equiv\frac{\varepsilon_H+P_H}{n_B}.
\label{eq:mu_avg}
\end{equation}
For cold charge-neutral matter in $\beta$ equilibrium, Eq.~\eqref{eq:mu_avg} reduces to the baryon chemical potential $\mu_B$.

Finally, the nucleonic RMF couplings
\begin{equation}
\left(g_{\sigma N},g_{\omega N},g_{\rho N}(\nsat),b,c,a_\rho\right),
\end{equation}
are fixed by the six nuclear empirical parameters
\begin{equation}
\left(\nsat,E_\mathrm{sat},K_\mathrm{sat},m^*/m,E_\mathrm{sym},L_\mathrm{sym}\right)
\end{equation}
defined at saturation in symmetric nuclear matter using the analytic mapping reported in~\cite{Passarella:2025zqb}.

\subsection{Quark sector}
\label{app:quark}

The quark phase is described by the zero-temperature CFL bag-model parametrization of Eq.~\eqref{eq:OmegaCFL_expanded}, following Refs.~\cite{Alford:2007xm,Weissenborn:2011qu}.
The pressure is
\begin{equation}
P_Q(\mu_q)
=
-\Omega_Q(\mu_q)
=
\frac{3a_4}{4\pi^2}\mu_q^4
-
\frac{3}{4\pi^2}
\left(m_s^2-4\gap^2\right)\mu_q^2
-
B .
\label{eq:PQ}
\end{equation}
The quark number density is $n_q=\partial P_Q/\partial\mu_q$.
Therefore the baryon number density is
\begin{equation}
n_{B}(\mu_q)
=
\frac{1}{3}\frac{\partial P_Q}{\partial\mu_q}
=
\frac{a_4}{\pi^2}\mu_q^3
-
\frac{1}{2\pi^2}
\left(m_s^2-4\gap^2\right)\mu_q .
\label{eq:nQ}
\end{equation}
The energy density follows from the zero-temperature Euler relation,
\begin{equation}
\varepsilon_Q(\mu_q)
=
-P_Q(\mu_q)+3\mu_q n_{B}(\mu_q),
\label{eq:epsQ}
\end{equation}
or equivalently
\begin{equation}
\varepsilon_Q(\mu_q)
=
\frac{9a_4}{4\pi^2}\mu_q^4
-
\frac{3}{4\pi^2}
\left(m_s^2-4\gap^2\right)\mu_q^2
+
B .
\label{eq:epsQ_explicit}
\end{equation}
The Gibbs energy per baryon of the CFL phase (which corresponds to the baryon chemical potential) is
\begin{equation}
\mu_{Q}
\equiv
\frac{\varepsilon_Q+P_Q}{n_{B}}
=
3\mu_q .
\label{eq:muBQ}
\end{equation}
At the surface of a self-bound quark star, $P_Q=0$. The corresponding value $\mu_{Q}(P_Q=0)$ enters the Bodmer--Witten stability condition discussed in Sec.~\ref{sec:EOS_quark}.


\section{Derivation and details of the mass--radius likelihood}
\label{app:likelihood}

In this appendix we derive the likelihood contribution associated with a mass--radius observation for a single source $i$, both in the one-family (1F) and in the two-families (2F) scenarios. For a given source, the stellar mass $M$ and radius $R$ are not known exactly and therefore they play the role of latent variables. The source likelihood is obtained by marginalizing over them. In the 2F scenario, the family label is an additional latent variable and it must also be marginalized over.

The publicly available posterior samples used in this work are posterior samples in the mass--radius plane. From these samples we reconstruct, by means of kernel density estimation (KDE), a smooth density that we denote by
\begin{equation}
q_i(M,R).
\end{equation}
As discussed below, this reconstructed density is not, strictly speaking, identical to the likelihood $\prob(\datumi\mid M,R)$, but can be used as a proxy for it under suitable assumptions on the prior adopted in the original parameter-estimation analysis.

\subsection{One-family scenario}

In the 1F scenario, the observed object is assumed to belong to the single hadronic family. The likelihood contribution associated with the data set $\datumi$ is therefore
\begin{align}
\mathcal{L}_i^{(\mathrm{1F})}(\theta)
&\equiv
\prob(\datumi\mid \theta,\mathrm{1F})
\nonumber\\
&=
\int dM \int dR\,
\prob(\datumi,M,R\mid \theta,\mathrm{1F})
\nonumber\\
&=
\int dM \int dR\,
\prob(\datumi\mid M,R,\theta,\mathrm{1F})\,
\prob(M,R\mid \theta,\mathrm{1F}).
\label{eq:L1F_start_appendix}
\end{align}
This expression is obtained by marginalizing the joint probability over the unobserved source parameters $M$ and $R$.

We assume that, once the source mass and radius are specified, the observational model is fully determined. Under this assumption,
\begin{equation}
\prob(\datumi\mid M,R,\theta,\mathrm{1F})=\prob(\datumi\mid M,R),
\end{equation}
so that Eq.~\eqref{eq:L1F_start_appendix} becomes
\begin{equation}
\mathcal{L}_i^{(\mathrm{1F})}(\theta)
=
\int dM \int dR\,
\prob(\datumi\mid M,R)\,
\prob(M,R\mid \theta,\mathrm{1F}).
\label{eq:L1F_data_term_appendix}
\end{equation}

For a fixed EOS parameter set $\theta$, the TOV equations determine a unique radius for each stellar mass along the stable hadronic branch,
\begin{equation}
R=R_{\rm HS}^{\rm TOV}(M;\theta).
\end{equation}
Hence the EOS-induced probability distribution in the $(M,R)$ plane has support only on the corresponding TOV curve and can be written as
\begin{equation}
\prob(M,R\mid \theta,\mathrm{1F})
=
w_{\rm HS}^{(\mathrm{1F})}(M;\theta)\,
\delta\!\left[R-R_{\rm HS}^{\rm TOV}(M;\theta)\right],
\label{eq:MRdist_1F_appendix}
\end{equation}
where
\begin{equation}
w_{\rm HS}^{(\mathrm{1F})}(M;\theta)\equiv \prob(M\mid \theta,\mathrm{1F})
\end{equation}
is the prior weight assigned to the stellar mass along the stable hadronic branch. Substituting Eq.~\eqref{eq:MRdist_1F_appendix} into Eq.~\eqref{eq:L1F_data_term_appendix} yields
\begin{align}
\mathcal{L}_i^{(\mathrm{1F})}(\theta)
&=
\int dM \int dR\,
\prob(\datumi\mid M,R)\,
w_{\rm HS}^{(\mathrm{1F})}(M;\theta)\,
\delta\!\left[R-R_{\rm HS}^{\rm TOV}(M;\theta)\right]
\nonumber\\
&=
\int dM\,
\prob\!\left[\datumi\mid M,R_{\rm HS}^{\rm TOV}(M;\theta)\right]\,
w_{\rm HS}^{(\mathrm{1F})}(M;\theta).
\label{eq:L1F_general_appendix}
\end{align}
The marginalization over $R$ restricts the likelihood to the theoretical mass--radius curve predicted by the EOS, while the marginalization over $M$ remains because the stellar mass is not known exactly.

A simple and commonly used choice is a constant prior on the mass within an interval. Up to normalization, this can be written as
\begin{equation}
w_{\rm HS}^{(\mathrm{1F})}(M;\theta)
\propto
\mathbf{1}\!\left(M_{\rm HS,min}^{\rm (1F)}(\theta)\le M\le M_{\rm HS,max}^{\rm (1F)}(\theta)\right),
\label{eq:flat_mass_prior_1F_appendix}
\end{equation}
where $\mathbf{1}(\cdots)$ denotes the indicator function. The normalized form would contain the factor
\begin{equation}
\left[M_{\rm HS,max}^{\rm (1F)}(\theta)-M_{\rm HS,min}^{\rm (1F)}(\theta)\right]^{-1}.
\end{equation}
Equation~\eqref{eq:L1F_general_appendix} then reduces to
\begin{equation}
\mathcal{L}_i^{(\mathrm{1F})}(\theta)
\propto
\int_{M_{\rm HS,min}^{\rm (1F)}(\theta)}^{M_{\rm HS,max}^{\rm (1F)}(\theta)} dM\,
\prob\!\left[\datumi\mid M,R_{\rm HS}^{\rm TOV}(M;\theta)\right].
\label{eq:L1F_flat_appendix}
\end{equation}

We now relate this formal expression to the reconstructed density $q_i(M,R)$. Let $I$ collectively denote the assumptions adopted in the original parameter-estimation analysis underlying the released samples, including the prior in the $(M,R)$ plane. Bayes' theorem then implies
\begin{equation}
\prob(M,R\mid \datumi,I)
=
\frac{\prob(\datumi\mid M,R)\,\pi_I(M,R)}{Z_i^I},
\label{eq:posterior_MR_appendix}
\end{equation}
where $\pi_I(M,R)$ is the prior used in the original analysis and
\begin{equation}
Z_i^I\equiv \prob(\datumi\mid I)
\end{equation}
is the corresponding evidence. Therefore,
\begin{equation}
\prob(\datumi\mid M,R)
=
Z_i^I\,
\frac{\prob(M,R\mid \datumi,I)}{\pi_I(M,R)}
\propto
\frac{\prob(M,R\mid \datumi,I)}{\pi_I(M,R)}.
\label{eq:q_from_post_appendix}
\end{equation}
The KDE-reconstructed density $q_i(M,R)$ is our approximation to the released posterior $\prob(M,R\mid \datumi,I)$. If the prior $\pi_I(M,R)$ is constant, or sufficiently slowly varying, over the relevant region of the $(M,R)$ plane, then
\begin{equation}
\prob(\datumi\mid M,R)\propto q_i(M,R).
\label{eq:flat_prior_approx_appendix}
\end{equation}
Under this approximation, Eq.~\eqref{eq:L1F_flat_appendix} becomes
\begin{equation}
\mathcal{L}_i^{(\mathrm{1F})}(\theta)
\propto
\int_{M_{\rm HS,min}^{\rm (1F)}(\theta)}^{M_{\rm HS,max}^{\rm (1F)}(\theta)} dM\,
q_i\!\left[M,R_{\rm HS}^{\rm TOV}(M;\theta)\right],
\label{eq:L1F_post_appendix}
\end{equation}
which is the form used in the main text.

In practice, the quality of the approximation $q_i(M,R)\propto \prob(\datumi\mid M,R)$ depends on the priors and assumptions adopted in the original parameter-estimation analysis. For the standard NICER mass--radius products of PSR~J0030+0451 and PSR~J0740+6620, this identification is reasonably well justified because the original analyses employ a jointly uniform prior in the $(M,R)$ plane within the prior support \cite{Raaijmakers:2021uju}. 
Although these NICER products are not model independent, since they still rely on pulse-profile modeling assumptions, they remain comparatively clean in the variables of interest.
By contrast, the posterior for PSR~J0614--3329 is more prior-informed along the mass direction because the inferred mass is essentially constrained by MeerKAT radio-timing priors \cite{Mauviard:2025dmd}. The baseline HESS~J1731--347 samples are obtained from spectral fits with a carbon-atmosphere model and Gaia parallax information; while the baseline analysis adopts flat priors on mass and radius, the inferred $(M,R)$ posterior is more model dependent than the standard NICER products because it also relies on specific spectral-atmosphere assumptions and distance priors \cite{Doroshenko:2022nwp}. Finally, the GW170817 source-properties samples are not mass--radius products but posteriors in the binary-parameter space $(M_1,M_2,\Lambda_1,\Lambda_2)$; in that space the priors are relatively simple, with $\Lambda_i$ taken to be uniform in the range $0\le \Lambda_i \le 5000$ and the source masses effectively uniform over the region relevant to the analysis, but the result still depends on the adopted waveform model and spin prior \cite{LIGOScientific:2018hze}. Accordingly, for PSR~J0614--3329, HESS~J1731--347, and GW170817 the reconstructed densities should be interpreted more cautiously as approximate, prior-informed compatibility densities rather than exact likelihoods.

\subsection{Two-families scenario}

In the 2F scenario, the family of the observed object is not known a priori. The family label is therefore an additional latent variable and must be marginalized over together with $M$ and $R$. The source likelihood is then
\begin{align}
\mathcal{L}_i^{(\mathrm{2F})}(\theta)
&\equiv
\prob(\datumi\mid \theta,\mathrm{2F})
\nonumber\\
&=
\sum_{f\in\{{\rm HS},{\rm QS}\}}
\int dM \int dR\,
\prob(\datumi,M,R,f\mid \theta,\mathrm{2F})
\nonumber\\
&=
\sum_{f\in\{{\rm HS},{\rm QS}\}}
\int dM \int dR\,
\prob(\datumi\mid M,R,f,\theta,\mathrm{2F})\,
\prob(M,R,f\mid \theta,\mathrm{2F}).
\label{eq:L2F_start_appendix}
\end{align}

The released observational product reconstructed by $q_i(M,R)$ constrains the source only in the $(M,R)$ plane and does not include any family label. We therefore assume that, once $M$ and $R$ are fixed, the observational information used in the present analysis does not further distinguish between the HS and QS interpretations. Under this assumption,
\begin{equation}
\prob(\datumi\mid M,R,f,\theta,\mathrm{2F})=\prob(\datumi\mid M,R).
\label{eq:conditional_independence_2F_appendix}
\end{equation}
This does not imply that the family label is determined by $(M,R)$: a HS and a QS may in principle correspond to the same values of $M$ and $R$. Rather, they are treated here as observationally degenerate at fixed $(M,R)$, and are distinguished only through their prior weights in the 2F model.

Moreover, for each branch $f\in\{{\rm HS},{\rm QS}\}$ the EOS predicts a deterministic TOV curve,
\begin{equation}
R=R_f^{\rm TOV}(M;\theta),
\end{equation}
so that
\begin{equation}
\prob(M,R,f\mid \theta,\mathrm{2F})
=
w_f^{(\mathrm{2F})}(M;\theta)\,
\delta\!\left[R-R_f^{\rm TOV}(M;\theta)\right],
\label{eq:MRdist_2F_appendix}
\end{equation}
where
\begin{equation}
w_f^{(\mathrm{2F})}(M;\theta)\equiv \prob(M,f\mid \theta,\mathrm{2F})
\end{equation}
is the joint prior weight assigned to mass and family. Substituting Eq.~\eqref{eq:MRdist_2F_appendix} into Eq.~\eqref{eq:L2F_start_appendix} gives
\begin{align}
\mathcal{L}_i^{(\mathrm{2F})}(\theta)
&=
\sum_{f\in\{{\rm HS},{\rm QS}\}}
\int dM \int dR\,
\prob(\datumi\mid M,R)\,
w_f^{(\mathrm{2F})}(M;\theta)\,
\delta\!\left[R-R_f^{\rm TOV}(M;\theta)\right]
\nonumber\\
&=
\sum_{f\in\{{\rm HS},{\rm QS}\}}
\int dM\,
\prob\!\left[\datumi\mid M,R_f^{\rm TOV}(M;\theta)\right]\,
w_f^{(\mathrm{2F})}(M;\theta).
\label{eq:L2F_general_appendix}
\end{align}

Equation~\eqref{eq:L2F_general_appendix} shows that the total likelihood in the 2F scenario is a \emph{sum} of the HS and QS contributions. This does \emph{not} mean that a given star is simultaneously a HS and a QS. Physically, each source belongs to one family only. The sum appears because the family assignment is not known a priori from the observation and must therefore be marginalized over. The two terms in Eq.~\eqref{eq:L2F_general_appendix} reflect our uncertainty about the family assignment of the source, not a physical coexistence of the two possibilities.

To recover the form used in the main text, we write, up to normalization,
\begin{equation}
w_f^{(\mathrm{2F})}(M;\theta)
\propto
\eta_f^{(\mathrm{2F})}(M;\theta)\,
\mathbf{1}\!\left(M_{f,\rm min}^{\rm (2F)}(\theta)\le M\le M_{f,\rm max}^{\rm (2F)}(\theta)\right),
\label{eq:wf_eta_appendix}
\end{equation}
where
\begin{equation}
\eta_f^{(\mathrm{2F})}(M;\theta)\equiv \prob(f\mid M,\theta,\mathrm{2F}),
\qquad
f\in\{{\rm HS},{\rm QS}\},
\end{equation}
encodes the relative population weight of each family at fixed mass. Equation~\eqref{eq:L2F_general_appendix} then becomes
\begin{equation}
\mathcal{L}_i^{(\mathrm{2F})}(\theta)
\propto
\sum_{f\in\{{\rm HS},{\rm QS}\}}
\int_{M_{f,\rm min}^{\rm (2F)}(\theta)}^{M_{f,\rm max}^{\rm (2F)}(\theta)} dM\,
\prob\!\left[\datumi\mid M,R_f^{\rm TOV}(M;\theta)\right]\,
\eta_f^{(\mathrm{2F})}(M;\theta).
\label{eq:L2F_flat_appendix}
\end{equation}

Using Eq.~\eqref{eq:flat_prior_approx_appendix}, this becomes
\begin{equation}
\mathcal{L}_i^{(\mathrm{2F})}(\theta)
\propto
\sum_{f\in\{{\rm HS},{\rm QS}\}}
\int_{M_{f,\rm min}^{\rm (2F)}(\theta)}^{M_{f,\rm max}^{\rm (2F)}(\theta)} dM\,
q_i\!\left[M,R_f^{\rm TOV}(M;\theta)\right]\,
\eta_f^{(\mathrm{2F})}(M;\theta),
\label{eq:L2F_post_appendix}
\end{equation}
which is the form used in the main text.

\section{Low-density and heavy-ion likelihoods}
\label{app:chieft_hic}

In this appendix we collect the explicit expressions used to implement the low-density $\chi$EFT input and the heavy-ion collision (HIC) constraint in the Bayesian likelihood. In both cases, we use \emph{soft} likelihood factors rather than hard cuts: EOSs compatible with the adopted reference bands are left unpenalized, whereas EOSs lying outside the bands are smoothly suppressed according to their distance from the corresponding boundaries. 
The $\chi$EFT and HIC inputs constrain different sectors of the hadronic EOS and are therefore complementary. The $\chi$EFT likelihood constrains the low-density neutron-rich regime through the energy per baryon of cold pure neutron matter, whereas the HIC likelihood constrains the pressure of nearly symmetric matter at supra-saturation density. In the full Bayesian analysis, these two likelihood factors are multiplied together with the astrophysical likelihood contributions described in the main text.

\subsection{Implementation of the $\chi$EFT likelihood}
\label{app:chieft_details}

At low baryon density, microscopic calculations based on chiral effective field theory ($\chi$EFT) provide a benchmark for the hadronic EOS of cold pure neutron matter (PNM). In the present work, following Ref.~\cite{Passarella:2025zqb}, we use exactly the same discretized $\chi$EFT band adopted there. More precisely, we consider the set of density points
\begin{equation}
\{n_B^k\}_{k=1,\dots,N_{\chi{\rm EFT}}},
\end{equation}
together with the corresponding reference values $\overline{E}_{\chi{\rm EFT}}^k$ and uncertainties $\sigma_{\chi{\rm EFT}}^k$ for the energy per baryon in cold PNM, as reported in Table II of Ref.~\cite{Passarella:2025zqb}.

For a given hadronic EOS parameter set $\theta$, we evaluate the energy per baryon in cold pure neutron matter as
\begin{equation}
E_H^{\rm PNM}(n_B^k;\theta)
\equiv
\frac{\varepsilon_H(n_B^k,Y_C=0,Y_S=0,Y_L=0,T=0;\theta)}{n_B^k}.
\label{eq:app_Epnm}
\end{equation}
Here $\varepsilon_H$ is the hadronic energy density computed from the RMF EOS for homogeneous matter with zero temperature, zero strangeness fraction, zero lepton fraction, and vanishing charge fraction, i.e.\ pure neutron matter.

We then quantify the distance of the EOS prediction from the adopted $\chi$EFT band through the dimensionless variable
\begin{equation}
d_{\chi{\rm EFT}}^k(\theta)
\equiv
\frac{
\left|E_H^{\rm PNM}(n_B^k;\theta)-\overline{E}_{\chi{\rm EFT}}^k\right|
}{
\sigma_{\chi{\rm EFT}}^k
}.
\label{eq:app_d_chieft}
\end{equation}
By construction, $d_{\chi{\rm EFT}}^k \le 1$ means that the model prediction lies inside the adopted $\chi$EFT band at the density point $n_B^k$, while $d_{\chi{\rm EFT}}^k > 1$ means that it lies outside the band.

The pointwise $\chi$EFT likelihood factor is defined as
\begin{equation}
\ell_{\chi{\rm EFT}}^k(\theta)=
\begin{cases}
1, & d_{\chi{\rm EFT}}^k(\theta)\le 1,\\[0.4em]
\exp\!\left[-\dfrac{1}{2}\left(d_{\chi{\rm EFT}}^k(\theta)-1\right)^2\right],
& d_{\chi{\rm EFT}}^k(\theta)>1.
\end{cases}
\label{eq:app_l_chieft}
\end{equation}
Thus, EOSs lying inside the band are not penalized, whereas EOSs outside the band are suppressed with a Gaussian penalty that depends on the distance from the band boundary.

The full $\chi$EFT likelihood is then taken as the product over all sampled density points,
\begin{equation}
\mathcal{L}_{\chi{\rm EFT}}(\theta)
=
\prod_{k=1}^{N_{\chi{\rm EFT}}}
\ell_{\chi{\rm EFT}}^k(\theta).
\label{eq:app_L_chieft_prod}
\end{equation}
Equivalently,
\begin{equation}
\mathcal{L}_{\chi{\rm EFT}}(\theta)
=
\exp\!\left[
-\frac{1}{2}
\sum_{k=1}^{N_{\chi{\rm EFT}}}
\left(
\max\!\left\{0,d_{\chi{\rm EFT}}^k(\theta)-1\right\}
\right)^2
\right].
\label{eq:app_L_chieft}
\end{equation}

This implementation therefore leaves unpenalized all EOSs that remain inside the adopted $\chi$EFT band over the full sampled density interval, while smoothly down-weighting those that deviate from it.

\subsection{Implementation of the heavy-ion likelihood}
\label{app:HIC_details}

Complementary information on the hadronic EOS at supra-saturation density is provided by collective-flow observables in intermediate-energy heavy-ion collisions. In the present work, we use the pressure band extracted in Ref.~\cite{Danielewicz:2002pu} from transverse and elliptic flow data as a phenomenological constraint on cold symmetric nuclear matter (SNM).

Since Ref.~\cite{Danielewicz:2002pu} provides the allowed region only graphically in the $(n_B,P)$ plane, the HIC input is implemented by digitizing the original pressure band from that paper. More precisely, we extract from the published figure a discrete set of points along the lower and upper boundaries of the band and use them to reconstruct the functions
\begin{equation}
P_{{\rm HIC,low}}(n_B), \qquad P_{{\rm HIC,up}}(n_B).
\end{equation}
In practice, we then evaluate these reconstructed boundaries on a common discrete density grid
\begin{equation}
\{n_B^k\}_{k=1,\dots,N_{\rm HIC}},
\end{equation}
restricted to the interval
\begin{equation}
n_B \le 3\nsat,
\end{equation}
thus obtaining the corresponding lower and upper pressure values
\begin{equation}
P_{{\rm HIC,low}}^k \equiv P_{{\rm HIC,low}}(n_B^k), \qquad
P_{{\rm HIC,up}}^k \equiv P_{{\rm HIC,up}}(n_B^k).
\end{equation}
We do not extend the constraint beyond $3\nsat$, since at larger densities the purely nucleonic interpretation underlying the flow extraction becomes increasingly questionable in the present hadronic model, where hyperons and $\Delta$ resonances may already appear in the $(2$--$3)\nsat$ range.

For a given parameter set $\theta$, we evaluate the hadronic pressure in cold symmetric nuclear matter as
\begin{equation}
P_H^{\rm SNM}(n_B;\theta)
\equiv
P_H(n_B,Y_C=0.5,Y_S=0,Y_L=0,T=0;\theta),
\label{eq:app_Psnm}
\end{equation}
where $Y_C=0.5$ corresponds to symmetric matter, with zero strangeness, no leptons, and zero temperature.

To quantify violations of the lower and upper boundaries of the flow band, we define the positive deviations
\begin{equation}
d_{{\rm HIC,low}}^k(\theta)
\equiv
\max\!\left[
0,\,
P_{{\rm HIC,low}}^k
-
P_H^{\rm SNM}(n_B^k;\theta)
\right],
\label{eq:app_d_HIC_low}
\end{equation}
\begin{equation}
d_{{\rm HIC,up}}^k(\theta)
\equiv
\max\!\left[
0,\,
P_H^{\rm SNM}(n_B^k;\theta)
-
P_{{\rm HIC,up}}^k
\right].
\label{eq:app_d_HIC_up}
\end{equation}
By construction, $d_{{\rm HIC,low}}^k$ is nonzero only when the EOS lies below the lower boundary of the allowed band, while $d_{{\rm HIC,up}}^k$ is nonzero only when the EOS lies above the upper boundary.

To account phenomenologically for the residual model dependence of the transport extraction, we associate to each boundary a tolerance proportional to its pressure value,
\begin{equation}
\sigma_{{\rm HIC,low}}^k
=
\alpha\,P_{{\rm HIC,low}}^k,
\qquad
\sigma_{{\rm HIC,up}}^k
=
\alpha\,P_{{\rm HIC,up}}^k,
\label{eq:app_sigma_HIC}
\end{equation}
and in the present work we adopt
\begin{equation}
\alpha = 0.15.
\end{equation}
This choice is meant to encode, at a phenomenological level, the residual systematic uncertainty associated with the transport-model dependence of the extracted flow band.

We then define the pointwise likelihood factors
\begin{equation}
\ell_{{\rm HIC,low}}^k(\theta)
=
\exp\!\left[
-\frac{1}{2}
\left(
\frac{d_{{\rm HIC,low}}^k(\theta)}
{\sigma_{{\rm HIC,low}}^k}
\right)^2
\right],
\label{eq:app_l_HIC_low}
\end{equation}
\begin{equation}
\ell_{{\rm HIC,up}}^k(\theta)
=
\exp\!\left[
-\frac{1}{2}
\left(
\frac{d_{{\rm HIC,up}}^k(\theta)}
{\sigma_{{\rm HIC,up}}^k}
\right)^2
\right].
\label{eq:app_l_HIC_up}
\end{equation}
These factors are equal to unity when the EOS lies inside the HIC band and decrease smoothly when the EOS violates either boundary.

The full heavy-ion likelihood is finally defined as the product over all density points of the lower- and upper-bound contributions,
\begin{equation}
\mathcal{L}_{\rm HIC}(\theta)
=
\prod_{k=1}^{N_{\rm HIC}}
\ell_{{\rm HIC,low}}^k(\theta)\,
\ell_{{\rm HIC,up}}^k(\theta).
\label{eq:app_L_HIC_prod}
\end{equation}
Equivalently,
\begin{equation}
\mathcal{L}_{\rm HIC}(\theta)
=
\exp\!\left[
-\frac{1}{2}
\sum_{k=1}^{N_{\rm HIC}}
\left(
\frac{d_{{\rm HIC,low}}^k(\theta)}
{\sigma_{{\rm HIC,low}}^k}
\right)^2
-
\frac{1}{2}
\sum_{k=1}^{N_{\rm HIC}}
\left(
\frac{d_{{\rm HIC,up}}^k(\theta)}
{\sigma_{{\rm HIC,up}}^k}
\right)^2
\right].
\label{eq:app_L_HIC}
\end{equation}

\section{Numerical details}
\label{app:numerics}

Evaluating the continuous two-dimensional integral in Eqs.~(\ref{eq:L_GW170817_1F},~\ref{eq:L_GW170817_2F_compact}) over a finely resolved mass grid at every stochastic sampling step is computationally prohibitive. Instead, we implement a highly optimized discrete Monte Carlo integration scheme that bypasses the construction of a continuous four-dimensional probability surface. 
Given $N$ discrete posterior samples $\{M_1^{(k)}, M_2^{(k)}, \Lambda_1^{(k)}, \Lambda_2^{(k)}\}_{k=1}^N$ provided by the posterior samples of GW170817 \cite{gw170817data}, the expected marginal likelihood is approximated by evaluating a two-dimensional Gaussian kernel directly at the sample nodes. For a specific EOS and parameter set $\theta$, the theoretical tidal deformabilities $\Lambda_{1,c}^{\rm TOV}$ and $\Lambda_{2,c}^{\rm TOV}$ are interpolated exactly at the discrete sample masses $M_1^{(k)}$ and $M_2^{(k)}$. The total numerical log-likelihood $\ln \mathcal{L}_{\rm GW170817}(\theta)$ is then computed as:
\begin{equation}
\begin{split}
    \ln \mathcal{L}_{\rm GW170817}(\theta) &\approx \ln \Biggl( \sum_{k=1}^N \sum_{c \in \mathcal{C}} 
     \exp \left[\ln W_c^{(k)}(\theta)
    + \ln K_c^{(k)}(\theta) \right] \Biggr)- \ln N,
\end{split}
\label{eq:L_GW170817_MC}
\end{equation}
where $\mathcal{C} = \{{\rm HS\text{-}HS}, {\rm HS\text{-}QS}, {\rm QS\text{-}HS}, {\rm QS\text{-}QS}\}$ represents the set of possible binary phase configurations, and $W_c^{(k)}(\theta)$ is the normalized prior weight for configuration $c$, which includes the phase assignment prior Eq.~\eqref{eq:eta_norm} and the mass prior Eq.~\eqref{eq:proper_norm}: 
\begin{equation}
    W_c^{(k)}(\theta) = \eta_{f_1-f_2}^{(2F)}(M_1^{(k)}, M_2^{(k)}; \theta) \times \frac{\mathbf{1}[M_{f_1,min}^{(2F)} (\theta)\le M_1^{(k)} \le M_{f_1,max}^{(2F)}(\theta)]}{M_{f_1,\rm max}^{(2F)}(\theta)-M_{f_1,\rm min}^{(2F)}(\theta)} \times \frac{\mathbf{1}[M_{f_2,\rm min}^{(2F)}(\theta) \le M_2^{(k)} \le M_{f_2,\rm max}^{(2F)}(\theta)]}{M_{f_2,\rm max}^{(2F)}(\theta)-M_{f_2,\rm min}^{(2F)}(\theta)}.
\end{equation}
The log-kernel $\ln K_c^{(k)}(\theta)$ isolates the tidal deformability deviations:
\begin{align}
    \ln K_c^{(k)}(\theta) =& -\frac{\left(\Lambda_{1,c}^{\rm TOV}(M_1^{(k)};\theta) - \Lambda_1^{(k)}\right)^2}{2\sigma_{\Lambda_1}^2}-\frac{\left(\Lambda_{2,c}^{\rm TOV}(M_2^{(k)};\theta) - \Lambda_2^{(k)}\right)^2}{2\sigma_{\Lambda_2}^2} - \ln(2\pi\sigma_{\Lambda_1}\sigma_{\Lambda_2}).
    \label{eq:GW_log_kernel}
\end{align}
The kernel bandwidths $\sigma_{\Lambda_i}$ are fixed dynamically using the standard deviation of the empirical samples scaled by $N^{-1/6}$. 
This discrete approach preserves the complex physical correlations inherent in the joint GW170817 posterior while maximizing computational throughput, and it has been used in recent multi-messenger EOS inferences (see, e.g., \cite{Huth:2021bsp,Raaijmakers:2021uju,Dietrich:2020efo,Cartaxo:2025jpi}).
To ensure rigorous numerical stability across the highly varied posterior landscape, the summation in Eq.~\eqref{eq:L_GW170817_MC} is executed in logarithmic space utilizing the \texttt{logsumexp} operation. 
\end{document}